\documentclass[12pt]{article}
\usepackage{amsmath}
\usepackage{amssymb}
\usepackage{amsthm}
\usepackage{amsfonts}
\usepackage{braket}
\usepackage[toc,page]{appendix}
\usepackage[utf8]{inputenc}
\usepackage{lipsum}
\usepackage{physics}
\usepackage{hyperref}
\usepackage{esint}
\usepackage{graphicx}
\usepackage{tgtermes} 
\usepackage{setspace} 

\usepackage{fancyhdr}
\usepackage{lipsum}

\usepackage[a4paper,
            bindingoffset=0.2in,
            left=1.5in,
            right=1in,
            top=1in,
            bottom=1in,
            footskip=.25in]{geometry}
\graphicspath{ {./Pic/} }
\usepackage[style=authoryear, backend=bibtex]{biblatex} 
\begin{document}

\begin{titlepage}

INTRODUCTION TO KOOPMAN-VON NEUMANN MECHANICS
\newline\newline
\centerline{AN HONORS THESIS}
\centerline{SUBMITTED ON THE FIFTH DAY OF DECEMBER, 2021}
\centerline{TO THE DEPARTMENT OF PHYSICS AND ENGINEERING PHYSICS}
\centerline{IN PARTIAL FULFILLMENT OF THE REQUIREMENTS}
\centerline{OF THE HONORS PROGRAM}
\centerline{OF NEWCOMB-TULANE COLLEGE}
\centerline{TULANE UNIVERSITY}
\centerline{FOR THE DEGREE OF}
\centerline{BACHELOR OF SCIENCES}
\centerline{WITH HONORS IN PHYSICS}
\centerline{BY}
\centerline{\rule{5cm}{0.15mm}}
\centerline{Daniel W. Piasecki}
\centerline{APPROVED:}            
\newline\newline
\rightline{\rule{5cm}{0.15mm}}
\rightline{Denys I. Bondar, Ph.D.}
\rightline{Director of Thesis}
\rightline{\rule{5cm}{0.15mm}}
\rightline{Ryan T. Glasser, Ph.D.}
\rightline{Second Reader}
\rightline{\rule{5cm}{0.15mm}}
\rightline{Frank J. Tipler, Ph.D}
\rightline{Third Reader}
\end{titlepage}

\newpage
\begin{figure}[!htb]
\centering
\makebox[\textwidth]{\includegraphics[width=0.9\paperwidth]{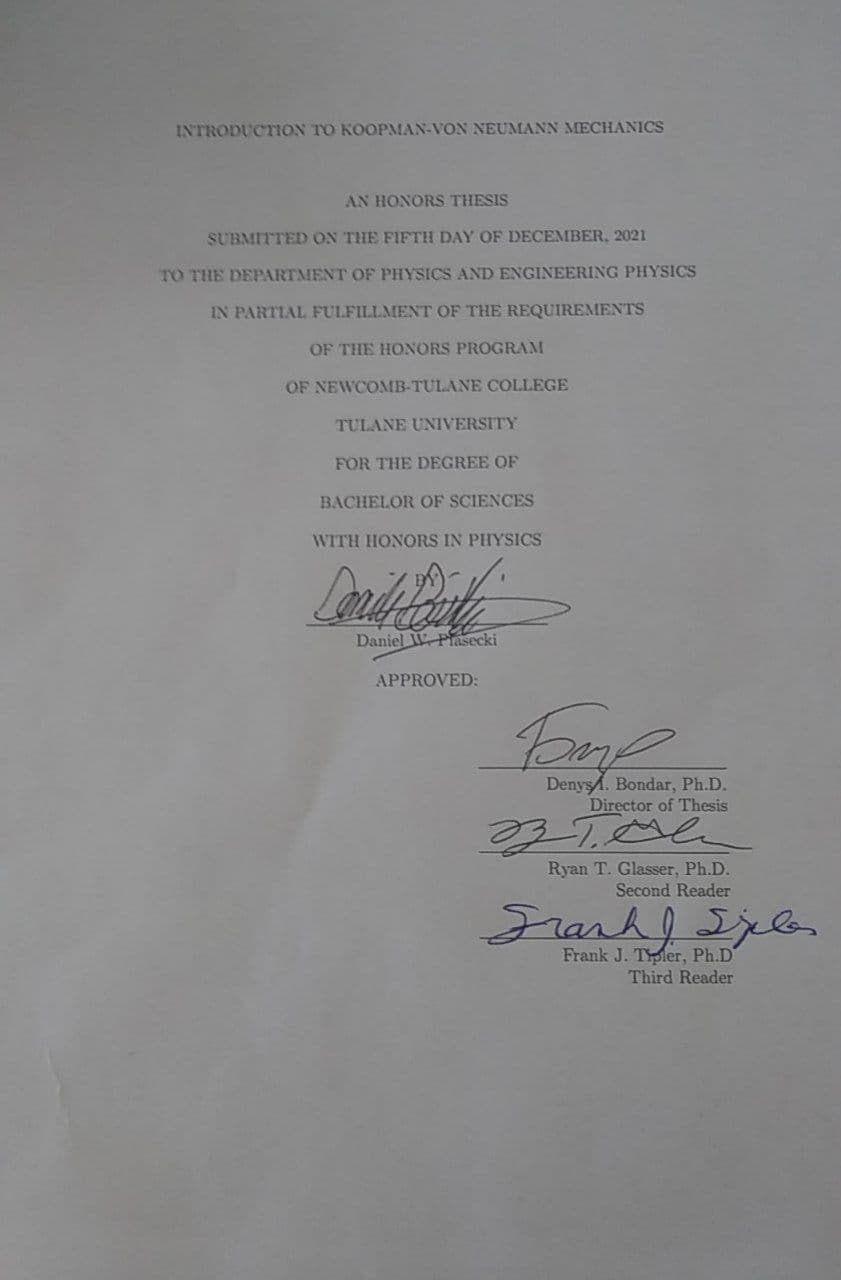}}
\end{figure}
\newpage
\newpage
\thispagestyle{empty}
\mbox{}
\newpage
\renewcommand{\thepage}{\roman{page}}
\setcounter{page}{2}

\begin{abstract}
This work is to consolidate current literature on Koopman-von Neumann (KvN) Mechanics into a simple and easy to understand text. KvN Mechanics is a branch of Classical Mechanics that has been recast into the mathematical language of Quantum Mechanics. KvN Mechanics utilizes a Hilbert phase space with operators to calculate the expectation values of observables of interest (expectation values such as position, momentum, etc.) It is an important tool in statistical physics and guide to illuminate the mysterious relationship between Quantum and Classical Physics. A lot of important applications of KvN Mechanics have been developed in the last decades. 
\end{abstract}
\newpage
\newpage

\section*{Acknowledgements}
Thanksgiving is approaching at the time I’m writing this, so there are some people who will get some messages of gratitude from me a few days early.

First, I would like to thank the Tulane Summer Research Program and the Tulane Honors Program. This project would not have been possible without their support.

Many thanks to the readers of this thesis: Dr. Ryan Glasser, Dr. Frank Tipler, and especially my advisor for this project, Dr. Denys Bondar. I am grateful for his advice on this paper and guidance for how I can take what I’ve learned from it and apply it to the rest of my career, both in terms of new knowledge about the subject matter and skills I’ve been practicing while working on this. His openness and directness in advising me made a big difference, and I’ve learned a lot from working on this project. Also, 80 pages is a lot to read, so I am grateful for the time all three of these professors have dedicated to reading my work.
\newline\newline

******************************************************************
\newline\newline

I am also very appreciative to my friends and family for their encouragement, prayers, and patience with me. This thesis was a huge commitment for me, and they were respectful of my time when I was busy and curious and uplifting during my few moments of downtime. One of them also helped look over some of it and did some proofreading, which I also am thankful for.
Thanks goes especially to my sweetheart Clover Robichaud for her continual advice, patience, and support while the project has been ongoing. Also, thanks goes to my good friend Kushal Singh for helping me prepare for the Thesis Defense itself.

Finally, I am grateful to all other physicists and aspiring physicists who read this; if you are one of these people, you are exploring a little known topic that can make a big difference, and I appreciate that you are likely branching out from the standard topics you are already taught in class. I am hopeful that my work contributed to their knowledge of Koopman–von Neumann Mechanics and that this will help them expand the field of Physics. This is currently a little known branch of Physics, and the goal of this project was to make this information more accessible to others unfamiliar with the topic, both in Physics and otherwise. If anyone could take what I did and have that help him or her in contributing to Physics, then I will know that I have made a difference.
\newpage
\newpage

\tableofcontents
\newpage
\newpage
\listoffigures

\newpage
\renewcommand{\thepage}{\arabic{page}}
\setcounter{page}{1}

\section{Introduction}
 Quantum Mechanics (QM) is known for its unique mathematical structure unlike that traditionally found in other branches of Physics. In QM, one has operators acting on the wavefunction in complex Hilbert Space in order to compute what the expectation values of observables seen in the laboratory are (\cite{QM}). Classical Mechanics (CM), on the other hand, is based on continuous, real valued functions of spatial and time dependent variables. One merely has to plug in the values of the variables at a particular point, and one knows the value of the function of interest. QM makes probabilistic predictions of the quantum particle in question, whereas CM makes deterministic predictions of all classical behavior. 

The large differences in mathematical form has historically made it difficult to compare QM to CM (\cite{kvn_thesis}). There have been many attempts to make QM appear more mathematically similar to CM (\cite{kvn_thesis}). For instance, the Born and Madelung approach to QM has a modified Hamilton-Jacobi equation as its basis (\cite{kvn_thesis}). But there have been few attempts to make CM more similar in form to QM. This changed in the 1930's, however, when a form of statistical CM was devised by Bernard Koopman and John von Neumann that had the same operatorial mathematical structure as QM (\cite{koopman}; \cite{vN}). Koopman-von Neumann Mechanics (KvNM), as it later became known, was subsequently used by mathematicians to formulate ergodic theory, while physicists mostly forgot about the existence of this field. Currently, there is a large renewal of interest by the physics community in this relatively obscure field of study.

In this Thesis I intend to articulate many significant contributions of KvNM to physics in a straightforward and easy to understand fashion. I am writing this text in such a way that even an undergraduate with limited training in Physics could understand. I will point out significant topics of research in KvNM and how they are changing physics as we know it. By making this topic more accessible to a variety of readers, I hope that many will find it of interest and continue to build upon this novel and fascinating avenue of study.

KvNM is statistical CM cast in the language of complex wavefunctions in Hilbert Space. Although wavefunctions in Hilbert Space are thought to be tools of QM, they can be used to exactly model the statistical properties of purely classical systems. Classical wavefunctions have been used to study a wide range of physical phenomena. For example, they have been used in studying chaotic systems in quantum and classical theory (\cite{chaos}), been used to develop a model of stochasticity from determinism  (\cite{time_arrow}), been used to study the interplay of connected quantum and classical systems (\cite{coupled_systems}), been used to study free fall and Fischer Information (\cite{free_fall}), been used to reproduce equations governing spin 1/2 relativistic particles (\cite{relativistic_kvn}), among many other areas of research. 

Of great interest, a field like KvNM might be able to decipher the true meaning of QM, which has been hotly debated for over a century.\footnote{Despite a century of progress, QM has no consensus interpretation of its equations. 

`Shut up and calculate' rings as true today as it did a century ago. Feynman gives us a glimmer of hope for the situation, however, as it always takes several generations to understand a great, mysterious truth: ``We have always had a great deal of difficulty understanding the world view that quantum mechanics represents. At least I do, because I’m an old enough man that I haven’t got to the point that this stuff is obvious to me... You know how it always is, every new idea, it takes a generation or two until it becomes obvious that there’s no real problem." (\cite{RF}) Perhaps we're getting closer to an understanding and perhaps a budding field like KvNM can hold clues to the answer.}

\section{General Review of Quantum Mechanics}
What follows is a brief review of Quantum Theory. QM depends on operations on the vector $\ket {\psi}$ in Hilbert Space. The wavefunction $\ket {\psi}$ is a complete description of the state of the system in QM (\cite{QM}). It contains probabilistic information of the system's behavior.

It has a number of convenient properties, such as the fact that it makes switching between one set of bases and another is a rather straightforward task. Hilbert Space useful for all sorts of mathematical procedures. It is, however, not the typical tool of CM.

\subsection{Fundamentals of Hilbert Space}
To understand QM, we must first be comfortable with the concept of a Hilbert Space. A Hilbert Space is an infinite dimensional vector space with a well defined inner product. One can conceive of the Hilbert Space as a coordinate system with an infinite number of perpendicular axes as a crude model. For QM, the Hilbert Space we use not only contains real numbers, but has also been extended to the complex numbers. Although there are Hilbert Spaces that only contain real numbers, we will not need to concern ourselves with those when working with QM or KvNM. 

To see why Hilbert Spaces are so useful in QM, imagine a generic function $f(x)$ which is continuous on its domain (\cite{QM}).
Since any continuous function can be represented by an infinite number of points, we can represent any continuous function in Hilbert Space with state vectors. Each \textit{point} along a generic function $f(x)$ can be represented by a \textit{vector} in Hilbert Space. Because there is an infinite number of points along the function $f(x)$ (even in a finite segment), we should not be surprised it will take an infinite number of vectors to represent $f(x)$. 

This collection of vectors can be represented in an orthogonal basis (i.e., each vector is perpendicular to any other vector in the collection or set, as there are an infinite number of perpendicular axes in Hilbert Space.) A vector can be represented with the following bra-ket notation (original to Dirac):

\begin{equation*}
\vec{v} =
\begin{pmatrix}
v_{1} \\
v_{2}  \\
v_{3}
\end{pmatrix} \Longrightarrow
\ket{v} =
\begin{pmatrix}
v_{1} \\
v_{2}  \\
v_{3}
\end{pmatrix}
,
\bra{v} =
\begin{pmatrix}
v_{1}^* & v_{2}^* & v_{3}^*.
\end{pmatrix}
\end{equation*}
where $\ket{v}$ is the ket vector and $\bra{v}$ is the bra vector (\cite{QM}). The asterisk (*) represents the complex conjugate. Vectors like $\vec{v} $ are ambiguous, because it is not often clear if to write them as columns (as depicted above) or rows of numbers. However, the Dirac bra-ket notation overcomes the ambiguity of typical vector notation in physics, by having that the ket is always equivalent to a column and the bra is equivalent to a row of values. Hilbert Space utilizes bra-ket notation in its infinite dimensional mathematics. A traditional Hilbert Space ket might have more than three entries therefore; it might be represented with an infinite number of entries.

With this in mind, the continuous function $f(x)$ would be represented in Hilbert Space as:

\begin{equation}
f(x) \Longrightarrow
\ket{f(x)} =
\begin{pmatrix}
f(x_{1}) \\
f(x_{2})  \\
f(x_{3})\\
\vdots\\
f(x_{n})\\
\vdots
\end{pmatrix}.
\end{equation}
You can think about this representation of $f(x)$ as the value of \textit{each point} along $f(x)$, i.e. $f(x_i)$, having a unique independent basis vector in Hilbert Space multiplying it. So according to the rules of Linear Algebra you may expand $\ket{f(x)}$, for example:

\begin{equation*}
\ket{f(x)} =
\begin{pmatrix}
1 \\
0  \\
0\\
\vdots\\
0\\
\vdots 
\end{pmatrix}
f(x_1) +
\begin{pmatrix}
0 \\
1  \\
0\\
\vdots\\
0\\
\vdots 
\end{pmatrix}
f(x_2) +
\begin{pmatrix}
0 \\
0  \\
1\\
\vdots\\
0\\
\vdots 
\end{pmatrix}
f(x_3) + ... +
\begin{pmatrix}
0 \\
0  \\
0\\
\vdots\\
1\\
\vdots 
\end{pmatrix}
f(x_n) +...
\end{equation*}
In the above expansion, notice that the column vectors form a unique set of linearly independent basis vectors. 

The way to switch between bras and kets in a complex Hilbert Space is to take the transpose of the vector and complex conjugate of each of the values, like so:

\begin{equation*}
\ket{v} = 
\begin{pmatrix}
v_{1} \\
v_{2}  \\
v_{3} \\
\vdots\\
\end{pmatrix}
\Longleftrightarrow 
\bra{v} =
\begin{pmatrix}
v_{1}^* & v_{2}^* & v_{3}^* & \hdots
\end{pmatrix}
\end{equation*}
The asterisks represents the complex conjugate. This operation will be represented by $\dagger$, so that $\ket{v}^ \dagger = \bra{v}$ and vice versa. Just like any ordinary vector, bra-ket vectors can be added component wise and multiplied together:
\begin{equation}\label{eq:D}
\begin{pmatrix}
v_{1} \\
v_{2} \\
v_{3} \\
\vdots\\
\end{pmatrix}
+
\begin{pmatrix}
w_{1} \\
w_{2}  \\
w_{3}\\
\vdots\\
\end{pmatrix}
=\ket{v} + \ket{w}
=\begin{pmatrix}
v_{1}+w_{1} \\
v_{2}+w_{2}  \\
v_{3} + w_{3}\\
\vdots\\
\end{pmatrix}
= \ket{v+w},
\end{equation}

\begin{equation*}
\begin{pmatrix}
w_{1}^* & w_{2}^* & w_{3}^* & \hdots
\end{pmatrix}
\begin{pmatrix}
v_{1} \\
v_{2}  \\
v_{3}\\
\vdots\\
\end{pmatrix}
= v_{1}w_{1}^*+ v_{2}w_{2}^*+ v_{3}w_{3}^*+ \hdots=\braket{w}{v},
\end{equation*}
where $\braket{w}{v}$ is a generalized ``dot product" or inner product for the vector space. Two vectors are said to be orthogonal if their inner product is 0. So in order for two vectors $\ket{v}$ and $\ket{w}$ in Hilbert Space to be orthogonal, we must have $\braket{w}{v} = 0$. Also, like for any common vector, you can rescale a vector by a complex constant $c$ to stretch or shrink its magnitude: $\ket{v'} = c\ket{v}$.

The vectors additionally obey the following properties:
\begin{equation*}
(\braket{w}{v})^\dagger = \braket{v}{w},
\end{equation*}
\begin{equation*}
\ket{c_{1}v+c_{2}w}^\dagger = (c_{1}\ket{v}+ c_{2}\ket{w} )^\dagger = (c_{1}\ket{v})^\dagger + (c_{2}\ket{w} )^\dagger = c_{1}^* \bra{v}+ c_{2}^* \bra{w}.
\end{equation*}
Since continuous functions like $f(x)$ can be represented in Hilbert Space, we can recast continuous functions of QM into the Hilbert Space formalism. We will shortly see why this is useful to do. 

The wavefunction $\psi$ is the backbone of the quantum mechanical description of reality. The wavefunction $\psi$ is an equation of state, i.e., it provides us with a description of the state of the system (\cite{QM}). \footnote{One way you can think of an `equation of state' is, for instance, the ideal gas law, $PV= NkT$. This equation will tell you what the state of the system, i.e., what the pressure $P$, volume $V$, number of gas particles $N$, and temperature $T$, for an ideal gas is. }The wavefunction got its name from the fact that its form resembles that of a wave equation. In configuration space, the wavefunction is often written as:
\begin{equation}\label{eq:configuration_space_wavefunction}
\psi(q_1,q_2,...,q_n,t) = R(q_1,q_2,...,q_n,t)e^{iS(q_1,q_2,...,q_n,t)/\hbar},
\end{equation}
where $q_1,q_2,...,q_n$ are positions in configuration space and $t$ is time. $S$ represents the Action and $R$ is the amplitude of the wave equation.  In light of the above discussion and since $\psi$ is a continuous function, we can represent $\psi$ as $\ket{\psi(t)}$ in complex Hilbert Space. Schr\"{o}dinger's equation, for example, becomes:
\begin{equation}\label{eq:analytical_schrodinger_equation}
i\hbar \frac{\partial\psi}{\partial t} =- \frac{\hbar^2 }{2m}\frac{\partial^2 \psi}{\partial q^2}+ V(q)\psi \Longrightarrow i\hbar \frac{\partial}{\partial t} \ket{\psi(t)} =- \frac{\hbar^2 }{2m}\frac{\partial^2 }{\partial q^2}\ket{\psi(t)}+ V(q)\ket{\psi(t)}.
\end{equation}
The benefits of using Hilbert Space instead of just utilizing $\psi$ is because Hilbert Space provides a rich mathematical, abstract structure where various manipulations are straightforward and easy to do (\cite{QM}; \cite{ODM}). We can very easily, for instance, change our basis of interest from one to another. We can `deconstruct' functions and put them in various representations. This is not a flexibility given by simply using $\psi$. As we will see, changing representations is a fundamental ingredient to QM, so the Hilbert Space form is absolutely necessary.

\subsection{Operators and Commutators}
An \textit{operator} is a mathematical object that ``acts on" a function, vector, or matrix to produce a new function (the derivative), a new vector, or a new matrix. For example, the derivative $\frac{d}{dt}$ acting on a function $f(t)$ is an operator, as is a matrix $M$ acting on a vector $\ket{v}$. The operator produces a new mathematical entity, such as the $\frac{df(t)}{dt} $ from $f(t)$ or a new vector $\ket{v'}$ from the original vector ($\ket{v'} = M\ket{v}$). All operators will be represented with triangular ``hats", for example, $\hat{A}$. 

Two important types of operators are Hermitian operators and anti-Hermitian operators, which obey the following rules respectively:

\begin{equation*}
\hat{A}^\dagger = \hat{A},
\end{equation*}
\begin{equation*}
\hat{A}^\dagger = - \hat{A},
\end{equation*}
where the $\dagger$ symbol is the transpose with complex conjugate as before. 
The operators we most use in QM are Hermitian operators. Hermitian operators are incredibly useful to utilize, because if we take the $\dagger$ of an expression with a Hermitian operator, the form of the Hermitian operator is left unchanged. For example, for two vectors $\ket{v}$ and $\ket{w}$ and Hermitian operator $\hat{H}$:

\begin{equation*}
(\bra{w}\hat{H}\ket{v})^\dagger =  \bra{v} \hat{H}^\dagger \ket{w} = \bra{v} \hat{H}\ket{w}.
\end{equation*}
An \textit{observable} in QM is a property of a system you would like to measure or observe. Examples of observables in QM would be the position of a particle, its momentum, its spin, etc. Observables in QM are given by Hermitian operators $\hat{A}$ acting on the state vector $\ket{\psi(t)}$, i.e., $\hat{A} \ket{\psi(t)}$. The state vector $\ket{\psi(t)}$ is a vector in Hilbert Space that contains complete dynamical information about the system in question. 

Hermitian operators relate directly to properties you measure inside the laboratory. For any generic operator $\hat{A}$, there exist special types of ket vectors which when acted on by an operator $\hat{A}$ they equal themselves rescaled by a constant. In other words, they obey the mathematical form of 
\begin{equation}\label{eq:eigenvalue_problem}
\hat{A}\ket{\psi(t)} = A\ket{\psi(t)},
\end{equation}
where `$A$' is always a real number (\cite{QM}). 

Note that this special relationship is not true for just \textit{any} vector $\ket{\psi(t)}$. This is only true for specific vectors which we will call \textit{eigenkets}. The rescaling real number `$A$' is known as the \textit{eigenvalue}, and the mathematical problem to identify for an operator $\hat{A}$ the correct eigenkets and eigenvalues is called the \textit{eigenvalue problem}(\cite{QM})\footnote{The function version of the eigenvalue problem is $\hat{A}\psi = A\psi$, where $\psi$ is called an eigenfunction. To recover the Hilbert Space version, remember the colloquial transformation $\psi \Longrightarrow \ket{\psi(t)}$. In the literature, terms like eigenvectors and eigenfunctions are used interchangeably, as they are technically the same thing, just one written in an abstract Hilbert Space and one written without.}. The eigenvalue problem is central to QM because it directly relates to the observables measured in the laboratory, as will be developed below. 

Perhaps one of the most important properties of a Hermitian operator is that its eigenvalues must necessarily always be real (\cite{QM}). Other types of operators might give, for example, imaginary eigenvalues. Hermitian operators will only give you real numbers, which are the types of values we would measure inside the laboratory. 

Two very important operators are the ones for the observables momentum and position. In coordinate representation (as other representations are possible), the momentum and coordinate operators are given respectively as follows: 

\begin{equation}
\hat{p} = \frac{\hbar}{i} \frac{\partial}{\partial q},
\end{equation}

 \begin{equation}
\hat{q} = q.
\end{equation}
These operators are important, because as we shall see, they are related to the average momentum and position of interest. \footnote{For example, in standard function form one can easily show that the $\hat{p}\psi = p\psi$ where $p$ is the momentum expectation value measurable in the laboratory and $\hat{p}$ is the above momentum operator. Earlier, we said that a common form of $\psi$ in configuration space is $\psi = Re^{iS/\hbar}$, where $R$ is the amplitude of the wavefunction and $S$ is the Action. Let's assume a simple, one dimensional case where $R = R(t)$ and $S = S(q,t)$. Then we would have:

 $$\hat{p}\psi = \frac{\hbar}{i} \frac{\partial}{\partial q}Re^{iS/\hbar} = \frac{\hbar }{i}Re^{iS/\hbar} \frac{i}{\hbar}\frac{\partial S}{\partial q} = \frac{\partial S}{\partial q} \psi.$$
From CM, we know that $p = \frac{\partial S}{\partial q}$, so the shoe fits. Please note that this exercise is a bit of an oversimplification. Generally, $R = R(t)$ is \textit{not} considered a legitimate choice because it can lead to nonnormalizable wavefunctions (see \cite{many_worlds} for further discussion.) For the sake of a simple demonstration of how operators acting on wavefunctions lead to eigenvalues measured in the laboratory, I chose this oversimplified approach to highlight connections between ideas .}

Commutators describe relations between operators. The principle that two numbers $A$ and $B$ obey $A\cdot B = B \cdot A$ is known as the \textit{principle of commutativity}. Operators do not need to, however, obey the principle of commutativity. We might have, for example, that $\hat{A}\hat{B} \neq \hat{B}\hat{A}$ for two operators $\hat{A}$ and $\hat{B}$. This might be most explicitly clear when we think about the example of matrix multiplication. The order matrices are multiplied in might give you different end products. We say therefore that matrix multiplication is noncommutative. 

In order to describe this property for operators, we use the idea of a commutator. For two operators $\hat{A}$ and $\hat{B}$, the commutator is defined as:

$$[\hat{A},\hat{B}] \equiv \hat{A}\hat{B} - \hat{B}\hat{A}. $$
If the commutator equals zero, the two operators commute:

$$[\hat{A},\hat{B}] = \hat{A}\hat{B} - \hat{B}\hat{A} = 0 \Longrightarrow  \hat{A}\hat{B} = \hat{B}\hat{A}.$$
However, if the commutator is nonzero, then obviously the operators do not commute. 

The position and momentum operators are an example of noncommuting operators. It can be demonstrated that: 

$$[\hat{q},\hat{p}] = \hat{q}\hat{p} - \hat{p}\hat{q} = q \frac{\hbar}{i} \frac{\partial}{\partial q} - \frac{\hbar}{i} \frac{\partial}{\partial q} q .$$
$$ [\hat{q},\hat{p}]\psi = q \frac{\hbar}{i} \frac{\partial \psi}{\partial q} - \frac{\hbar}{i} \frac{\partial}{\partial q} (q\psi) = q \frac{\hbar}{i} \frac{\partial \psi}{\partial q} - \frac{\hbar}{i} \frac{\partial q}{\partial q} \psi - q \frac{\hbar}{i} \frac{\partial \psi}{\partial q} = - \frac{\hbar}{i} \psi.$$
Ergo, the commutator relation for $\hat{q}$ and $\hat{p}$ is

\begin{equation}\label{eq:canonical_comm}
[\hat{q},\hat{p}] \psi = - \frac{\hbar}{i} \psi = i \hbar \psi \Longrightarrow [\hat{q},\hat{p}] = i \hbar. 
\end{equation}
This is the famous Heisenberg Canonical Commutator (\cite{QM}). In section 5.2 we will derive the Heisenberg Uncertainty Principle from the Canonical Commutator. The following are some important commutator relationships given below with derivation for brevity:

\begin{equation}\label{eq:comm_rule_1}
[\hat{A}, \hat{B}] = - [\hat{B}, \hat{A}].
\end{equation}

\begin{equation}\label{eq:comm_rule_2}
[c\hat{A}, \hat{B}] = [\hat{A}, c\hat{B}] = c[\hat{A}, \hat{B}] .
\end{equation}

\begin{equation}\label{eq:comm_rule_3}
[\hat{A}\hat{B},\hat{C}] = \hat{A}[\hat{B},\hat{C}] + [\hat{A},\hat{C}] \hat{B}.
\end{equation}

\begin{equation}\label{eq:comm_rule_4}
[\hat{A},\hat{B}\hat{C}] =\hat{B} [\hat{A},\hat{C}] +  [\hat{A},\hat{B}] \hat{C}.
\end{equation}

\begin{equation}\label{eq:comm_rule_5}
[\hat{A},\hat{B}+\hat{C}] = [\hat{A},\hat{B}] + [\hat{A},\hat{C}].
\end{equation}

\subsection{Expectation Values and the Eigenvalue Problem of Quantum Mechanics}
What exactly are the eigenvalues that we get out of the eigenvalue problem? I have stated before that they are linked to measured observables, but in what way exactly? Observables can be anything you would like to measure, such as the momentum or position of a quantum particle.

Eigenvalues are going to be the numbers you measure directly inside the laboratory. If you conduct an experiment on a quantum system, the eigenvalues will be the exact numbers you record. Remember that for a Hermitian operator, its eigenvalues will always be real numbers. This is critical, because any number you measure in an experiment will be real instead of, say, complex. 

Eigenvalues are connected to finding the \text{expectation values} of an experiment. Imagine you conduct an experiment on a quantum system and you measure the position or momentum. Instead of predicting individual momenta or position, QM can predict the \textit{weighted mean} value that you should expect. If you took many measurements on identical systems, and took the weighted average of all those measurements, the value you get out of this is the expectation value.

I will demonstrate how expectation values arise in QM in the following. I will use both the Hilbert Space formulism and the function $\psi$ that some might be more comfortable with. Both derivations will end up with identical conclusions. 

First, start with the eigenvalue expression $\hat{A}\psi = A\psi$ and multiply both sides by the complex conjugate, like so:
$$\psi^* \hat{A}\psi = \psi^* A\psi = A\psi^* \psi$$ 
In Hilbert Space, the same can be achieved by multiplying the eigenvalue expression $\hat{A}\ket{\psi(t)} = A\ket{\psi(t)} $ with a bra vector from the left, so that the operator is `sandwiched':
$$\bra{\psi(t)}\hat{A}\ket{\psi(t)} = \bra{\psi(t)}A\ket{\psi(t)}$$
For the typical function form, we are going to multiply both sides by the `operator' $\int_{-\infty}^{\infty} dx$, in order to get:
\begin{equation}\label{eq:A}
\int_{-\infty}^{\infty} \psi^* \hat{A}\psi dx =\int_{-\infty}^{\infty} A\psi^* \psi dx  
\end{equation}

We now introduce a fundamental postulate of QM, namely, the Born Rule. The Born Rule states that $\rho = \psi^* \psi$ where $\rho$ represents the probability density for continuous $\psi$. From probability theory, we know that if a probability density is integrated over all possible states, it must be equal to 1 (the renormalization condition): 
\begin{equation}\label{eq:renormalization}
\int_{-\infty}^{\infty} \psi^* \psi dx = 1 
\end{equation}
In other words, the probability of something happening is certain (= 1) if all possible outcomes occur. 
The equivalent renormalization condition in Hilbert Space is that $\braket{\psi(t)} = 1$.\footnote{Both renormalization conditions can be shown to be equivalent. From Appendix A, we have the important completeness theorem, which says:

$$I = \int dx \ket{x}\bra{x}$$ 
Since $I$ is just the identity operator (operator where $I\hat{A} = \hat{A}$ or $I\ket{\psi} = \ket{\psi}$, just like multiplying something by the number `1'), we can do the following operations:

$$\braket{\psi(t)} = \bra{\psi(t)}I\ket{\psi(t)} = \int \braket{\psi(t)}{x}\braket{x}{\psi(t)} dx $$
Recall from before that we have said that $\braket{x}{\psi(t)} = \psi(x,t)$, so we get:
$$\braket{\psi(t)} = \int \psi^* \psi dx$$}

Equation \eqref{eq:A} is defined to be the expectation value of the observable given by operator $\hat{A}$, as can be seen from the similarity of both:

\begin{equation*}
\langle f(x) \rangle = \int_{-\infty}^{\infty} f(x) \rho ~ dx \Longleftrightarrow \int_{-\infty}^{\infty} \psi^* \hat{A} \psi dx = \int_{-\infty}^{\infty} A\psi^* \psi dx ,
\end{equation*}
where you can see that the eigenvalue $A$ takes on the role of $f(x)$ in the above expression and $\psi^* \psi$ is the probability density. Please keep in mind that there could be many different values you can measure for observable $A$, so it makes sense it is equivalent to the function $f(x)$. Therefore the expectation value of observable $A$ is given by 

\begin{equation}\label{eq:expectation_value}
\langle A \rangle = \int_{-\infty}^{\infty} \psi^* \hat{A} \psi dx = \bra{\psi(t)}\hat{A}\ket{\psi(t)}.
\end{equation}
The expectation values measured in the laboratory happen to be the exact ones predicted by the above equation. The expectation value expression was strictly derived from the eigenvalue problem and the Born Rule.

\subsection{Review of Dirac Delta Functional}

The Dirac delta functional is important to QM for the renormalization of all the basis vectors. For any Hilbert Space, one can find an orthogonal set of bases to represent it. But we can do one better - one can always find an \textit{orthonormal} basis. An orthonormal basis is a basis where not only are all the basis vectors orthogonal, but they are also normalized to unit length 1:
\begin{equation}
\ket{A}_{normalized} = \frac{\ket{A}}{\sqrt{\braket{A}}} 
\end{equation}
Any two generic vectors $\ket{A}$ and $\ket{A'}$ in Hilbert Space are orthonormal if and only if they satisfy
\begin{equation*}
\braket{A}{A'} = \delta_{AA'} = 
\begin{cases}
1 \quad &\text{if} \, \ket{A} = \ket{A'} \\
0 \quad &\text{if} \, \ket{A} \neq \ket{A'} \\
     \end{cases}
\end{equation*}
where $\delta_{AA'}$ is the Kronecker delta. The Kronecker delta functional is however only appropriate when dealing with \textit{strictly} discrete systems. Many problems in QM, however, can be \textit{either} discrete or continuous. We therefore need a functional that can encompass \textit{both} discrete and continuous systems to describe an orthonormal basis. This is done by using the Dirac delta functional\footnote{Please note that the $A$ here is a generic placeholder variable with arbitrary units. It could be a placeholder for position $q$, momentum $p$, or some other measurable quantity. So you could, for example, have $\braket{q}{q'} = \delta(q-q')$ for position and $\braket{p}{p'} = \delta(p-p')$ for momentum, etc.}:
\begin{equation}\label{eq:dirac_delta_1}. 
\braket{A}{A'} = \delta(A-A')
\end{equation}
The Dirac delta functional is a generalization of the Kronecker delta (\cite{QM}).
The Dirac delta functional obeys the following properties:

$$\delta(A - A') = 0 ~ \text{if}~ A \neq A'.$$

$$\int_a^b \delta(A-A') ~dA' = 1 ~\text{if}~ a \leq A \leq b.$$
The functional is zero everywhere but at $A'$ (\cite{QM}).

The Dirac delta \textit{functional} should not be treated as a regular function. It has its own unique properties and form, distinct from how functions behave. For example, the area under a single point when $A = A'$ should be zero (the area under a point is zero). But for the Dirac delta functional, it is precisely 1. This and other properties of this functional highlight the fact that it should \textit{not} by treated like the functions you have seen up to this point. It is a \textit{functional}, not a function. It is, however, fundamental to the renormalization of vectors in Hilbert Space.

Other important properties of the functional we will utilize in the text of this article include the following:

\begin{equation}\label{eq:int_dirac_delta_functional}
\int \delta(A-A')f(A') dA' = f(A).
\end{equation}

\begin{equation}
\delta(A-A') = \delta(A' -A).
\end{equation}

\begin{equation}
\frac{d}{dA}\delta(A-A') = -\frac{d}{dA'}\delta(A -A').
\end{equation}
Some useful relations for QM:
\begin{equation}\label{eq:C}
\frac{d^n}{dA^n} \delta(A-A') = \int \frac{d\omega}{2\pi}(-i\omega)^n e^{-i\omega (A-A')}.
\end{equation}
\begin{equation}
\int  \frac{d^n}{dA^n}  \delta(A-A') f(A')~dA' = (-1)^n \frac{d^n f(A)}{dA^n}.
\end{equation}

\begin{equation}\label{eq:dirac_delta_a}
 \delta(\alpha A) = \frac{\delta(A)}{|\alpha|},
\end{equation}
where $\alpha$ is a constant.

\subsection{Summary of Quantum Mechanics and its Postulates}
%
The postulates of QM can be reviewed as follows (\cite{QM}; \cite{ODM}):
\begin{enumerate}
			\item The state of a quantum system (particle or collection of particles) is represented by a vector $\ket{\psi}$ in complex Hilbert Space.
 
\item For \textit{any} observable $A$, there is an associated Hermitian operator $\hat{A}$ with eigenvalue problem $\hat{A}\ket{A}= A\ket{A}$ where $\ket{A}$ is the specific eigenket and $A$ is the eigenvalue which is measured in the laboratory. Examples of common observables are position of a particle $q$ and momentum $p$. The variables position $q$ and momentum $p$ are represented by Hermitian operators $\hat{q}$ and $\hat{p}$ which obey the following properties:  $\bra{q}\hat{q}\ket{q'} = q\delta(q - q')$ and $\bra{q}\hat{p}\ket{q'} = - i\hbar \frac{d}{dq}\delta(q - q')$.

\item \textit{Born Rule}: For a continuous distribution, the probability density for a particle is $\rho = \psi^* \psi$. Equivalently in Dirac Notation, it can be written that the probability density of observing $A$ is $\rho(A) \propto \braket{\psi}{A}\braket{A}{\psi} = \|\braket{A}{\psi}\|^2$ where ``$\propto$" is the proportionality sign. More precisely we could say that the Born Rule dictates that
\begin{equation*}
\rho(A) = \frac{ \braket{\psi}{A}\braket{A}{\psi}}{\braket{\psi}}
\end{equation*}
As long as the above equation meets the renormalization condition $\braket{\psi} = 1$ (see equation \eqref{eq:renormalization} and associated footnote), the Born Rule in its final form is often written as
$$\rho(A) = \braket{\psi}{A}\braket{A}{\psi}$$
For a discrete system, the Born Rule is written as:
$$P(A) = \braket{\psi}{A}\braket{A}{\psi},$$
assuming renormalization, where $P$ is the probability. For both the continuous and discrete cases, the state of the system collapses from the initial state $\ket{\psi(t)}$ to the final state $\ket{A}$ whenever a measurement of the quantum system is undertaken. Under the Copenhagen Interpretation of QM, the collapse of the wavefunction $\ket{\psi}$ to the specific eigenket $\ket{A}$ is instantaneous and irreversible. If $A$ is observed in experiment, the wavefunction collapses to the associated eigenket $\ket{A}$. 

\item  The state space of a composite system is the tensor product of the subsystem's state spaces 
\end{enumerate}
There are a couple other important things to say about the above overview of QM. Firstly, although not included as a postulate, the expectation value relationship \eqref{eq:expectation_value} plays a critical role inside of QM. They are of fundamental importance as we reviewed in section 2.3.

Not discussed thoroughly is the centrality of the Schr\"{o}dinger equation to QM. The state of the quantum system $\ket{\psi}$ obeys the Schr\"{o}dinger equation: 
\begin{equation}\label{eq:schrodinger_equation}
i\hbar \frac{\partial}{\partial t} \ket{\psi} = \Big[- \frac{\hbar^2 }{2m}\frac{\partial^2 }{\partial q^2}+ V(q)\Big]\ket{\psi} = \hat{H}\ket{\psi}
\end{equation}
where $\hat{H}$ is the quantum Hamiltonian that bears great similarity to the classical Hamiltonian\footnote{That is, a classical Hamiltonian where the potential energy has no velocity or time dependence, which takes on the form $H(q,p) = \frac{p^2}{2m} + V(q)$. Note that $p$ is momentum and $V(q)$ is a one dimensional potential energy.}, except that the momentum and position are replaced by the momentum and position operators of QM:

\begin{equation*}
H(q,p) = \frac{p^2}{2m} + V(q) \Longrightarrow \hat{H}(\hat{q},\hat{p}) = \frac{\hat{p}^2}{2m} + V(\hat{q}) = - \frac{\hbar^2 }{2m}\frac{\partial^2 }{\partial q^2}+ V(\hat{q})
\end{equation*}
Of important note, the Hamiltonian operator which acts on the wavefunction in equation \eqref{eq:schrodinger_equation} is Hermitian (\cite{QM}). Since it is Hermitian, its eigenvalues will always be real instead of complex or imaginary, as required for numbers measured in the laboratory. 

There will be a slight adjustment of notation from the previous couple sections. Although it was convenient to previously write $\hat{A}\ket{\psi} = A\ket{\psi}$ as the eigenvalue problem, it is more precise to write the eigenvalue problem as $\hat{A}\ket{A} = A\ket{A}$. As described before, for an operator $\hat{A}$ there are \textit{specific} eigenvalues and eigenkets that satisfy the above equation. Although $\ket{\psi}$ might be used to describe the \textit{specific} eigenket in the context of the eigenvalue problem, $\ket{\psi}$ is often the generic state vector of the quantum system, so this might lead to some confusion. To highlight the fact that only specific eigenkets $\ket{A}$ satisfy this relationship, $\hat{A}\ket{A} = A\ket{A}$ will be used from now on. 

Two very important eigenvalue problems are $\hat{q}\ket{q} = q\ket{q}$ for position $q$ and $\hat{p}\ket{p} = p\ket{p}$ for momentum $p$. The expectation values of the position and momentum measured in the laboratory will be the eigenvalues $q$ and $p$. For any observable $A$, there is a relationship $\hat{A}\ket{A} = A\ket{A}$ for which combined with the Born Rule axiom, you can deduce the expectation value of $A$ (as described in section 2.3). 

One of the initially confusing things about the Dirac notation is that the symbols inside the bras and kets are \textit{just labels}. So you might see vector bras and kets with labels such as $\ket{1st}$, $\ket{2nd}$, $\ket{I}$, $\ket{II}$, $\ket{\omega = 2.5}$, etc. All of these have valid notation, as the labels simply describe the associated vector, and that is all. Usually, the eigenvalue is used to label the associated eigenket. Hence the eigenvalue `$A$' being associated with eigenket $\ket{A}$ in the eigenvalue problem. It is why in equation \eqref{eq:D}, we could make a statement such as $\ket{v} + \ket{w} = \ket{v+w}$, because both $v$ and $w$ are just labels.

Finally, the collapse of the wavefunction describes one of the ongoing mysteries of QM. It can be seen that the Schr\"{o}dinger equation is fully deterministic from equation \eqref{eq:schrodinger_equation}, but the collapse of the wavefunction is (as far as we know) indeterministic and random. There is no known formula to describe \textit{how} the wavefunction collapses. More will be said about wavefunction collapse in section 5.3 of this article, in the context of QM and CM.

\subsection{The Density Operator}
Before we delve straight into KvNM, it is critical to briefly look at one more tool that will be of use to us in this work. Instead of representing the state of a system as a wavefunction vector in Hilbert Space $\ket{\psi}$, we can instead use something called a \textit{density matrix} for certain systems, represented by $\hat{\rho}(t)$ (not to be confused with the previously used $\rho$ for probability density.) The density matrix is defined by
\begin{equation}\label{eq:density_matrix_def}
\hat{\rho}(t) = \sum_j p_j\ket{\psi_j(t)}\bra{\psi_j(t)},
\end{equation}
for the discrete case.
The density matrix is used when you are not certain of the configuration of your system at some point in time. You can represent it as statistical ensemble of \textit{possible} wavefunction configurations $\ket{\psi_j(t)}$, each with probability of occurrence $p_j$. Note with probability, we must have

$$\sum_j p_j = 1.$$
We will see an example of the density matrix used in action later in section 5.3. Density matrices are good for describing the above described so-called ``mixed states" of a system (with the ``pure states" being the individual wavefunctions $\ket{\psi_j}$). If $j = 1$ in \eqref{eq:density_matrix_def} the density matrix represents a pure state, but if it is greater than 1 it represents a mixed state. 

The probability of measuring $A$, or $P(A)$, is given by the density matrix formulation as follows:

\begin{equation}\label{eq:density_matrix_born_rule}
P(A) = \Tr[\hat{\rho}(t)\ket{a}\bra{a}],
\end{equation}
where $\Tr[...]$ refers to the trace of the matrix. You can represent the trace by
\begin{equation}\label{eq:trace_formula}
\Tr[...] = \sum_n \bra{n} ... \ket{n},
\end{equation}
where $\ket{n}$ are arbitrary basis of your choice.
This is the density matrix version of the Born Rule. It is easy to see that this is fully equivalent to the Born Rule, for instance, when we plug \eqref{eq:density_matrix_def} with $j =1$ into \eqref{eq:density_matrix_born_rule}:

\begin{equation}
P(A) = \bra{a}\ket{\psi_1}\bra{\psi_1}\ket{a}\bra{a}\ket{a} = | \bra{A}\ket{\psi_1(t)}|^2,
\end{equation}
since $\bra{a}\ket{a} = 1$ assuming orthonormality. 
The expectation value of operator $\hat{A}$ is given by 

\begin{equation}\label{eq:more_eq_1120}
\langle A \rangle = \Tr[\hat{A}\hat{\rho}(t)],
\end{equation}
where $\hat{A}\hat{\rho}(t) $ form a matrix and $\Tr[...]$ refers to the trace of the matrix. You can reduce \eqref{eq:more_eq_1120} to \eqref{eq:expectation_value} just by applying the trace formula \eqref{eq:trace_formula}, easily demonstrated for the pure state $j =1$:

\begin{equation*}
\Tr[\hat{A}\hat{\rho}(t)] = \sum_n \bra{n} \hat{A}\ket{\psi_1(t)}\bra{\psi_1(t)} \ket{n} = \sum_n \bra{\psi_1(t)} \ket{n}\bra{n} \hat{A}\ket{\psi_1(t)}.
\end{equation*}
Applying resolution of identity (Theorem II of Appendix A) we get the standard expectation value formula:
$$ \Tr[\hat{A}\hat{\rho}(t)] =  \bra{\psi_1(t)} \hat{A}\ket{\psi_1(t)} = \langle A \rangle .$$
Other useful properties of the density matrix:

\begin{enumerate}
\item $\Tr[\hat{\rho}(t)] = 1$, which is the normalization condition.
\item $\Tr[\hat{\rho}(t)^2] = 1$ for pure state.
\item $\Tr[\hat{\rho}(t)^2] < 1 $ for mixed state.
\item $\hat{\rho}(t) = \hat{\rho}(t)^\dagger,$ in other words, $\hat{\rho}(t) $ is Hermitian. 
\end{enumerate}
(Based on \cite{density_matrix}.)
We will utilize the density operator in future sections.

\section{Introduction to KvN Mechanics}

Koopman-von Neumann Mechanics is unique because it is a form of classical statistical mechanics that truly takes on a lot of the mathematical structure of QM (\cite{koopman}; \cite{vN}; \cite{ODM}). It was invented in the 1930's by Bernard Koopman and John von Neumann. Subsequently forgotten for some time by most physicists, its resurgence raises intriguing questions about the nature of QM to CM. Just like with QM, we have a set of operators that act on vectors in Hilbert Space in order to retrieve eigenvalues of observables (\cite{koopman}; \cite{vN}; \cite{ODM}). In order to have a proper Hilbert Space, we first must define the inner product of that Hilbert Space (\cite{koopman}):

\begin{equation*}
\braket{\psi}{\phi}  = \int dA \braket{\psi}{A}\braket{A}{\phi} = \int dA~ \psi^*(A) \phi(A).
\end{equation*} 
CM always has distinct trajectories that particles travel along. In QM, we have the Heisenberg Uncertainty Principle which states you cannot have both a well defined position of a particle and traveling momentum (or velocity) at the same time. Under the Copenhagen Interpretation, this is not due to some principle of measurement, rather, \textit{this is a fundamental property of the fabric of reality} (\cite{QM}). Reality forbids quantum particles from having a precise momentum and position simultaneously. In QM, trajectories could be viewed as being ``fuzzy" instead of well defined and localized like in CM. You can no longer draw nice curves to describe how a particle travels in space under QM.

Because of this fact, both position and momentum would have a common set of eigenstates in KvN Hilbert Space (\cite{wigner}; \cite{time_arrow}).  The state vectors for KvN would therefore have the following properties (\cite{koopman}):

\begin{equation*}
\ket{A} = \ket{q, p} = \ket{q}\otimes \ket{p},
\end{equation*}

\begin{equation*}
\hat{q}\ket{q,p} = q\ket{q,p},
\end{equation*}

\begin{equation*}
\hat{p}\ket{q,p} = p\ket{q,p},
\end{equation*}

\begin{equation*}
\braket{q, p}{q', p'} = \delta(q - q')\delta(p - p'),
\end{equation*}

\begin{equation}\label{eq:basic_properties_KvN}
\int dp ~ dq \ket{q,p}\bra{q,p} = I.
\end{equation}
The observable $A$'s eigenket $\ket{A}$ is the KvN ket $\ket{q,p}$ because of the common set of eigenstates between $q$ and $p$. Because it contains information about $q$ and $p$, this is providing information in phase space instead of the previously discussed configuration space (\cite{koopman}). Previously for configuration space we had $\braket{q}{\psi(t)} = \psi(q,t)$,\footnote{Which can be generalized into $\braket{q_1, q_2, q_3,...,q_n}{\psi(t)} = \psi(q_1, q_2, q_3,...,q_n,t)$ for configuration space.} whereas here we will have $\braket{q,p}{\psi(t)} = \psi(q,p,t)$ in phase space. 

In QM, we have the Schr\"{o}dinger equation \eqref{eq:schrodinger_equation} to dictate to us the time evolution of the quantum particle described by $\ket{\psi}$. To estimate the probability of an observable, we utilize $\rho(A, t)= \psi^*(A,t) \psi(A,t) = |\braket{A}{ \psi(t)}|^2$ (Born Rule). In CM, we have the following equation known as the Louiville Equation to describe classical trajectories:

\begin{equation}\label{eq:louiville_equation}
\frac{\partial \rho}{\partial t} + \sum \Big(\frac{\partial H}{\partial p_j}\frac{\partial \rho}{\partial q_j} - \frac{\partial H}{\partial q_j}\frac{\partial \rho}{\partial p_j}\Big) = 0 
\end{equation}
The Louiville Equation is important because it tells you the probability density $\rho$ of finding a particle in observable point $\omega = (q,p)$ in phase space. In other words, it defines the probabilistic time evolution of an ensemble of particles. We can define the Louiville operator to be (\cite{kvn_thesis}):

\begin{equation}\label{eq:L_operator}
\hat{L} = \sum \Big( -i\frac{\partial H}{\partial p_j}\frac{\partial}{\partial q_j} + i\frac{\partial H}{\partial q_j}\frac{\partial}{\partial p_j}\Big)
\end{equation}
By doing so, we can rewrite equation \eqref{eq:louiville_equation} in operational terms:
\begin{equation}\label{eq:operatorial_louiville_equation}
i\frac{\partial \rho}{\partial t} = \hat{L} \rho
\end{equation}
Notice each term in equation \eqref{eq:L_operator} has been multiplied by the imaginary unit $i$. 
Just like in QM, we postulate a Born Rule for this probability density. If $\rho(A, t) = \psi^* \psi $ (where $A$ is the observable), then we can easily show that \footnote{Demonstrated as follows. If $\rho = \psi^* \psi$, then for a simple case with index $j =1$ we plug it into equation \eqref{eq:operatorial_louiville_equation}: 

$$i\frac{\partial \psi^* \psi}{\partial t} + i\frac{\partial H}{\partial p}\frac{\partial \psi^* \psi}{\partial q}- i\frac{\partial H}{\partial q}\frac{\partial \psi^* \psi}{\partial p} = 0 $$

$$i\frac{\partial \psi^*}{\partial t}\psi + i\frac{\partial \psi}{\partial t}\psi^*  + i\frac{\partial H}{\partial p}\frac{\partial \psi^*}{\partial q}\psi  + i\frac{\partial H}{\partial p}\frac{\partial \psi}{\partial q}\psi^* - i\frac{\partial H}{\partial q}\frac{\partial \psi^*}{\partial p}\psi - i\frac{\partial H}{\partial q}\frac{\partial \psi}{\partial p}\psi^* = 0 $$

$$ \Big[i\frac{\partial \psi^*}{\partial t}  + i\frac{\partial H}{\partial p}\frac{\partial \psi^*}{\partial q} - i\frac{\partial H}{\partial q}\frac{\partial \psi^*}{\partial p}\Big]\psi + \Big[i\frac{\partial \psi}{\partial t}  + i\frac{\partial H}{\partial p}\frac{\partial \psi}{\partial q}- i\frac{\partial H}{\partial q}\frac{\partial \psi}{\partial p}\Big]\psi^* = 0 $$
Since generally neither $\psi$ nor $\psi^*$ are zero, this means the expressions in square brackets must equal zero. The bracketed expressions are, of course, equation \eqref{eq:KvN_L_wavefunction} and its complex conjugate.}

\begin{equation}\label{eq:KvN_L_wavefunction}
i\frac{\partial \psi}{\partial t} = \hat{L} \psi
\end{equation}

This equation, of course, bears a striking resemblance to equation \eqref{eq:schrodinger_equation}, the Schr\"{o}dinger equation in its operatorial form. The classical wavefunction $\psi$ here is interpreted as a probability density amplitude, like under most interpretations of QM (\cite{QM}). Just like with QM, KvN depends on the generic state vector $\ket{\psi(t)}$ containing statistical information about the system. Colloquially, just as before, $\psi \Longrightarrow \ket{\psi(t)}$. 

An important property of KvN Mechanics is that the Louiville operator $\hat{L}$ is Hermitian, just like the Hamiltonian operator of the Schr\"{o}dinger equation. 
The norm of $\braket{\psi(t)} = \int d\omega \psi^*(\omega) \psi(\omega)$ is therefore conserved, further strengthening the Born's Rule postulated for KvN Mechanics (\cite{kvn_thesis}). 

The Axioms of KvN Mechanics can be summarized as follows (based off \cite{ODM}): 
\begin{enumerate}
\item The state of a system is represented by a vector $\ket {\psi}$ in a complex Hilbert Space.

\item For any observable $A$, there is an associated Hermitian operator $\hat{A}$ with eigenvalue problem $\hat{A}\ket {A} = A\ket{A}$ where $\ket{A}$ is the specific eigenket and $A$ is the eigenvalue measured in the laboratory. 

\item \textit{Born Rule:} The probability of measuring $A$ is given by $P(A) =|\braket{A}{\psi(t)}|^2 $.
The wavefunction $\ket{\psi}$ describing the system instantaneously collapses into the eigenket $\ket{A}$ associated with the observed eigenvalue $A$. 

\item The state space of a composite system is the tensor product of the subsystems' state space
\end{enumerate}
Notice, these are exactly the same postulates of QM in section 2.5, just with anything ``quantum" removed (\cite{ODM}). The formula for the expectation value of the system is still the same, since Axioms 2 and 3 remain unchanged, and therefore the derivation in section 2.3 is the same. The very important expectation value rule \eqref{eq:expectation_value} therefore still applies in KvNM.

Given these four `universal' postulates, one can derive both QM and KvNM from the Ehrenfest Theorems. The Ehrenfest Theorems are as follows:

\begin{equation}\label{eq:ehrenfest_theorem_1}
\frac{d}{dt}\langle q \rangle = \frac{\langle p \rangle}{m},
\end{equation}
\begin{equation}\label{eq:ehrenfest_theorem_2}
\frac{d}{dt}\langle p \rangle = \langle -V'(q)\rangle,
\end{equation}
and are usually considered strictly within the context of QM. 

Under QM, the fabric of reality forbids the mutual existence of a well defined position \textit{and} momentum. Under CM, we can know the position and momentum to an arbitrary precision. This difference is encoded with the commutator relationship. In section 5.3, we will show that the Canonical Commutator of QM \eqref{eq:canonical_comm} directly produces the Heisenberg Uncertainty Principle. We can also write a commutator relationship for CM under the axioms of KvNM. We can say:

\begin{equation}\label{eq:classical_comm}
[\hat{q}, \hat{p}] = 0
\end{equation}
I shall call this the ``Classical Commutator''. This commutator relationship would produce no uncertainty relationship between the classical position $q$ and momentum $p$. Ergo, position and momentum can be arbitrarily known. 

For QM, one can show that the Canonical Commutator combined with the Ehrenfest Theorem produces the traditional Schr\"{o}dinger equation with quantum Hamiltonian. We will demonstrate this as follows. According to the common set of Axioms, we can say that the expectation value of observable $A$ is simply $\bra{\psi(t)}\hat{A}\ket{\psi(t)}$ \eqref{eq:expectation_value}. Therefore we can easily say that

$$\langle q \rangle = \bra{\psi(t)}\hat{q}\ket{\psi(t)}$$

$$\langle p \rangle = \bra{\psi(t)}\hat{p}\ket{\psi(t)}$$
Then from the Ehrenfest Theorems one can carry out the following operations:
\begin{equation*}
\frac{d}{dt}\bra{\psi(t)}\hat{q}\ket{\psi(t)}= \frac{1}{m}\bra{\psi(t)}\hat{p}\ket{\psi(t)},
\end{equation*}
\begin{equation*}
\frac{d}{dt}\bra{\psi(t)}\hat{p}\ket{\psi(t)} = \bra{\psi(t)}-V'(\hat{q})\ket{\psi(t)}.
\end{equation*}

We are using the Schr\"{o}dinger picture of QM, so we assume time dependence solely resides in $\ket{\psi(t)}$ and any operators are independend of time. \footnote{This is the opposite of the Heisenberg picture, which assumes no time dependence in the wavefunction vector ket and puts all the time dependence inside the operators!} Using the chain rule of calculus, we write:

\begin{equation*}
\bra{\frac{d\psi(t)}{dt}}\hat{q}\ket{\psi(t)}  + \bra{\psi(t)}\hat{q}\ket{\frac{d\psi(t)}{dt}} = \frac{1}{m}\bra{\psi(t)}\hat{p}\ket{\psi(t)},
\end{equation*}
\begin{equation}\label{eq:F}
\bra{\frac{d\psi(t)}{dt}}\hat{p}\ket{\psi(t)}  + \bra{\psi(t)}\hat{p}\ket{\frac{d\psi(t)}{dt}} = \bra{\psi(t)}-V'(\hat{q})\ket{\psi(t)}.
\end{equation}
Stone's Theorem (see Appendix B) dictates that there exists a unique  Schr\"{o}dinger-like equation $i\frac{\partial}{\partial t} \ket{\psi} = \hat{G}\ket{\psi}$ where $\hat{G}$ is the Hermitian generator of motion in operator form. We deduce there must be a $\hat{G}$ that governs the evolution of CM, just as the Hamiltonian operator does for QM. The only thing is to figure out what this generator $\hat{G}$ is. 

For QM, we can derive what the generator of motion is by using the commutator relationships. We can take Stone's Theorem and then derive:

$$\ket{\frac{d\psi(t)}{dt}} = \frac{\hat{H}}{i\hbar}\ket{\psi(t)} \iff \bra{\frac{d\psi(t)}{dt}} = - \bra{\psi(t)} \frac{\hat{H}}{i\hbar}$$
Which means:
\begin{equation*}
- \bra{\psi(t)} \frac{\hat{H}}{i\hbar}\hat{q}\ket{\psi(t)}  + \bra{\psi(t)}\hat{q}\frac{\hat{H}}{i\hbar}\ket{\psi(t)} = \frac{1}{m}\bra{\psi(t)}\hat{p}\ket{\psi(t)}
\end{equation*}
\begin{equation*}
- \bra{\psi(t)} \frac{\hat{H}}{i\hbar}\hat{p}\ket{\psi(t)}  + \bra{\psi(t)}\hat{p}\frac{\hat{H}}{i\hbar}\ket{\psi(t)} = \bra{\psi(t)}-V'(\hat{q})\ket{\psi(t)}
\end{equation*}
You can expand the vector `sandwich' and rewrite it in terms of a commutator to get:
\begin{equation*}
\bra{\psi(t)} -\hat{H}\hat{q} + \hat{q}\hat{H}\ket{\psi(t)} = - \bra{\psi(t)} [\hat{H},\hat{q}]\ket{\psi(t)}= \frac{i\hbar}{m}\bra{\psi(t)}\hat{p}\ket{\psi(t)}
\end{equation*}
\begin{equation*}
\bra{\psi(t)} -\hat{H}\hat{p}  + \hat{p}\hat{H}\ket{\psi(t)} = - \bra{\psi(t)} [\hat{H},\hat{p}]\ket{\psi(t)}= \bra{\psi(t)}-i\hbar V'(\hat{q})\ket{\psi(t)}
\end{equation*}
Since the Ehrenfest Theorems are true regardless of the particular state of the system $\ket{\psi(t)}$ (i.e., they are true for \textit{any} state you might find), we can remove the wavefunction vector from both sides to get:
\begin{equation*}
 - [\hat{H},\hat{q}] = \frac{i\hbar}{m}\hat{p}
\end{equation*}
\begin{equation}\label{eq:E}
  [\hat{H},\hat{p}]= i\hbar V'(\hat{q})
\end{equation}
We will make the assumption that $\hat{H} = H(\hat{q},\hat{p})$ and utilize Theorem VI from Appendix A. By this Theorem, we have $\hat{q} = \hat{A}_1$, $\hat{p} = \hat{A}_2$, and $H(\hat{q},\hat{p}) = f(\hat{A}_1, \hat{A}_2)$. Therefore:

$$[\hat{H},\hat{q}] = [H(\hat{q},\hat{p}),\hat{q}] =  \sum_{k = 1} ^{n=2}  [\hat{A}_k, \hat{q}] \frac{\partial H(\hat{A}_1, \hat{A}_2)}{\partial \hat{A}_k} = [\hat{q},\hat{q}] \frac{\partial H(\hat{q}, \hat{p})}{\partial \hat{q}} + [\hat{p},\hat{q}] \frac{\partial H(\hat{q}, \hat{p})}{\partial \hat{p}}$$

$$[\hat{H},\hat{p}] = [H(\hat{q},\hat{p}),\hat{p}] =  \sum_{k = 1} ^{n=2}  [\hat{A}_k, \hat{p}] \frac{\partial H(\hat{A}_1, \hat{A}_2)}{\partial \hat{A}_k} = [\hat{q},\hat{p}] \frac{\partial H(\hat{q}, \hat{p})}{\partial \hat{q}} + [\hat{p},\hat{p}] \frac{\partial H(\hat{q}, \hat{p})}{\partial \hat{p}}$$
A commutator with itself will always equal zero:

$$[\hat{q},\hat{q}] = \hat{q}\hat{q} - \hat{q}\hat{q} = 0 $$
$$[\hat{p},\hat{p}] = \hat{p}\hat{p} - \hat{p}\hat{p} = 0 $$
Ergo, the above expressions based in Theorem VI reduce to:
$$[\hat{H},\hat{q}] = [\hat{p},\hat{q}] \frac{\partial H(\hat{q}, \hat{p})}{\partial \hat{p}}$$

$$[\hat{H},\hat{p}] = [\hat{q},\hat{p}] \frac{\partial H(\hat{q}, \hat{p})}{\partial \hat{q}}$$
Plugging these expressions into \eqref{eq:E} gives us:
$$  - [\hat{p},\hat{q}] \frac{\partial H(\hat{q}, \hat{p})}{\partial \hat{p}} = \frac{i\hbar}{m}\hat{p}$$
\begin{equation}\label{eq:G}
[\hat{q},\hat{p}] \frac{\partial H(\hat{q}, \hat{p})}{\partial \hat{q}} = i\hbar V'(\hat{q}) 
\end{equation}
From the Canonical  Commutator we know that $[\hat{q},\hat{p}] = i\hbar$.\footnote{We will also use the fact that $[\hat{q},\hat{p}] = - [\hat{p},\hat{q}]$. It's not hard to prove, just takes a couple lines of algebra:

$$[\hat{q},\hat{p}] = \hat{q}\hat{p} - \hat{p}\hat{q} = - ( -\hat{q}\hat{p} + \hat{p}\hat{q})  = - (  \hat{p}\hat{q} -\hat{q}\hat{p}) = - [\hat{p},\hat{q}]$$} 
This step is absolutely critical. It is in fact the step that is going to distinguish QM from classical KvNM - the usage of the Canonical Commutator. We are going to plug in the Canonical Commutator to the above expressions, so that we will get:
\begin{equation*}
\frac{\partial H(\hat{q},\hat{p})}{\partial \hat{p}} = \frac{ \hat{p}}{m}
\end{equation*}
\begin{equation*}
\frac{\partial H(\hat{q},\hat{p})}{\partial \hat{q}} = \frac{dV(\hat{q})}{d\hat{q}}
\end{equation*}
Theorem VII of Appendix A will be useful, as we can now translate this expression with operators into more familiar analytical functions in order to do operations with derivations. We will have the following analytical functions as a result of Theorem VII:

\begin{equation*}
\frac{\partial H(q,p)}{\partial p} = \frac{ p}{m}
\end{equation*}
\begin{equation*}
\frac{\partial H(q,p)}{\partial q} = \frac{dV(q)}{dq}
\end{equation*}
Some basic calculus reveals:
\begin{equation*}
\int \frac{\partial H(q,p)}{\partial p} dp = H(q,p) = \int \frac{ p}{m} dp = \frac{p^2}{2m} + C(q)
\end{equation*}
where $C(q)$ is some function solely in terms of $q$ so that it disappears when the partial with respect to $p$ is taken. Taking the partial derivative with respect to $q$ of the above expression reveals:

\begin{equation*}
\frac{\partial H(q,p)}{\partial q} = \frac{d C(q)}{dq} = \frac{dV(q)}{dq}
\end{equation*}
Ergo, $C(q) = V(q)$ so that the final solution is $H(\hat{q},\hat{p}) = \frac{\hat{p}^2}{2m} + V(\hat{q})$, which happens to be equivalent to the quantum Hamiltonian. Therefore, plugging this solution of $\hat{H}$ into the above version of Stone's Theorem we reclaim the Schr\"{o}dinger equation\eqref{eq:schrodinger_equation}:

\begin{equation*}
i\hbar \frac{\partial}{\partial t} \ket{\psi} = \Big[\frac{\hat{p}^2 }{2m}+ V(\hat{q})\Big]\ket{\psi} 
\end{equation*}

The same series of steps can be carried out to derive KvNM. The only difference in the derivation is that we will assume the Classical Commutator instead of the Canonical Commutator. 
Once again we start with the Ehrenfest Theorems and follow a similar series of steps. Starting from \eqref{eq:F}, we will utilize Stone's Theorem as $i\frac{\partial}{\partial t}\ket{\psi} = \hat{K}\ket{\psi}$ to get:
\begin{equation*}
- \bra{\psi(t)} \frac{\hat{K}}{i}\hat{q}\ket{\psi(t)}  + \bra{\psi(t)}\hat{q}\frac{\hat{K}}{i}\ket{\psi(t)} = \frac{1}{m}\bra{\psi(t)}\hat{p}\ket{\psi(t)}
\end{equation*}
\begin{equation*}
- \bra{\psi(t)} \frac{\hat{K}}{i}\hat{p}\ket{\psi(t)}  + \bra{\psi(t)}\hat{p}\frac{\hat{K}}{i}\ket{\psi(t)} = \bra{\psi(t)}-V'(\hat{q})\ket{\psi(t)}
\end{equation*}
Following a similar series of steps, we can derive that:
\begin{equation*}
 - [\hat{K},\hat{q}] = \frac{i}{m}\hat{p}
\end{equation*}
\begin{equation}
  [\hat{K},\hat{p}]= i V'(\hat{q})
\end{equation}
Now we run into a problem. If we assume $\hat{K} = K(\hat{q},\hat{p})$ as before, then we will derive once again \eqref{eq:G}. However, now we will utilize the Classical Commutator instead of the Canonical Commutator, since we want well defined classical trajectories. By imposing the Classical Commutator $[\hat{q},\hat{p}] = 0$, we run into a contradiction, namely:

$$  - [\hat{p},\hat{q}] \frac{\partial K(\hat{q}, \hat{p})}{\partial \hat{p}} = (0) \frac{\partial K(\hat{q}, \hat{p})}{\partial \hat{p}} = 0 =\frac{i}{m}\hat{p} \Longrightarrow p = 0$$
$$ [\hat{q},\hat{p}] \frac{\partial K(\hat{q}, \hat{p})}{\partial \hat{q}} = (0)\frac{\partial K(\hat{q}, \hat{p})}{\partial \hat{q}}  = 0 = i V'(\hat{q}) \Longrightarrow \frac{dV}{dq} = 0$$
Obviously, this cannot be true for all systems, because many classical systems definitely have a nonzero momentum. This contradiction between theory and observation suggests that we must approach this from a different angle. One such angle is to propose two new operators $\hat{\theta}$ and $\hat{\lambda}$ (\cite{ODM}). These operators obey the eigenvalue problems, $\hat{\theta}\ket{\theta} = \theta\ket{\theta}$ and $\hat{\lambda}\ket{\lambda} = \lambda\ket{\lambda}$, by postulate 2 of the `universal axioms', but are not tied to any direct physically observable features of reality (as noted in \cite{harmonic_oscillator}).\footnote{Even though these operators are not directly physically observable, their existence has been seen through indirect means. These operators are the famous Bopp Operators of the phase-space formulation of QM (\cite{wigner}). We will explore the Wigner Distribution in section 6.} They serve as an analogue to Lagrange Multipliers (\cite{time_arrow}), serving to constrain the realm of possible behavior to a very specific subset.

For KvNM, we will postulate the following behavior for these operators:

$$[\hat{q},\hat{\theta}] = [\hat{p},\hat{\lambda}] = i$$
\begin{equation}\label{eq:koopman_algebra}
[\hat{q}_c, \hat{p}_c] = [\hat{q}, \hat{\lambda}] = [\hat{p}, \hat{\theta}] = [\hat{\theta}, \hat{\lambda}] = 0
\end{equation}
where $i$ is the imaginary unit. Equations \eqref{eq:koopman_algebra} make up what is known as the \textit{Koopman-von Neumann Algebra}. If we do this, then we can assume $\hat{K} = K(\hat{q},\hat{p}, \hat{\theta},\hat{\lambda})$. By Theorem VI of Appendix A again, we will get:

$$[K(\hat{q},\hat{p}, \hat{\theta},\hat{\lambda}),\hat{q}] =  \sum_{k = 1} ^{n=4}  [\hat{A}_k, \hat{q}] \frac{\partial K(\hat{A}_1, \hat{A}_2, \hat{A}_3, \hat{A}_4)}{\partial \hat{A}_k} = [\hat{q},\hat{q}] \frac{\partial K(\hat{q}, \hat{p},\hat{\theta},\hat{\lambda})}{\partial \hat{q}} + [\hat{p},\hat{q}] \frac{\partial K(\hat{q}, \hat{p},\hat{\theta},\hat{\lambda})}{\partial \hat{p}} $$

$$+ [\hat{\theta},\hat{q}] \frac{\partial K(\hat{q}, \hat{p},\hat{\theta},\hat{\lambda})}{\partial \hat{\theta}}+ [\hat{\lambda},\hat{q}] \frac{\partial K(\hat{q}, \hat{p},\hat{\theta},\hat{\lambda})}{\partial \hat{\lambda}}$$

$$[K(\hat{q},\hat{p}, \hat{\theta},\hat{\lambda}),\hat{p}] =  \sum_{k = 1} ^{n=4}  [\hat{A}_k, \hat{p}] \frac{\partial K(\hat{A}_1, \hat{A}_2, \hat{A}_3, \hat{A}_4)}{\partial \hat{A}_k} = [\hat{q},\hat{p}] \frac{\partial K(\hat{q}, \hat{p},\hat{\theta},\hat{\lambda})}{\partial \hat{q}} + [\hat{p},\hat{p}] \frac{\partial K(\hat{q}, \hat{p},\hat{\theta},\hat{\lambda})}{\partial \hat{p}}$$

$$+ [\hat{\theta},\hat{p}] \frac{\partial K(\hat{q}, \hat{p},\hat{\theta},\hat{\lambda})}{\partial \hat{\theta}}+ [\hat{\lambda},\hat{p}] \frac{\partial K(\hat{q}, \hat{p},\hat{\theta},\hat{\lambda})}{\partial \hat{\lambda}}$$
Three terms from each expression will disappear because of the KvN Algebra above and the Classical Commutator. The equations of motion will end up being:

\begin{equation*}
  [\hat{\theta},\hat{q}] \frac{\partial K(\hat{q}, \hat{p},\hat{\theta},\hat{\lambda})}{\partial \hat{\theta}} = -\frac{i}{m}\hat{p}
\end{equation*}
\begin{equation}
  [\hat{\lambda},\hat{p}] \frac{\partial K(\hat{q}, \hat{p},\hat{\theta},\hat{\lambda})}{\partial \hat{\lambda}} = i V'(\hat{q})
\end{equation}
Once we substitute in $[\hat{q},\hat{\theta}] = [\hat{p},\hat{\lambda}] = i$ we will simply have:

\begin{equation*}
  \frac{\partial K(\hat{q}, \hat{p},\hat{\theta},\hat{\lambda})}{\partial \hat{\theta}} = \frac{\hat{p}}{m}
\end{equation*}
\begin{equation*}
  \frac{\partial K(\hat{q}, \hat{p},\hat{\theta},\hat{\lambda})}{\partial \hat{\lambda}} = - V'(\hat{q})
\end{equation*}
Utilizing Theorem VII again, we get the differential equations:
\begin{equation*}
  \frac{\partial K(q,p,\theta,\lambda)}{\partial \theta} = \frac{p}{m}
\end{equation*}
\begin{equation*}
  \frac{\partial K(q, p,\theta,\lambda)}{\partial \lambda} = - V'(q)
\end{equation*}
Which we proceed to solve:
\begin{equation*}
\int \frac{\partial K(q,p,\theta,\lambda)}{\partial \theta} d\theta=  K(q, p,\theta,\lambda) = \int \frac{p}{m} d\theta = \frac{p\theta}{m} + C(q,p,\lambda)
\end{equation*}
\begin{equation*}
\frac{\partial}{\partial \lambda} K(q,p,\theta, \lambda) = \frac{\partial}{\partial \lambda} (\frac{p\theta}{m} + C(q,p,\lambda)) = \frac{\partial C(q,p,\lambda)}{\partial \lambda} = -\frac{V(q)}{dq}
\end{equation*}
\begin{equation*}
\int \frac{\partial C(q,p,\lambda)}{\partial \lambda} d\lambda = C(q,p,\lambda) = - \int \frac{V(q)}{dq} d\lambda = -\frac{V(q)}{dq}\lambda + C(q,p)
\end{equation*}
We can therefore derive

\begin{equation}\label{eq:koopman_generator}
\hat{K} = K(\hat{q}, \hat{p},\hat{\theta},\hat{\lambda}) = \frac{\hat{p}\hat{\theta}}{m} - V'(\hat{q})\hat{\lambda} + C(\hat{q},\hat{p})
\end{equation}
which is the important \textit{Koopman Generator} of KvNM. The Koopman Generator will become very important when we discuss the Path Integral Formulation of QM and CM (section 4). 

We can show that the Koopman Generator gives us the Louiville Equations again and produces the KvN classical wavefunction. To show this is quite straightforward. Since the Born Rule for KvNM postulates that $\rho = |\braket{q,p}{\psi(t)}|^2$, let us try to use the Koopman Generator to find an expression that governs the evolution of the classical probability density. 

First, refer to \eqref{eq:basic_properties_KvN}. From these basic properties, one can easily show the following:

\begin{equation}\label{eq:J}
\bra{q,p}\hat{q}\ket{\psi(t)} = [\bra{q,p}\hat{q}]\ket{\psi(t)} = \bra{q,p}q\ket{\psi(t)} = q\braket{q,p}{\psi(t)}
\end{equation}

\begin{equation}\label{eq:K}
\bra{q,p}\hat{p}\ket{\psi(t)} = [\bra{q,p}\hat{p}]\ket{\psi(t)} = \bra{q,p}p\ket{\psi(t)} = p\braket{q,p}{\psi(t)}
\end{equation}
Using Theorem IV from Appendix A, we can demonstrate that the following are true. Since $[\hat{q},\hat{\theta}] = i$, we can set $\kappa = 1$, $\ket{A} = \ket{q,p}$, and $B = \theta$ for the parameters in Theorem IV. We can say that the following is true then:

\begin{equation}\label{eq:H}
\bra{q, p}\hat{\theta}\ket{\psi(t)} = -i\frac{\partial}{\partial q}\braket{q,p}{\psi(t)}
\end{equation}
Notice that although $A = (q,p)$, we do not write a partial derivative with respect to $p$, and that is because $[\hat{p},\hat{\theta}] = 0$ per the Koopman Algebra. 

The same can be done with $[\hat{p},\hat{\lambda}] = i$. Using Theorem IV with parameters $\kappa = 1$, $\ket{A} = \ket{q,p}$, and $B = \lambda$, we get:
\begin{equation}\label{eq:I}
\bra{q, p}\hat{\lambda}\ket{\psi(t)} = -i\frac{\partial}{\partial p}\braket{q,p}{\psi(t)}
\end{equation}
Again, there is no partial with respect to $q$ here, since $[\hat{q},\hat{\lambda}] = 0$ per the Koopman Algebra. 

Plugging the Koopman Generator into Stone's Theorem gives us:
$$ i \frac{\partial}{\partial t} \ket{\psi(t)} = \hat{K}\ket{\psi(t)} = \Big[\frac{\hat{p}\hat{\theta}}{m} - V'(\hat{q})\hat{\lambda} + C(\hat{q},\hat{p}) \Big] \ket{\psi(t)}$$
And now we will `sandwich' it from the left hand side with the vector bra $\bra{q,p}$ and work out the algebra:

$$\bra{q,p} i \frac{\partial}{\partial t} \ket{\psi(t)} = \bra{q,p}\Big[\frac{\hat{p}\hat{\theta}}{m} - V'(\hat{q})\hat{\lambda} + C(\hat{q},\hat{p}) \Big] \ket{\psi(t)}$$

$$ i \frac{\partial}{\partial t} \braket{q,p}{\psi(t)} = \bra{q,p}\frac{\hat{p}\hat{\theta}}{m}  \ket{\psi(t)} - \bra{q,p}V'(\hat{q})\hat{\lambda} \ket{\psi(t)} + \bra{q,p}C(\hat{q},\hat{p}) \ket{\psi(t)}$$
Utilizing Theorems I of Appendix A\footnote{This Theorem tells us that $V'(\hat{q})\ket{q,p} = V'(q)\ket{q,p}$ and $C(\hat{q},\hat{p})\ket{q,p} = C(q,p)\ket{q,p}$}and equations \eqref{eq:J},\eqref{eq:K},\eqref{eq:H} and \eqref{eq:I}, we can derive:
$$ i \frac{\partial}{\partial t} \braket{q,p}{\psi(t)} = \frac{p}{m} \bra{q,p}\hat{\theta}\ket{\psi(t)} - V'(q)\bra{q,p}\hat{\lambda}\ket{\psi(t)} + C(q,p)\braket{q,p}{\psi(t)}$$

\begin{equation}\label{eq:L}
 i \frac{\partial}{\partial t} \braket{q,p}{\psi(t)} = -i\frac{p}{m}\frac{\partial}{\partial q}\braket{q,p}{\psi(t)} +iV'(q)\frac{\partial}{\partial p}\braket{q,p}{\psi(t)} + C(q,p)\braket{q,p}{\psi(t)}
\end{equation}
We can derive the time evolution equation for the probability density by utilizing equation \eqref{eq:L} above. 
This will be done in this fashion:

\begin{equation*}
 i \frac{\partial}{\partial t} |\braket{q,p}{\psi(t)}|^2 = i \frac{\partial}{\partial t} [\braket{\psi(t)}{q,p} \braket{q,p}{\psi(t)}] =
\end{equation*}
$$ = i \frac{\partial}{\partial t} [\braket{\psi(t)}{q,p}]\braket{q,p}{\psi(t)} +  i\braket{\psi(t)}{q,p} \frac{\partial}{\partial t} [\braket{q,p}{\psi(t)}] $$
Into this equation plug in \eqref{eq:L} and its complex conjugate:
$$ i \frac{\partial}{\partial t} |\braket{q,p}{\psi(t)}|^2 = \Big[-i\frac{p}{m}\frac{\partial}{\partial q}\braket{\psi(t)}{q,p} +iV'(q)\frac{\partial}{\partial p}\braket{\psi(t)}{q,p} - C(q,p)\braket{\psi(t)}{q,p}\Big]\braket{q,p}{\psi(t)}$$
$$+  \braket{\psi(t)}{q,p} \Big[-i\frac{p}{m}\frac{\partial}{\partial q}\braket{q,p}{\psi(t)} +iV'(q)\frac{\partial}{\partial p}\braket{q,p}{\psi(t)} + C(q,p)\braket{q,p}{\psi(t)}\Big] =$$
\newline
$$ = -i\frac{p}{m}\braket{q,p}{\psi(t)}\frac{\partial}{\partial q}\braket{\psi(t)}{q,p} +iV'(q)\braket{q,p}{\psi(t)}\frac{\partial}{\partial p}\braket{\psi(t)}{q,p} + $$
$$ - \braket{q,p}{\psi(t)}C(q,p)\braket{\psi(t)}{q,p} -i\braket{\psi(t)}{q,p}\frac{p}{m}\frac{\partial}{\partial q}\braket{q,p}{\psi(t)}$$
$$ +i\braket{\psi(t)}{q,p}V'(q)\frac{\partial}{\partial p}\braket{q,p}{\psi(t)} + \braket{\psi(t)}{q,p}C(q,p)\braket{q,p}{\psi(t)} =$$
\newline
$$ = -i\frac{p}{m}\braket{q,p}{\psi(t)}\frac{\partial}{\partial q}\braket{\psi(t)}{q,p}  -i\braket{\psi(t)}{q,p}\frac{p}{m}\frac{\partial}{\partial q}\braket{q,p}{\psi(t)} $$
$$+iV'(q)\braket{q,p}{\psi(t)}\frac{\partial}{\partial p}\braket{\psi(t)}{q,p} + i\braket{\psi(t)}{q,p}V'(q)\frac{\partial}{\partial p}\braket{q,p}{\psi(t)}$$
Recognizing the involvement of chain rule applied to the wavefunction:
$$ i \frac{\partial}{\partial t} |\braket{q,p}{\psi(t)}|^2 = -i\frac{p}{m}\frac{\partial}{\partial q}|\braket{q,p}{\psi(t)}|^2 +  iV'(q)\frac{\partial}{\partial p}  |\braket{q,p}{\psi(t)}|^2 $$
Since $ |\braket{q,p}{\psi(t)}|^2  = \rho$, the probability density, we can see right away that this equation is fully equivalent to the Louiville equation \eqref{eq:louiville_equation}\footnote{Recall from Classical Mechanics that 

$$ F = - \frac{\partial H}{\partial q} = - V'(q)$$
$$ \frac{dq}{dt} = \frac{\partial H}{\partial p} = \frac{p}{m} $$}, and that $ \braket{q,p}{\psi(t)} = \psi(q,p,t)$ is the previously identified KvN classical wavefunction in phase space. 

In summary, KvNM is a classical theory of physics that utilizes complex Hilbert Spaces, just like in QM. Like QM, it is probabilistic in nature, so it can only make probabilistic predictions of classical systems. Like in QM, you have a wavefunction that completely describes the state of the system \eqref{eq:KvN_L_wavefunction}. This wavefunction is in phase space, and KvNM is a phase space theory. It operates on the same axioms as QM, but is a fully classical theory. It is not an extension of classical theory, but fully equivalent to it, recast in a non-traditional form. The mathematical conveniences of QM exist in KvNM. For example, one reason QM has its odd Hilbert Space form is to conveniently change between vector bases. One can do this in KvNM with equal ease. Just like in QM, the expectation value of observable $A$ is given by $\langle A \rangle = \bra{\psi(t)} \hat{A} \ket{\psi(t)}$. 

The backbone of KvNM lies in the standard Hilbert Space behaviors \eqref{eq:basic_properties_KvN}, the Koopman Generator \eqref{eq:koopman_generator}, and the Koopman Algebra \eqref{eq:koopman_algebra}, which lets one derive the Louiville Equation governing classical probabilities via usage of the KvNM classical wavefunction. One can derive KvNM from the Ehrenfest Theorems by assuming the Classical Commutator. The main difference between Classical and Quantum Physics appears to lie in the choice of the commutator for position $q$ and momentum $p$ (\cite{ODM}).

In the pages that follow, we will demonstrate the many applications of KvNM. Even though technically it is strictly equivalent to any classical theory, it allows for a wide variety of applications and uses. KvNM is a very important and convenient tool for statistical physics.

\section{Path Integral Formulation of Quantum and Classical Mechanics}

\subsection{Feynman Path Integral of Quantum Mechanics}

Feynman Path Integral is another formulation of QM that is completely equivalent to the Schr\"{o}dinger equation. It was invented by the famous physicist Richard Feynman in 1948 and has applications many usages in QM, some which you will see in this Thesis.  Whereas the Schr\"{o}dinger equation is more analogous to Hamiltonian Dynamics (for instance, see section 5.1 for the similarity to Hamilton-Jacobi Theory), the Path Integral is more analogous to Lagrangian Dynamics (\cite{QM}).

Under Lagrangian Dynamics in Classical Theory, you select two points, a starting point and an endpoint point. In the allowed time interval, the classical particle would take the path where the function $S$, called the Action, would be minimized. From CM, the term $T-V $ is called the Lagrangian (where $T$ is kinetic energy and $V$ is potential energy) and the Action is defined as the time integral of the Lagrangian:

\begin{equation}\label{eq:action}
S = \int_0^t dt (T - V) 
\end{equation}
This function has to be minimized to define the classical path. The minimization rule for classical particles - called the Least Action Principle or Hamilton's Principle - is often written as:

$$ \delta  \int_0^t dt (T - V) = 0 $$
Things are not so simple in Quantum Theory. 

We begin with a simple derivation of the Feynman Path Integral to understand the differences between QM and Lagrangian Theory. We start with the following expression, which is simply a rewrite of the Schr\"{o}dinger equation:

\begin{equation}\label{eq:propogator_relationship}
\ket{\psi(t)} = e^{-\frac{i}{\hbar}t\hat{H}}\ket{\psi(0)}
\end{equation}
In Appendix C, one can see how this expression is completely equivalent to the Schr\"{o}dinger equation. One might also notice that the operator $\hat{H}$ is exponated, but this is not unusual in QM (more information about exponated operators and what they mean is found in the same Appendix). The term $\exp{-\frac{i}{\hbar}t\hat{H}}$ is called the unitary time-evolution operator and defines how the quantum particle will evolve in time from its starting condition $\ket{\psi(0)}$ to its final condition $\ket{\psi(t)}$ at time $t$ (see Appendix B for more on unitary operators).

To begin, we insert a position $q$ bra on both sides of \eqref{eq:propogator_relationship}:

\begin{equation}
\bra{q}\ket{\psi(t)} = \psi(q,t) = \bra{q} e^{-\frac{i}{\hbar}t\hat{H}}\ket{\psi(0)}
\end{equation}
And then insert resolution of identity (Appendix A Theorem II) for initial position $q'$:

\begin{equation}\label{eq:feynman_path_integral_formulism_1}
\bra{q}\ket{\psi(t)} = \psi(q,t) = \int dq' \bra{q} e^{-\frac{i}{\hbar}t\hat{H}}\ket{q'}\bra{q'}\ket{\psi(0)} = \int dq' \bra{q} e^{-\frac{i}{\hbar}t\hat{H}}\ket{q'}\psi(q',0)
\end{equation}
The term $\bra{q} e^{-\frac{i}{\hbar}t\hat{H}}\ket{q'}$ is called the \textit{kernell} of the Path Integral. The kernell represents the probability amplitude while traveling from the starting point to the final point. Think of the $\ket{q'} = \ket{q_0}$ as the starting point in the Lagrangian formulism and $\ket{q} =\ket{q_f}$ as the final prespecified point.

Let $\Delta t$ be a very small change in time, where $\Delta t = t_{j+1} - t_j$. We will break the total allowed time $t$ into $N$ equal increments, so that we have $\Delta t = t/N$. We can therefore write:
\begin{equation*}
 \bra{q} e^{-\frac{i}{\hbar}t\hat{H}}\ket{q'} = \bra{q} e^{-\frac{i}{\hbar}(\Delta t_1 + \Delta t_2 + ... + \Delta t_N)\hat{H}}\ket{q'} =  \bra{q} e^{-\frac{i}{\hbar}\Delta t_1 \hat{H} -\frac{i}{\hbar}\Delta t_2\hat{H} + ... -\frac{i}{\hbar}\Delta t_N\hat{H}}\ket{q'}
\end{equation*}
\begin{equation*}
 \bra{q} e^{-\frac{i}{\hbar}t\hat{H}}\ket{q'}  =  \bra{q} e^{-\frac{i}{\hbar}\Delta t_1 \hat{H}}e^{ -\frac{i}{\hbar}\Delta t_2\hat{H}} \cdot \cdot \cdot e^{-\frac{i}{\hbar}\Delta t_N\hat{H}}\ket{q'}
\end{equation*}
The subscripts on each of the $\Delta t$ are there just to make clear that we are breaking the total time $t$ into a large number of $N$ increments. Since the $\Delta t$ here are very small, we will drop the subscripts from here on out. Just keep in mind there are now a total $N$ exponential terms inside the above expression. 

Using Theorem II of Appendix A we can insert between each exponential the resolution of identity: 

$$ \bra{q_f} e^{-\frac{i}{\hbar}\Delta t \hat{H}}e^{ -\frac{i}{\hbar}\Delta t\hat{H}}\cdot \cdot \cdot e^{-\frac{i}{\hbar}\Delta t\hat{H}}\ket{q_0} = \bra{q_f} e^{-\frac{i}{\hbar}\Delta t \hat{H}}Ie^{ -\frac{i}{\hbar}\Delta t\hat{H}}I \cdot \cdot \cdot Ie^{-\frac{i}{\hbar}\Delta t\hat{H}}\ket{q_0} =$$

\begin{equation*}
 = \int \int \cdot \cdot \cdot \int  dq_{N-1} dq_{N-2} \cdot \cdot \cdot dq_1 \bra{q_f} e^{-\frac{i}{\hbar}\Delta t \hat{H}}\ket{q_{N-1}}\bra{q_{N-1}}e^{ -\frac{i}{\hbar}\Delta t\hat{H}} \ket{q_{N-2}}\times 
\end{equation*}
\begin{equation}\label{eq:very_long_eq}
\times \bra{q_{N-2}}\cdot \cdot \cdot \ket{q_{1}}\bra{q_{1}}e^{-\frac{i}{\hbar}\Delta t\hat{H}}\ket{q_0} 
\end{equation}

We will take one of the ``bra-ket'' sandwiches above and do some further operations on it. We can represent each ``sandwich'' as $\bra{q_{j+1}} e^{-\frac{i}{\hbar}\Delta t\hat{H}}\ket{q_{j}}$ and plug in the value of the Hamiltonian operator $\hat{H}$:

\begin{equation*}
 \bra{q_{j+1}} e^{-\frac{i}{\hbar}\Delta t\hat{H}}\ket{q_{j}}  =   \bra{q_{j+1}} e^{-\frac{i}{\hbar}\Delta t [\frac{\hat{p}^2}{2m} + V(\hat{q})]}\ket{q_{j}} = \bra{q_{j+1}} e^{-\frac{i}{\hbar}\Delta t \frac{\hat{p}^2}{2m}} e^{-\frac{i}{\hbar}\Delta t  V(\hat{q})}\ket{q_{j}}
\end{equation*}
From Theorem I of Appendix A  we know that:

\begin{equation*}
 e^{-\frac{i}{\hbar}\Delta t  V(\hat{q})}\ket{q_{j}} =  e^{-\frac{i}{\hbar}\Delta t  V({q_j})}\ket{q_{j}}
\end{equation*}
Next we apply the completeness theorem with momentum $p$:

\begin{equation*}
 \bra{q_{j+1}} e^{-\frac{i}{\hbar}\Delta t \frac{\hat{p}^2}{2m}} I e^{-\frac{i}{\hbar}\Delta t  V(\hat{q})}\ket{q_{j}} =  \int dp \bra{q_{j+1}} e^{-\frac{i}{\hbar}\Delta t \frac{\hat{p}^2}{2m}}\ket{p}\bra{p} e^{-\frac{i}{\hbar}\Delta t  V({q}_j)}\ket{q_{j}}
\end{equation*}
Again applying Theorem I we know:

\begin{equation*}
 e^{-\frac{i}{\hbar}\Delta t \frac{\hat{p}^2}{2m}}\ket{p} =  e^{-\frac{i}{\hbar}\Delta t \frac{{p}^2}{2m}}\ket{p}
\end{equation*}
so that the whole thing becomes

\begin{equation*}
 \bra{q_{j+1}} e^{-\frac{i}{\hbar}\Delta t \frac{\hat{p}^2}{2m}} e^{-\frac{i}{\hbar}\Delta t  V(\hat{q})}\ket{q_{j}} =  \int dp \bra{q_{j+1}}\ket{p}\bra{p}\ket{q_{j}}  e^{-\frac{i}{\hbar}\Delta t [\frac{p^2}{2m} + V({q_j})]}
\end{equation*}

Since QM obeys the Canonical Commutator $[\hat{q},\hat{p}] = i\hbar$, we know what $\bra{q_{j+1}}\ket{p}$ and $\bra{p}\ket{q_j}$ are from Theorem V of Appendix A where we set $\kappa = \hbar$. Plugging these results into the above equation we get:

\begin{equation*}
 \bra{q_{j+1}} e^{-\frac{i}{\hbar}\Delta t \frac{\hat{p}^2}{2m}} e^{-\frac{i}{\hbar}\Delta t  V(\hat{q})}\ket{q_{j}} =  \int dp~ e^{-\frac{i}{\hbar}\Delta t [\frac{p^2}{2m} + V({q_j})]}\frac{1}{2\pi \hbar}e^{\frac{i}{\hbar}pq_{j+1}}e^{-\frac{i}{\hbar}pq_{j}} = 
\end{equation*}

$$ = \int dp~ e^{-\frac{i}{\hbar}\Delta t [\frac{p^2}{2m} + V({q_j})]}\frac{1}{2\pi \hbar}e^{\frac{i}{\hbar}p(q_{j+1}-q_{j})} $$
This is known as a Gaussian Integral. From Theorem VIII of Appendix A, we know how to compute Gaussian Integrals. We simply set $a = \frac{i}{\hbar}\frac{\Delta t}{2m}$, $b = \frac{i}{\hbar}(q_{j+1} - q_{j})$, and $c = - \frac{i}{\hbar}\Delta t V(q_j)$ in this Theorem and conduct the following steps:

$$ \frac{1}{2\pi \hbar} \int dp~ e^{-\frac{i}{\hbar}\frac{\Delta t}{2m} p^2 + \frac{i}{\hbar}(q_{j+1}-q_{j})p  -\frac{i}{\hbar}\Delta t V({q_j})} = \frac{1}{2\pi \hbar}\sqrt{\frac{\pi}{a}}e^{\frac{b^2}{4a}+c} = $$
$$ = \frac{1}{2\pi \hbar}\sqrt{\frac{\pi}{(\frac{i}{\hbar}\frac{\Delta t}{2m})}}\exp[\frac{(\frac{i}{\hbar}(q_{j+1} - q_{j}))^2}{4(\frac{i}{\hbar}\frac{\Delta t}{2m})}- \frac{i}{\hbar}\Delta t V(q_j)]  $$

\begin{equation}\label{eq:Kernel_portion}
 \therefore \bra{q_{j+1}} e^{-\frac{i}{\hbar}\Delta t \hat{H}} \ket{q_j} = \frac{1}{2\pi \hbar}\sqrt{\frac{2\pi m\hbar}{i\Delta t }}\exp[\frac{im(q_{j+1} - q_{j})^2}{2\hbar\Delta t}- \frac{i}{\hbar}\Delta t V(q_j)] 
\end{equation}

Plugging equation \eqref{eq:Kernel_portion} into \eqref{eq:very_long_eq} we get the following for the total time $t$:

\begin{equation*}
\bra{q_f} e^{-\frac{i}{\hbar}t\hat{H}}\ket{q_0} = \Big(\frac{1}{2\pi \hbar}\Big)^N \Big({\frac{2\pi m\hbar}{i\Delta t }}\Big)^{\frac{N}{2}} \int \int \cdot \cdot \cdot \int  dq_{N-1} dq_{N-2} \cdot \cdot \cdot dq_1 \times
\end{equation*}
$$ \times  \exp[\sum_{j = 0} ^{N-1} \Big(\frac{im(q_{j+1} - q_{j})^2}{2\hbar\Delta t}- \frac{i}{\hbar}\Delta t V(q_j)\Big)] $$
This hideous looking expression can be notationally simplified a bit by treating all the integrals with respect to $q_j$ as a product of integrals

$$\int \int \cdot \cdot \cdot \int  dq_{N-1} dq_{N-2} \cdot \cdot \cdot dq_1 = \prod_{j = 0}^{N-1}\int dq_j$$
where $\Pi$ just means product from $j = 0, 1,..., N-1$, similarly like $\Sigma$ means sum from $j = 0, 1,..., N-1$. Compactly written:
\begin{equation}\label{eq:discrete_quantum_propogator}
\bra{q_f} e^{-\frac{i}{\hbar}t\hat{H}}\ket{q_0} = \Big(\frac{1}{2\pi \hbar}\Big)^N \Big({\frac{2\pi m\hbar}{i\Delta t }}\Big)^{\frac{N}{2}} \prod_{j = 0}^{N-1}\int dq_j \exp[\sum_{j = 0} ^{N-1}\Big(\frac{i}{\hbar}\frac{m(q_{j+1} - q_{j})^2}{2\Delta t}- \frac{i}{\hbar}\Delta t V(q_j)\Big)] 
\end{equation}
Factoring out a factor of $\frac{i\Delta t}{\hbar}$ from the exponent:
\begin{equation}
\bra{q_f} e^{-\frac{i}{\hbar}t\hat{H}}\ket{q_0} = \Big(\frac{1}{2\pi \hbar}\Big)^N \Big({\frac{2\pi m\hbar}{i\Delta t }}\Big)^{\frac{N}{2}} \prod_{j = 0}^{N-1}\int dq_j   \exp[\sum_{j = 0} ^{N-1}\Big(\frac{i\Delta t}{\hbar}\Big(\frac{m(q_{j+1} - q_{j})^2}{2\Delta t^2}- V(q_j)\Big)\Big)] 
\end{equation}
Now, let us take the limit as $\Delta t \rightarrow 0$. Since $\Delta t = t/N$, this is equivalent to taking $N \rightarrow \infty$. 

\begin{equation}
\lim_{\Delta t \rightarrow 0} \Big[\frac{q_{j+1} - q_{j}}{\Delta t}\Big]^2 = \Big(\frac{dq}{dt}\Big)^2
\end{equation}
This is simply the definition of a derivative. And: 

\begin{equation}
\lim_{N \rightarrow \infty} \sum_{j = 0}^{N - 1} = \int_0^t
\end{equation}
We end up with:
\begin{equation}
\bra{q_f} e^{-\frac{i}{\hbar}t\hat{H}}\ket{q_0} = \Big(\frac{1}{2\pi \hbar}\Big)^N \Big({\frac{2\pi m\hbar}{i\Delta t }}\Big)^{\frac{N}{2}} \prod_{j = 0}^{N-1}\int dq_j  \exp[\frac{i}{\hbar} \int_0^t dt \Big(\frac{m}{2}\Big(\frac{dq}{dt}\Big)^2- V(q)\Big)] 
\end{equation}

Since $q$ is the position, $\frac{dq}{dt}$ is obviously the velocity. A quick glance will reveal that the exponated term $\frac{m}{2}(\frac{dq}{dt})^2$ is simply the Kinetic Energy $T$. 
Ergo, if you compare the exponent to \eqref{eq:action}, the last term becomes $e^{iS/\hbar}$, which should be familiar. It is exactly the same term that appears in equation \eqref{eq:configuration_space_wavefunction}, the wavefunction represented in radial form. 

We can write some of the messy, ugly terms into a nice format. 
We say that 
$$ \lim_{N \rightarrow \infty} \Big(\frac{-i2\pi m}{\Delta t}\Big)^{\frac{N}{2}} \prod_{j = 0}^{N - 1} \int dq_j = \int Dq(t) $$
And therefore the entire thing can be written simply as:

\begin{equation*}
\bra{q_f} e^{-\frac{i}{\hbar}t\hat{H}}\ket{q_0} =  \int Dq(t) ~ \exp[\frac{i}{\hbar}\int_0^t dt \Big(\frac{m}{2}\Big(\frac{dq}{dt}\Big)^2- V(q)\Big)] 
\end{equation*}

\begin{equation}\label{eq:finished_Kernel}
\bra{q_f} e^{-\frac{i}{\hbar}t\hat{H}}\ket{q_0} =  \int Dq(t) ~ e^{iS/\hbar} 
\end{equation}
Equation \eqref{eq:finished_Kernel} represents the final Kernel of the Feynman Path Integral. This integral is fascinating because it says that for a fixed starting point $\ket{q_0}$ and ending point $\ket{q_f}$, the quantum particle takes \textit{all possible} paths inbetween within the allotted time interval. This is like saying that if you are throwing darts at a dart board, the flying dart \textit{takes all possible paths from your hand to the bullseye}. This shocking outcome highlights another strange property of quantum particles. This goes hand in hand with the earlier comments about quantum particles having ``fuzzy" trajectories.

\subsection{Koopman-von Neumann Classical Path Integral}

Just like a Path Integral can be formulated for Quantum Theory, one can also formulate a path integral for classical theory. The path integral, however, must enforce Newton's Laws and particles must travel along well defined paths instead of the ``fuzzy'' trajectories of QM. This classical path integral is possible thanks to the KvN formulism (\cite{kvn_thesis}; \cite{time_arrow}).

First, in order to develop the KvN Path Integral, let us develop some initial useful tools. These will be ingredients we will use in a future step in making the Path Integral. Using Theorem V of Appendix A, you can compute what expressions like $\bra{q, \lambda}\ket{q' , p} $ are. Let $A = p$, $B = \lambda$, and $\kappa = 1$. Utilizing the Koopman-von Neumann Algebra \eqref{eq:koopman_algebra}, you can get:

\begin{equation}\label{eq:ingredient_1}
\bra{q, \lambda}\ket{q' , p} = \bra{q} \otimes \bra{\lambda} ~ \ket{q'} \otimes \ket{p} = \braket{q}{q'}\braket{\lambda}{p} = \frac{1}{\sqrt{2\pi}}\delta(q - q') e^{-ip\lambda}
\end{equation}
Note that there is no need to evaluate $\braket{q}{p}$ and $\braket{\lambda}{q'}$ since $[\hat{q}, \hat{p}] = 0$ and $[\hat{\lambda}, \hat{q}] = 0$. From \eqref{eq:dirac_delta_1}, recall $\braket{q}{q'} = \delta(q - q')$. You can follow a similar procedure to calculate the following two expressions:

\begin{equation}\label{eq:ingredient_2}
\bra{\theta, p}\ket{q , p'} = \bra{\theta} \otimes \bra{p} ~ \ket{q} \otimes \ket{p'} = \braket{\theta}{q}\braket{p}{p'} = \frac{1}{\sqrt{2\pi}}\delta(p - p') e^{-i\theta q}
\end{equation}

\begin{equation}\label{eq:ingredient_3}
\bra{\theta, p}\ket{q , \lambda} = \bra{\theta} \otimes \bra{p} ~ \ket{q} \otimes \ket{\lambda} = \braket{\theta}{q}\braket{p}{\lambda} = \frac{1}{{2\pi}}e^{ip\lambda - i\theta q}
\end{equation}
The three expressions above are critical ingredients which we will utilize soon.

Now we can begin with the proper formulation of the Path Integral. Similar to \eqref{eq:propogator_relationship}, we will utilize 
\begin{equation}\label{eq:koopman}
\ket{\psi(t)} = e^{-it\hat{K}}\ket{\psi(0)}
\end{equation}
with the Koopman Generator $\hat{K}$ \eqref{eq:koopman_generator} instead of the quantum Hamiltonian $\hat{H}$.\footnote{More information about exponated operators in Appendix C.} Since KvNM deals with phase space, we will attack this expression with a $q,p$ bra on the left to get the following:

\begin{equation}
\bra{q, p}\ket{\psi(t)} = \psi(q,p,t) = \bra{q,p} e^{-it\hat{K}}\ket{\psi(0)}
\end{equation}
And like before, we will introduce the resolution of identity, this time with respect to $q',p'$:
\begin{equation*}
\bra{q, p}\ket{\psi(t)} = \int dq' ~ dp' \bra{q,p} e^{-it\hat{K}}\ket{q', p'}\bra{q',p'}\ket{\psi(0)} =
\end{equation*}
\begin{equation}\label{eq:koopman_path_integral_1}
 = \int dq' ~ dp' \bra{q, p} e^{-\frac{i}{\hbar}t\hat{K}}\ket{q', p' }\psi(q',p', 0)
\end{equation}
The KvN Kernel is $\bra{q, p} e^{-\frac{i}{\hbar}t\hat{K}}\ket{q', p' }$ with $\ket{q',p'} = \ket{q_0, p_0}$ being the initial state of the system in phase space and $\ket{q,p} =\ket{q_f,p_f}$ being the final state of the system.
Again, let $\Delta t$ be a small change in time. As before in the quantum case, it is a very small increment where we will have $\Delta t = t/N$. We will get:

\begin{equation}
 \bra{q,p} e^{-it\hat{K}}\ket{q',p'}  =  \bra{q,p} e^{-i\Delta t_1 \hat{K}}e^{ -i\Delta t_2\hat{K}} \cdot \cdot \cdot e^{-i\Delta t_N\hat{K}}\ket{q',p'}
\end{equation}
The subscripts on each of the $\Delta t$ are there just to make clear that we are breaking the total time $t$ into a large number of $N$ increments. Since the $\Delta t$ here are very small, we will drop the subscripts from here on out. Just keep in mind there are now a total $N$ exponential terms inside the above expression. 

Using Theorem II we can insert between each exponential a completeness relationship, as before:

\begin{equation*}
\bra{q_f,p_f} e^{-i\Delta t \hat{K}}e^{ -i\Delta t\hat{K}}\cdot \cdot \cdot e^{-i\Delta t\hat{K}}\ket{q_0,p_0} = 
\end{equation*}
\begin{equation*} = \int \cdot \cdot \cdot \int  dq_{N-1} dp_{N-1} \cdot \cdot \cdot dq_1 dp_1 \bra{q_f, p_f} e^{-i\Delta t \hat{K}}\ket{q_{N-1}, p_{N-1}} \times
\end{equation*}
\begin{equation}\label{eq:main_equation}
\times \bra{q_{N-1}, p_{N-1}}e^{ -i\Delta t\hat{K}} \ket{q_{N-2},p_{N-2}}\bra{q_{N-2},p_{N-2}}\cdot \cdot \cdot \ket{q_{1}, p_1}\bra{q_{1},p_1}e^{-i\Delta t\hat{K}}\ket{q_0,p_0}
\end{equation}
We will take one of the ``bra-ket'' sandwiches above and do some further operations on it. We can represent each ``sandwich'' as $\bra{q_{j+1}, p_{j+1}} e^{-i\Delta t\hat{K}}\ket{q_{j},p_j}$ and plug in the value of the Koopman Generator $\hat{K}$:
\begin{equation*}
 \bra{q_{j+1}, p_{j+1}} e^{-i\Delta t\hat{K}}\ket{q_{j},p_j}  =   \bra{q_{j+1}, p_{j+1}} \exp(-i\Delta t \Big[ \frac{\hat{p}\hat{\theta}}{m} - V'(\hat{q})\hat{\lambda}\Big])\ket{q_{j}, p_j} = 
\end{equation*}
$$ = \bra{q_{j+1}, p_{j+1}} e^{-i\Delta t\frac{\hat{p}\hat{\theta}}{m}}e^{i\Delta t V'(\hat{q})\hat{\lambda}}\ket{q_{j}, p_j}$$
Please note, in keeping with \cite{time_arrow}, we have omitted the constant of integration $C(\hat{q},\hat{p})$ from \eqref{eq:koopman_generator} in the following calculations. It will not affect the dynamics of the experiment for classical systems (\cite{ODM}; \cite{time_arrow}).  
Next we apply the completeness theorem twice with $\theta, p$ and $q, \lambda$:

\begin{equation*}
 \bra{q_{j+1}, p_{j+1}} e^{-i\Delta t\hat{K}}\ket{q_{j},p_j}  =   \bra{q_{j+1}, p_{j+1}} e^{-i\Delta t\frac{\hat{p}\hat{\theta}}{m}}e^{i\Delta t V'(\hat{q})\hat{\lambda})}\ket{q_{j}, p_j} = 
\end{equation*}
$$ = \bra{q_{j+1}, p_{j+1}} e^{-i\Delta t\frac{\hat{p}\hat{\theta}}{m}}IIe^{i\Delta t V'(\hat{q})\hat{\lambda}}\ket{q_{j}, p_j} = $$
\begin{equation}\label{eq:step}
= \iiiint d\theta ~dp~dq ~ d\lambda \bra{q_{j+1}, p_{j+1}} e^{-i\Delta t\frac{\hat{p}\hat{\theta}}{m}}\ket{\theta,p}\bra{\theta,p}\ket{q, \lambda}\bra{q, \lambda}e^{i\Delta t V'(\hat{q})\hat{\lambda}}\ket{q_{j}, p_j} 
\end{equation}
Note that by Theorem I, we have that
$$ \exp({-i\Delta t\frac{\hat{p}\hat{\theta}}{m}})\ket{\theta,p} = \exp({-i\Delta t\frac{p\theta}{m}})\ket{\theta,p} $$
And also that
$$ \bra{q, \lambda}\exp({i\Delta t V'(\hat{q})\hat{\lambda}}) = \bra{q, \lambda}\exp({i\Delta t V'(q){\lambda}})$$
Therefore, plugging into the expression \eqref{eq:step} we get:
$$\iiiint d\theta ~dp~dq ~ d\lambda \bra{q_{j+1}, p_{j+1}} \ket{\theta,p}\bra{\theta,p}\ket{q, \lambda}\bra{q, \lambda}\ket{q_{j}, p_j}e^{-i\Delta t\frac{p\theta}{m}}e^{i\Delta t V'(q)\lambda}$$
For the expressions $\bra{q_{j+1}, p_{j+1}} \ket{\theta,p}$, $\bra{\theta,p}\ket{q, \lambda}$, and $\bra{q, \lambda}\ket{q_{j}, p_j}$, we substitute in our ingredients we derived earlier \eqref{eq:ingredient_1}, \eqref{eq:ingredient_2}, and \eqref{eq:ingredient_3}:

$$\iiiint d\theta ~dp~dq ~ d\lambda \frac{1}{\sqrt{2\pi}}\delta(p - p') e^{i\theta q_{j+1}} \frac{1}{{2\pi}}e^{ip_{j+1}\lambda - i\theta q_j}\frac{1}{\sqrt{2\pi}}\delta(q - q') e^{-ip_j\lambda}e^{-i\Delta t\frac{p\theta}{m}}e^{i\Delta t V'(q)\lambda}$$
Remember that according to \eqref{eq:int_dirac_delta_functional}, the integrals of Dirac delta functionals will reduce to the following, neater expression:

$$= \iint d\theta ~ d\lambda \frac{1}{\sqrt{2\pi}}e^{i\theta q_{j+1}} \frac{1}{{2\pi}}e^{ip_{j+1}\lambda - i\theta q_j}\frac{1}{\sqrt{2\pi}} e^{-ip_j\lambda}e^{-i\Delta t\frac{p\theta}{m}}e^{i\Delta t V'(q)\lambda}$$
And simplifying:
$$ = \frac{1}{(2\pi)^2}\iint d\theta ~ d\lambda ~ e^{i\theta q_{j+1} - i\theta q_j -i\Delta t\frac{p\theta}{m}} e^{ip_{j+1}\lambda -ip_j\lambda + i\Delta t V'(q)\lambda}$$

Factoring out the observables $\theta$ and $\lambda$ from each exponent we get:
$$ \frac{1}{(2\pi)^2}\iint d\theta ~ d\lambda ~ \exp(i\theta \Delta t \Big[\frac{q_{j+1} - q_j}{\Delta t} -\frac{p}{m}\Big]) \exp(i\lambda \Delta t \Big[\frac{p_{j+1} -p_j}{\Delta t} + V'(q)\Big])$$
Inserting this important expression into equation \eqref{eq:main_equation} we will get that:
\begin{equation*}
\bra{q_f,p_f} e^{-it \hat{K}}\ket{q_0,p_0} = 
\end{equation*}
\begin{equation*} 
= \int \cdot \cdot \cdot \int  dq_{N-1} dp_{N-1} \cdot \cdot \cdot dq_1 dp_1 \bigg[\frac{1}{(2\pi)^2}\iint d\theta_{N-1} ~ d\lambda_{N-1} ~ \exp\Big(i\theta_{N-1} \Delta t \Big[\frac{q_{j+1} - q_j}{\Delta t} -\frac{p_j}{m}\Big]\Big)\times  
\end{equation*}
\begin{equation*}
\times \exp\Big(i\lambda_{N-1} \Delta t \Big[\frac{p_{j+1} -p_j}{\Delta t} + V'(q_j)\Big]\Big)\bigg] \cdot \cdot \cdot \bigg[\frac{1}{(2\pi)^2}\iint d\theta_1 ~ d\lambda_1 ~ \exp\Big(i\theta_1 \Delta t \Big[\frac{q_{j+1} - q_j}{\Delta t} -\frac{p_j}{m}\Big]\Big) \times
\end{equation*}
\begin{equation*}
\times \exp\Big(i\lambda_1 \Delta t \Big[\frac{p_{j+1} -p_j}{\Delta t} + V'(q_j)\Big]\Big)\bigg]
\end{equation*}
We can utilize some of the simplifying notation we used for Feynman's Path Integral:
\begin{equation*}
\bra{q_f,p_f} e^{-it \hat{K}}\ket{q_0,p_0} = \frac{1}{(2\pi)^{2N}}\iiiint \prod_{j = 1}^{N-1} dq_j ~ dp_j ~ d\theta_j ~d\lambda_j \times
\end{equation*}
\begin{equation}\label{eq:discrete_equation}
\times \exp\Big(i \Delta t \sum_{j = 0}^{N-1} \Big[\theta_j \Big(\frac{q_{j+1} - q_j}{\Delta t} -\frac{p_j}{m}\Big)\Big]\Big)  \exp\Big(i \Delta t \sum_{j = 0}^{N-1} \Big[\lambda_j \Big(\frac{p_{j+1} - p_j}{\Delta t} +V'(q_j)\Big)\Big]\Big) 
\end{equation}
Just as before, let $\Delta t \rightarrow 0$ (equivalent to $N \rightarrow \infty$) to get:

$$ \lim_{\Delta t \rightarrow 0} \frac{q_{j+1} - q_j}{\Delta t} = \frac{dq}{dt}$$

$$\lim_{\Delta t \rightarrow 0} \frac{p_{j+1} -p_j}{\Delta t} = \frac{dp}{dt}$$

\begin{equation*}
\lim_{N \rightarrow \infty} \sum_{j = 0}^{N - 1} = \int_0^t
\end{equation*}
One more piece of notation. Just as before, let us define that for a variable $f$, that we will have:

\begin{equation}
Df = \lim_{N \rightarrow \infty} \prod_{n}^{N} \frac{df_n}{\sqrt{2\pi}}
\end{equation}
And so the classical analog of the path integral can be written:

\begin{equation}\label{eq:equation_to_keep_track_of}
\bra{q_f,p_f} e^{-it \hat{K}}\ket{q_0,p_0} = \iiiint Dq ~Dp ~ D\theta ~ D\lambda~ \exp\bigg(i \int_0^t dt \Big[ \lambda(t) \Big(\frac{dq}{dt} - \frac{p}{m}\Big)+\theta(t)\Big(\frac{dp}{dt} + V'(q)\Big)\Big]\bigg)
\end{equation}
What is the relevance of this? This encodes the equations of CM (\cite{time_arrow}). The path integral is a kind of redundancy that can be shown to enforce the laws of CM. We can show that it leads to a Dirac delta functional that enforces Newton's Laws. 

Recall that the Dirac delta functional $\delta(A - A')$ is nonzero when $A = A'$ but zero if $A \neq A'$. Let us use equation \eqref{eq:discrete_equation} above, as it is written in a slightly easier to mathematically manipulate form (following the footsteps of \cite{time_arrow}). Let us look at the portion of the integral with respect to $\theta_j$:

$$\int d\theta_j \exp\Big(i \Delta t \Big[\theta_j \Big(\frac{q_{j+1} - q_j}{\Delta t} -\frac{p_j}{m}\Big)\Big]\Big) = \int d\theta_j  \exp\Big(i \frac{\Delta t}{m} \Big[\theta_j \Big(m\frac{q_{j+1} - q_j}{\Delta t} -p_j\Big)\Big]\Big)$$

Next, we will utilize equation \eqref{eq:C} for our convenience. We will take the 0th derivative of the Dirac delta functional in the formula (which is the same as taking no derivative at all): 

$$\frac{d^0}{dA^0} \delta(A-A') = \delta(A-A') = \int \frac{d\omega}{2\pi}(-i\omega)^0 e^{-i\omega (A-A')} = \int \frac{d\omega}{2\pi} e^{-i\omega (A-A')}$$

If you set $\theta = \omega/{2\pi}$, you can plug into the integral with respect to $\theta_j$ to get:

$$ \int d\theta_j  \exp\Big(i \frac{\Delta t}{m} \Big[\theta_j \Big(m\frac{q_{j+1} - q_j}{\Delta t} -p_j\Big)\Big]\Big) =  \int \frac{d\omega}{2\pi}  \exp\Big(-i \frac{\Delta t}{m} \Big[2\pi \omega \Big(p_j - m\frac{q_{j+1} - q_j}{\Delta t}\Big)\Big]\Big) = $$
$$ = \int \frac{d\omega}{2\pi}  \exp\Big(-i\omega  \Big[2\pi\frac{\Delta t}{m}\Big(p_j - m\frac{q_{j+1} - q_j}{\Delta t}\Big)\Big]\Big)$$
By comparing it to the Dirac delta expression above, we see right away that $(A - A') = \Big[2\pi\frac{\Delta t}{m}(p_j -m\frac{q_{j+1} - q_j}{\Delta t})\Big]$, and so:

\begin{equation*}
\int d\theta_j \exp\Big(i \Delta t \Big[\theta_j \Big(\frac{q_{j+1} - q_j}{\Delta t} -\frac{p_j}{m}\Big)\Big]\Big) = \delta\Big(\frac{\Delta t}{2\pi m}(p_j -m\frac{q_{j+1} - q_j}{\Delta t})\Big) =
\end{equation*}
\begin{equation}\label{eq:delta_enforcing_CM_1}
= 2\pi \frac{m}{\Delta t}\delta\Big(p_j -m\frac{q_{j+1} - q_j}{\Delta t}\Big)
\end{equation}
utilizing equation \eqref{eq:dirac_delta_a} for the last step. This reinforces the dynamical equation $p_j -m\frac{q_{j+1} - q_j}{\Delta t} = 0$, which is simply the discretized  version of the momentum relation, $p = mv$. If this delta functional is plugged back into equation \eqref{eq:equation_to_keep_track_of}, then you can reduce the total number of variables integrated over from four to just $q$ and $\lambda$ (\cite{time_arrow}). The two variable version of the Classical Path Integral ensures CM holds (\cite{kvn_thesis}; \cite{time_arrow}).

The above integration over $\theta$ can also be conducted over the variable $\lambda$ instead to get

\begin{equation*}
\int d\lambda_j \exp\Big(i \Delta t \Big[\lambda_j \Big(\frac{p_{j+1} - p_j}{\Delta t} +V'(q_j)\Big)\Big]\Big)  = \int \frac{d\omega}{2\pi} \exp\Big(-i\omega \Big[\frac{\Delta t}{2\pi}\Big(-\frac{p_{j+1} - p_j}{\Delta t} - V'(q_j)\Big)\Big]\Big) = 
\end{equation*}
\begin{equation}\label{eq:delta_enforcing_CM_2}
 = \frac{2\pi}{\Delta t} \delta\Big(\frac{p_{j+1} - p_j}{\Delta t} +V'(q_j)\Big)
\end{equation}
This reinforces the dynamical equation $\frac{p_{j+1} - p_j}{\Delta t} +V'(q_j) = 0$, which is simply the discrete version of Newton's 2nd Law of Motion, $\frac{dp}{dt} = -V'(q) = F$. 

All the above equations are elaborate formulations of the laws of CM. These are different ways of enforcing the laws of CM hold true. We will go over some practical usages of the Classical Path Integral in the pages that follow.

\section{Classical and Quantum Comparison}
\subsection{The Phase in Koopman-von Neumann and Quantum Mechanics}
In QM, the interaction of the quantum phase $S$ with the wavefunction amplitude $R$ gives rise to quantum mechanical behavior (\cite{kvn_thesis}). Recall that one way of expressing the wavefunction $\psi(t)$ is \eqref{eq:configuration_space_wavefunction}. If we plug this expression into the Schr\"{o}dinger equation \eqref{eq:analytical_schrodinger_equation} we get two equations:

\begin{equation}\label{eq:M}
\frac{\partial S}{\partial t} +\frac{ (\grad S)^2}{2m} - \frac{\hbar^2}{2mR} \laplacian{R} + V(q) = 0
\end{equation}
which is the classical Hamilton-Jacobi theory with quantum potential\footnote{As an interesting aside, one can trivially show that the quantum potential disappears if one assumes the Classical Commutator. For instance, equation \eqref{eq:M} can be rewritten since we know that $(i\hbar)^2 = -\hbar^2$:

$$ \frac{\partial S}{\partial t} +\frac{ (\grad S)^2}{2m} + \frac{[\hat{q},\hat{p}]^2}{2mR} \laplacian{R} + V(q) = 0$$
If you assume the Canonical Commutator, you retrieve the Schr\"{o}dinger equation as is. If you assume the Classical Commutator, you retrieve the completely classical Hamilton-Jacobi equation. This result is trivial, but still interesting to consider from the perspective of unity of ideas.}, and 

\begin{equation}
\frac{\partial R^2}{\partial t} + \grad\cdot(R^2 \frac{\grad S}{m}) = 0
\end{equation}
which is the continuity equation for density $R^2$. \footnote{Keep in mind that if $\rho = \psi^* \psi$, then $\rho = \psi^* \psi = Re^{-iS/\hbar}Re^{iS/\hbar} = R^2$.} From the above expression, you can see that the $S$ and $R$ are coupled together and cannot be separated from each other in QM.

This is not the case in CM. From the KvNM equation  \eqref{eq:KvN_L_wavefunction}, you can see if you plug in $\psi(q,p,t) = R(q, p, t)e^{iG(q,p,t)}$ you will get complete decoupling of the amplitude $R$ from the phase $G$. All one has to do is plug the phase space wavefunction $\psi(q,p,t)$ into \eqref{eq:KvN_L_wavefunction}, like so:

$$ i\frac{\partial \psi}{\partial t} = \hat{L}\psi =i\frac{\partial}{\partial t}R(q, p, t)e^{iG(q,p,t)} = \Big[-i\frac{p}{m}\frac{\partial }{\partial q} + i\frac{dV}{dq}\frac{\partial }{\partial p}\Big]R(q, p, t)e^{iG(q,p,t)} $$
This process begets:

\begin{equation}
i \frac{\partial R}{\partial t} = \hat{L}R
\end{equation}

\begin{equation}
i \frac{\partial G}{\partial t} = \hat{L}G
\end{equation}
The amplitude and the phase are completely seperate. Quantum mechanics has therefore been described as the theory that tells how the phase interacts with the amplitude (\cite{kvn_thesis}). 

If the phase and the amplitude evolve separately in KvNM, one may ask what is the practical use of the KvNM formulism? Why the use of complex wavefunctions if the exponated phase is completely independent of the amplitude? While it is true that \textit{in the momentum position representation} $\ket{q,p}$ of KvNM the formulism can be redundant, one can always use the convenient properties of Hilbert Spaces to change to \textit{other} representations, which gives us new insights into CM (\cite{ODM}; \cite{wigner};  \cite{time_arrow}, etc.) We will see examples of this change in basis utilized in section 6. In order to have the most general form of CM, it is important to write the classical wavefunction in this complex form (\cite{kvn_thesis}).

\subsection{Uncertainty Principle in Koopman-von Neumann and Quantum Mechanics}

One of the hallmarks of QM is the existence of an uncertainty principle between the position and momentum. The uncertainty principle tells you that the more precisely you measure where a wavefunction is localized, the less precisely you know the momentum distribution. This contrasts with CM, where you can measure the values of both the momentum and position to an arbitrary precision at the same time. This trade off is another re-articulation of the strange property of ``fuzzy" trajectories in QM and well-defined trajectories in CM. 

One can naturally derive the Heisenberg Uncertainty Principle from the Canonical  Commutator. Through the same series of steps, one can also demonstrate the lack of an uncertainty principle between position and momentum for the KvN classical case. The procedure is quite straightforward.

 Firstly, from statistics we know that the standard deviation of some measured quantity $A$ is given by:

\begin{equation}\label{eq:stand_deviation}
\sigma_A = \sqrt{\langle A^2 \rangle - \langle A \rangle ^2}
\end{equation}
And this will be a crucial ingredient in deriving the Heisenberg Uncertainty Principle. From equation \eqref{eq:expectation_value}, we know that 
$$\langle A \rangle = \bra{\psi(t)}\hat{A}\ket{\psi(t)}$$
$$\langle A^2 \rangle = \bra{\psi(t)}\hat{A}^2\ket{\psi(t)}$$
Treat $\langle A \rangle$ as a constant. Then we can see the following important relationship:

$$(\hat{A} - \langle A \rangle)^2 = \hat{A}^2 - 2\langle A \rangle \hat{A} + \langle A \rangle^2,$$
\newline
$$\bra{\psi(t)}[\hat{A}^2 - 2\hat{A}\langle A \rangle + \langle A \rangle^2 ]\ket{\psi(t)} = \bra{\psi(t)}\hat{A}^2\ket{\psi(t)} - \bra{\psi(t)}2\langle A \rangle \hat{A}\ket{\psi(t)} + \bra{\psi(t)}\langle A \rangle^2 \ket{\psi(t)} = $$
$$= \bra{\psi(t)}\hat{A}^2\ket{\psi(t)} - 2\langle A \rangle\bra{\psi(t)}\hat{A}\ket{\psi(t)} + \langle A \rangle^2 \braket{\psi(t)}{\psi(t)} = $$
$$= \bra{\psi(t)}\hat{A}^2\ket{\psi(t)} - 2\langle A \rangle ^2 + \langle A \rangle^2 = \langle A^2 \rangle - \langle A \rangle^2 .$$
utilizing \eqref{eq:expectation_value} and $\braket{\psi(t)}{\psi(t)} = 1$.

We have all the ingredients we need to derive the Heisenberg Uncertainty Principle. First, let us compute the variance $\sigma^2$ for both the position and momentum. Definitionally following \eqref{eq:stand_deviation} and our above result, they would be 

$$\sigma_q^2 = \langle q^2 \rangle - \langle q \rangle^2 =  \bra{\psi(t)} (\hat{q} - \langle q \rangle)^2 \ket{\psi(t)}$$
$$\sigma_p^2 = \langle p^2 \rangle - \langle p \rangle^2 =  \bra{\psi(t)} (\hat{p} - \langle p \rangle)^2 \ket{\psi(t)}$$
We can define a new vector to simplify our calculation. We can define $\ket{Q} = (\hat{q} - \langle q \rangle) \ket{\psi(t)}$ and $\ket{P} = (\hat{p} - \langle p \rangle) \ket{\psi(t)}$ in order to write the variance of position and momentum simply as 

$$\sigma_q^2 = \braket{Q}$$
$$\sigma_p^2 = \braket{P}$$
Then we apply the Cauchy-Schwarz Inequality (Theorem III of Appendix A) which tells us that:

\begin{equation}\label{eq:useful_equation_2}
\sigma_p^2  \sigma_q^2  = \braket{P}\braket{Q} \ge |\braket{P}{Q}|^2 
\end{equation}
We know that for any complex number $z = a + bi$, that

$$|z|^2 = a^2 + b^2 \ge b^2$$
where we know from complex analysis that $b = \frac{z - z^*}{2i}$. 
Ergo, 
$$|z|^2 = |\braket{P}{Q}|^2 \ge \Big(\frac{z - z^*}{2i}\Big)^2 = \Big(\frac{\braket{P}{Q} - \braket{Q}{P}}{2i}\Big)^2,$$
where

\begin{equation*}
\braket{P}{Q} = \bra{\psi(t)} (\hat{p} - \langle p \rangle) (\hat{q} - \langle q \rangle) \ket{\psi(t)} = \bra{\psi(t)} [\hat{p}\hat{q}-\hat{q}\langle p \rangle - \langle q \rangle\hat{p} + \langle q \rangle\langle p \rangle] \ket{\psi(t)} =
\end{equation*}
\begin{equation*}
= \bra{\psi(t)} \hat{p}\hat{q}\ket{\psi(t)}-\bra{\psi(t)}\hat{q}\langle p \rangle\ket{\psi(t)} - \bra{\psi(t)}\langle q \rangle\hat{p}\ket{\psi(t)} + \bra{\psi(t)}\langle q \rangle\langle p \rangle] \ket{\psi(t)},
\end{equation*}
and
\begin{equation*}
\braket{Q}{P} = \bra{\psi(t)} \hat{q}\hat{p}\ket{\psi(t)}-\bra{\psi(t)}\hat{q}\langle p \rangle\ket{\psi(t)} - \bra{\psi(t)}\langle q \rangle\hat{p}\ket{\psi(t)} + \bra{\psi(t)}\langle q \rangle\langle p \rangle] \ket{\psi(t)}.
\end{equation*}
\newline
With the calculations above, we plug into \eqref{eq:useful_equation_2} and derive:
\begin{equation*}
 |\braket{P}{Q}|^2 \ge \Big(\frac{\braket{P}{Q} - \braket{Q}{P}}{2i}\Big)^2 = \Big(\frac{\bra{\psi(t)} \hat{p}\hat{q}\ket{\psi(t)} - \bra{\psi(t)} \hat{q}\hat{p}\ket{\psi(t)}}{2i}\Big)^2,
\end{equation*}

\begin{equation}
 |\braket{P}{Q}|^2 \ge \Big(\frac{\bra{\psi(t)}[\hat{p},\hat{q}]\ket{\psi(t)}}{2i}\Big)^2.
\end{equation}
From this, it is simple to see that:
\begin{equation}\label{eq:uncertainty_and_commutator}
 \sigma_p \sigma_q \ge \frac{|\bra{\psi(t)}[\hat{p},\hat{q}]\ket{\psi(t)}|}{2}
\end{equation}
If we plug in the Canonical Commutator for $[\hat{p},\hat{q}]$ we derive the Heisenberg Uncertainty Principle,
\begin{equation}
 \sigma_p \sigma_q \ge \hbar/2
\end{equation}
If we plug in the Classical Commutator for $[\hat{p},\hat{q}]=0$ there is no uncertainty relation between the position and momentum, hence there is no theoretical restriction to knowledge of precision of both at the same time. Depending on your choice of the Commutator, you derive either Classical or Quantum Physics (\cite{ODM}).

This does not, however, mean there are no uncertainty relations for KvNM. 
Because the Koopman Algebra \eqref{eq:koopman_algebra} tells us that $[\hat{q},\hat{\theta}] = i$ and $[\hat{p},\hat{\lambda}] = i$, we can develop uncertainty relationships for $q$-$\theta$ and $p$-$\lambda$ by plugging the commutators into expressions analogous to \eqref{eq:uncertainty_and_commutator}:
 
$$ \sigma_p \sigma_\lambda \ge \frac{|\bra{\psi(t)}[\hat{p},\hat{\lambda}]\ket{\psi(t)}|}{2} = \frac{1}{2}$$
$$  \sigma_q \sigma_\theta \ge \frac{|\bra{\psi(t)}[\hat{q},\hat{\theta}]\ket{\psi(t)}|}{2} = \frac{1}{2}$$
These would be the `classical' uncertainty relationships. Since neither $\theta$ nor $\lambda$ are directly observable (\cite{harmonic_oscillator}), these uncertainty relations will not impact our observations of classical systems. 
For KvNM, many ``quantum" phenomena are tucked into the not directly observable operators $\hat{\theta}$ and $\hat{\lambda}$.

\subsection{Measurement in Koopman-von Neumann and Quantum Mechanics} 

QM is known to have the strange property that it makes probabilistic predictions instead of deterministic ones. One strange feature related to this fact that tucked into the previously mentioned postulates of QM you have a phenomenon termed \textit{the collapse of the wavefunction}. Any wavefunction can be expanded across a basis using the completeness theorem (Theorem II of Appendix A):

\begin{equation*}
\ket{\psi} = I\ket{\psi},
\end{equation*}
where $I$ is the identity operator (analogous to the number `1'.) Inserting the discrete\footnote{Note that in \eqref{eq:expansion} I used the discrete case instead of the continuous completeness identity. This is simply due to the ease of highlighting certain concepts. The continuous completeness identity would do the same thing as the discrete case, but might obscure for some readers results we will discuss. Remember that the integral is just a continuous \textit{summation} of infinitely small things.}
 completeness relation we get:

\begin{equation}\label{eq:expansion}
\ket{\psi} = \sum_j \ket{j}\bra{j}\ket{\psi} = \ket{A}  \bra{A}\ket{\psi}  + \ket{B}  \bra{B}\ket{\psi} + \ket{C} \bra{C}\ket{\psi} +\cdot\cdot\cdot,
\end{equation}
where the collection of all $\ket{j}$ vectors ($j = A, B, C,...$) are your basis vectors according to the principles of Linear Algebra and $ \bra{j}\ket{\psi}$ are scalar values or functions (recalling that $ \bra{j}\ket{\psi} = \psi(j)$). This is simply expanding the state vector $\ket{\psi}$ along a particular set of basis vectors $\ket{j}$. $A$, $B$, $C$, etc., are different eigenvalues associated with each `eigenbasis', for example, $\hat{A}\ket{A}=A\ket{A}$.  

If you measure $B$ then the initial wavefunction $\ket{\psi}$ instantaneously collapses the entire quantum system to state $\ket{B}$. \textit{How} the state of the system $\ket{\psi}$ instantaneously changes to $\ket{B}$ when $B$ is measured is still a modern day quantum mystery. As described by Postulate 3 in section 2, the probability for observing $B$ is $P(B) =  |\braket{B}{\psi}|^2$ for discrete systems and the probability density is $\rho(B) = |\braket{B}{\psi}|^2$ in continuous systems. 

The strange phenomenon of collapse is preserved in KvNM. You have a KvN classical wavefunction that again can be expanded over a particular basis of your choice 
\begin{equation*}
\ket{\psi} = \ket{A}  \bra{A}\ket{\psi}  + \ket{B}  \bra{B}\ket{\psi} + \ket{C} \bra{C}\ket{\psi} +\cdot\cdot\cdot,
\end{equation*}
and that when a particular eigenvalue is observed, the system collapses to its associated eigenbasis. The same Born Rule applies. You have the same postulate for collapse from the set of `universal axioms' . 

The common behavior of the wavefunction collapsing during measurement in QM and KvNM is mystifying, and perhaps one day further light can be shed on it. Maybe it is a common feature of probabilistic models built in Hilbert Space, or perhaps there will be some other explanation developed. As far as we know, the collapse is instant, irreversible, and no causal process exists to explain it. 

There is an important difference though in the collapse of the wavefunction during measurement for the KvNM versus QM cases. The wavefunction will not collapse for \textit{non-selective measurements} in the KvN case but will in the QM case (\cite{minimal_coupling}; \cite{kvn_thesis}). Non-selective measurements are measurements conducted on the system where the results are not `read out loud', meaning the system would be hypothetically collapsed but we do not `record', remember, or consider which eigenbasis it collapses to. We remain completely ignorant of whether the system state $\ket{\psi}$ hypothetically collapses into $\ket{A}$, $\ket{B}$, $\ket{C}$, or some other state. This difference in collapse for KvN and QM wavefunctions will be demonstrated (following the work of \cite{minimal_coupling}; \cite{kvn_thesis}). 

First, let us consider the QM case. Consider a Hamiltonian $\hat{H}$ which obeys the following eigenvalue relationships:

\begin{equation}\label{eq:woah_1}
\hat{H}\ket{+} = \hbar\omega\ket{+},
\end{equation}

\begin{equation}\label{eq:woah_2}
\hat{H}\ket{-} = -\hbar\omega\ket{-},
\end{equation}
where $\ket{+}$ and $\ket{-}$ are the eigenvectors and $\hbar\omega$ and $-\hbar\omega$ are the energy eigenvalues of the operator $\hat{H}$, respectively. \footnote{This should bring to mind, for instance, that the energy of a photon is given by $E = \hbar\omega$, where $\omega$ is the angular frequency.} Let us consider an observable with operator $\hat{\Omega}$, which has the eigenvectors $\ket{a}$ and $\ket{b}$ defined by:

\begin{equation}\label{eq:eigen_eq_1}
\ket{a} = \frac{1}{\sqrt{2}}\Big[\ket{+}+\ket{-}\Big],
\end{equation}

\begin{equation}\label{eq:eigen_eq_2}
\ket{b} = \frac{1}{\sqrt{2}}\Big[\ket{+}-\ket{-}\Big].
\end{equation}
For the sake of our analysis, we can choose the initial state of the system to be the following superposition of states (\cite{kvn_thesis}):

\begin{equation}\label{eq:original_eq}
\ket{\psi(t=0)} = \frac{1}{2}\ket{+} + \sqrt{\frac{3}{4}}\ket{-}
\end{equation}
We will utilize the Schr\"{o}dinger equation in the form of \eqref{eq:propogator_relationship} (see also Appendix C) order to evolve the above wavefunction, $\ket{\psi(0)}$. We want to evolve the system to $t=2\tau$ and see what the probability of measuring $a$, that is $P(a)$, will be (\cite{kvn_thesis}). 

\begin{equation}
\ket{\psi(t = 2\tau)} = e^{-\frac{i}{\hbar}\hat{H}2\tau}\ket{\psi(0)} = \exp[-\frac{i}{\hbar}\hat{H}2\tau]\Big[\frac{1}{2}\ket{+} + \sqrt{\frac{3}{4}}\ket{-}\Big].
\end{equation}
We will utilize the very useful Theorem I from Appendix A alongside \eqref{eq:woah_1} and \eqref{eq:woah_2}. This will give us:

\begin{equation*}
\ket{\psi(t = 2\tau)} =\frac{1}{2}\exp[-\frac{i}{\hbar}\hat{H}2\tau]\ket{+} + \sqrt{\frac{3}{4}}\exp[-\frac{i}{\hbar}\hat{H}2\tau]\ket{-} = 
\end{equation*}
\begin{equation*}
 =\frac{1}{2}\exp[-\frac{i}{\hbar}2\tau(\hbar\omega)]\ket{+} + \sqrt{\frac{3}{4}}\exp[-\frac{i}{\hbar}2\tau(-\hbar\omega)]\ket{-}=  
\end{equation*}
\begin{equation}
=  \frac{1}{2}\exp[-2i\tau\omega]\ket{+} + \sqrt{\frac{3}{4}}\exp[2i\tau\omega]\ket{-}.
\end{equation}
To find the probability of observing $a$ at $t = 2\tau$ we will use the Born Rule: 
$$P(a) = \bra{\psi(2\tau)}\ket{a}\bra{a}\ket{\psi(2\tau)}=|\bra{a}\ket{\psi(2\tau)}|^2,$$
where we can evaluate:

$$\bra{a}\ket{\psi(2\tau)} =  \frac{1}{2}\exp[-2i\tau\omega]\bra{a}\ket{+} + \sqrt{\frac{3}{4}}\exp[2i\tau\omega]\bra{a}\ket{-}=$$
$$ = \frac{1}{2}\exp[-2i\tau\omega]\frac{1}{\sqrt{2}}\Big[\bra{+}+\bra{-}\Big]\ket{+} + \sqrt{\frac{3}{4}}\exp[2i\tau\omega]\frac{1}{\sqrt{2}}\Big[\bra{+}+\bra{-}\Big]\ket{-}. $$
Using the orthonormality condition, we know that $\braket{+}{-} =0$ while $\braket{+}{+} = \braket{-}{-} =1$, so that the above expression reduces to 

$$\bra{a}\ket{\psi(2\tau)} =  \frac{1}{2\sqrt{2}}\exp[-2i\tau\omega] + \sqrt{\frac{3}{8}}\exp[2i\tau\omega].$$
Therefore, taking $(\bra{a}\ket{\psi(2\tau)} )^\dagger = \bra{\psi(2\tau)}\ket{a}$ and plugging into the Born Rule:

$$ P(a) = \bra{\psi(2\tau)}\ket{a}\bra{a}\ket{\psi(2\tau)}= \Big(\frac{1}{2\sqrt{2}}\exp[2i\tau\omega] + \sqrt{\frac{3}{8}}\exp[-2i\tau\omega] \Big) \times $$
$$ \times \Big(\frac{1}{2\sqrt{2}}\exp[-2i\tau\omega] + \sqrt{\frac{3}{8}}\exp[2i\tau\omega] \Big) = $$
\newline
$$ = \frac{1}{2\sqrt{2}}\exp[2i\tau\omega]\frac{1}{2\sqrt{2}}\exp[-2i\tau\omega] +  \frac{1}{2\sqrt{2}}\exp[2i\tau\omega]\sqrt{\frac{3}{8}}\exp[2i\tau\omega] + $$
$$+  \sqrt{\frac{3}{8}}\exp[-2i\tau\omega]\frac{1}{2\sqrt{2}}\exp[-2i\tau\omega] +  \sqrt{\frac{3}{8}}\exp[-2i\tau\omega]\sqrt{\frac{3}{8}}\exp[2i\tau\omega] =$$
\newline
\begin{equation}\label{eq:step_along_the_way}
= \frac{1}{8} + \frac{\sqrt{2}}{4}\sqrt{\frac{3}{8}}\exp[4i\omega\tau] + \frac{\sqrt{2}}{4}\sqrt{\frac{3}{8}}\exp[-4i\omega\tau] + \frac{3}{8}. 
\end{equation}
There are one important trigonometric identity we can make use of. It is:

\begin{equation}\label{eq:cosine_important}
\cos(\theta) = (1/2)(e^{i\theta} + e^{-i\theta})
\end{equation}
Taking the identity \eqref{eq:cosine_important} with \eqref{eq:step_along_the_way}, we can calculate that 

\begin{equation}\label{eq:fascinating_method_1}
P(a) = \frac{1}{2}\Big(1 + \sqrt{\frac{3}{4}}\cos(4\omega\tau)\Big). 
\end{equation}
The above usage of the wavefunction is for ``pure states" (term defined in section 2.6). 

Now, instead of letting the quantum system evolve continuous from $t=0$ to $t=2\tau$, let us introduce a non-selective measurement at $t = \tau$. How will this affect the probability of measuring $a$ at $t= 2\tau$? We will start with the same $\ket{\psi(0)}$. We will, however, use the density operator formulism to describe the system (see section 2.6) instead of strictly using only the `pure' wavefunction. This is because we have here what is called a ``mixed state". When we take the measurement at $t = \tau$, the original system given by \eqref{eq:original_eq} collapses, but we do not know whether it collapses into $\ket{a}$ or $\ket{b}$, as we do not `read out' or record what the collapse leads to. We can use \eqref{eq:density_matrix_def} to write down:

\begin{equation}\label{eq:first_application_of_density_operator}
\hat{\rho}(\tau) = p_a(\tau) \ket{a}\bra{a} + p_b(\tau) \ket{b}\bra{b},
\end{equation}
where $p_a(\tau)$ and $p_b(\tau)$ are the probabilistic weights to the `statistical mixture' of possible states $\ket{a}$ and $\ket{b}$ of the system. The factor $p_a(\tau)$ refers to the probability that we observe $a$ at $t= \tau$ (and hence the system collapsed to $\ket{a}$) and the term $p_b(\tau)$ refers to the probability that we observe $b$ at $t= \tau$ (and hence the system collapsed to $\ket{b}$). Since we are ignorant of which way the wavefunction collapses, the probabilities became weights of likelihood in \eqref{eq:first_application_of_density_operator}. 

We calculate what the factors $p_a(\tau)$ and $p_b(\tau)$ are in a similar fashion to calculating \eqref{eq:fascinating_method_1}:

$$p_a(\tau) = \bra{\psi(\tau)}\ket{a}\bra{a}\ket{\psi(\tau)} = \Big(\frac{\sqrt{2}}{4}\exp[-i\omega\tau] + \sqrt{\frac{3}{8}}\exp[i\omega\tau]\Big)\times $$
$$\times \Big(\frac{\sqrt{2}}{4}\exp[i\omega\tau] + \sqrt{\frac{3}{8}}\exp[-i\omega\tau]\Big) =$$
\begin{equation}\label{eq:super_3}
=\frac{1}{2}\Big(1 + \sqrt{\frac{3}{4}}\Big[\frac{1}{2}\exp[-2i\omega\tau] + \frac{1}{2}\exp[2i\omega\tau] \Big]\Big)= \frac{1}{2}\Big(1 + \sqrt{\frac{3}{4}}\cos(2\omega\tau)\Big)
\end{equation}
\newline
$$p_b(\tau) = \bra{\psi(\tau)}\ket{b}\bra{b}\ket{\psi(\tau)} = \Big(\frac{\sqrt{2}}{4}\exp[-i\omega\tau] - \sqrt{\frac{3}{8}}\exp[i\omega\tau]\Big) \times$$
$$\times \Big(\frac{\sqrt{2}}{4}\exp[i\omega\tau] - \sqrt{\frac{3}{8}}\exp[-i\omega\tau]\Big) =$$
\begin{equation}\label{eq:super_4}
=\frac{1}{2}\Big(1 - \sqrt{\frac{3}{4}}\Big[\frac{1}{2}\exp[-2i\omega\tau] + \frac{1}{2}\exp[2i\omega\tau] \Big]\Big)= \frac{1}{2}\Big(1 - \sqrt{\frac{3}{4}}\cos(2\omega\tau)\Big)
\end{equation}
We will use \eqref{eq:propogator_relationship} again as before to evolve the system then from \eqref{eq:first_application_of_density_operator} at $t = \tau$ to $t=2\tau$, in the following fashion:

\begin{equation}
\ket{\psi_a(2\tau)} = e^{-\frac{i}{\hbar}\hat{H}(2\tau - \tau)}\ket{a} = \exp[-\frac{i}{\hbar}\hat{H}\tau]\frac{1}{\sqrt{2}}\Big[\ket{+}+\ket{-}\Big],
\end{equation}

\begin{equation}
\ket{\psi_b(2\tau)} = e^{-\frac{i}{\hbar}\hat{H}(2\tau - \tau)}\ket{b} = \exp[-\frac{i}{\hbar}\hat{H}\tau] \frac{1}{\sqrt{2}}\Big[\ket{+}-\ket{-}\Big],
\end{equation}
utilizing the expressions \eqref{eq:eigen_eq_1} and \eqref{eq:eigen_eq_2} above. 

As before, utilizing the eigenvalue expressions for $\hat{H}$ \eqref{eq:woah_1} and \eqref{eq:woah_2}, and Theorem I we get:
\begin{equation}\label{eq:super_1}
\ket{\psi_a(2\tau)}  = \frac{1}{\sqrt{2}}\exp[-\frac{i}{\hbar}\tau(\hbar\omega)]\ket{+}+\frac{1}{\sqrt{2}}\exp[-\frac{i}{\hbar}\tau(-\hbar\omega)]\ket{-},
\end{equation}

\begin{equation}\label{eq:super_2}
\ket{\psi_b(2\tau)} = \frac{1}{\sqrt{2}}\exp[-\frac{i}{\hbar}\tau(\hbar\omega)]\ket{+}-\frac{1}{\sqrt{2}}\exp[-\frac{i}{\hbar}\tau(-\hbar\omega)]\ket{-}.
\end{equation}
The density operator therefore at $t=2\tau$ can be represented as the following using the above expressions:

\begin{equation}\label{eq:new_density_operator_1121}
\hat{\rho}(2\tau) =  p_a(\tau) \ket{\psi_a(2\tau)}\bra{\psi_a(2\tau)} + p_b(\tau) \ket{\psi_b(2\tau)}\bra{\psi_b(2\tau)}
\end{equation}
And we can use this to evaluate the probability of measuring $a$ by using \eqref{eq:density_matrix_born_rule}:

$$P(a) = \Tr[\hat{\rho}(2\tau)\ket{a}\bra{a}] = \bra{a}\hat{\rho}(2\tau)\ket{a}\bra{a}\ket{a} + \bra{b}\hat{\rho}(2\tau)\ket{a}\bra{a}\ket{b}.$$
Remember that $\bra{a}\ket{b} = 0$ and $\bra{a}\ket{a} = 1$, so:

$$P(a)  = \bra{a}\hat{\rho}(2\tau)\ket{a}.$$
Plugging in \eqref{eq:new_density_operator_1121} we get:

\begin{equation}
P(a) = p_a(\tau) \bra{a}\ket{\psi_a(2\tau)}\bra{\psi_a(2\tau)}\ket{a} + p_b(\tau) \bra{a}\ket{\psi_b(2\tau)}\bra{\psi_b(2\tau)}\ket{a}.
\end{equation}
And now we can substitute the expressions \eqref{eq:super_1}, \eqref{eq:super_2}, \eqref{eq:super_3}, and \eqref{eq:super_4} to get after some algebra our final expression:

\begin{equation}\label{eq:comparison_2_1121}
P(a) = \frac{1}{2}\Big(1 + \sqrt{\frac{3}{4}}\cos^2(2\omega\tau) \Big).
\end{equation}
Obviously, comparing \eqref{eq:fascinating_method_1} with \eqref{eq:comparison_2_1121} shows that they are two different equations. This shows us that in the quantum case making a non-selective measurement will change the outcome of a later measurement. A non-selective measurement will modify the probability of outcome at a later point in time $t$. 

So far, what we looked at is an abstract example to demonstrate non-selective measurements in QM. Let us look at a more concrete example in QM, and then figure out what the implications are for KvNM (following \cite{kvn_thesis}). 
Starting off, let us measure the probability of a quantum particle's position from $t = 0$ to $t = \tau$. 

This will be given by \eqref{eq:density_matrix_def} except we are no longer dealing with a discrete case. Since position is continuous, we replace the summation sign with an integral (as an integral is simply continuous summation of infinitely small things). We start with a pure state with wavefunction $\ket{\psi(0)}$ at $t = 0$. The probability  density $\rho$ (not to be confused with the density matrix) of observing $q$ at time $t = \tau$ will be given by

\begin{equation}\label{eq:to_consider_1}
\rho(q,\tau) = \Tr[\hat{\rho}(\tau)\ket{q}\bra{q}] = |\psi(q,\tau)|^2,
\end{equation}
where $\hat{\rho}(\tau) = \ket{\psi(\tau)}\bra{\psi(\tau)}$ (pure state density matrix). 

Let us say that we make a non-selective measurement \textit{immediately} after $t =0$, which we will represent as time $t=0_+$. The pure state density matrix gets replaced with a mixed density matrix. We utilize a continuous version of \eqref{eq:density_matrix_def} to write the density matrix

$$ \hat{\rho}(0_+) = \int dq_0 |\psi(q_0,0)|^2 \ket{q_0}\bra{q_0}.$$
To represent a state of the system \textit{immediately} before $t =0$, we can write time $t=0_-$ and say that:

$$ \hat{\rho}(0_-) = \ket{\psi(0_-)}\bra{\psi(0_-)}$$
The system goes from $ \hat{\rho}(0_-)$ to $ \hat{\rho}(0_+)$ as a non-selective measurement is made at $t = 0$. 

In QM, the measurement right before and the measurement right after will observe the same thing (\cite{kvn_thesis}). You can convince yourself of this by computing the following:

$$\rho(q,0_-) =  \Tr[\hat{\rho}(0_-)\ket{q}\bra{q}]  = \psi^*(q,0)\psi(q,0)$$
$$\rho(q,0_+) =  \frac{\Tr[\hat{\rho}(0_+)\ket{q}\bra{q}]}{\Tr[\hat{\rho}(0_+)]}  = \psi^*(q,0)\psi(q,0)$$
Remember that the denominator, $\Tr[\hat{\rho}(0_+)]$, just represents the normalization condition (described in section 2.6).
We can evolve $\psi^*(q,0)\psi(q,0)$ to see that in the long term the non-selective measurement does affect the quantum system, just like in the above worked out example (\cite{kvn_thesis}). The probability density will be the same instantaneously right before and right after measurement, but will alter in the long run in QM.

Under KvNM, however, we can observe some differences arise. If we start with a pure state again

$$ \hat{\rho}(0_-) = \ket{\psi(0_-)}\bra{\psi(0_-)},$$
and conduct no measurements, we will have the KvN wavefunction evolve to 
\begin{equation}\label{eq:future_kvn_function}
\rho(q,p,\tau) = \Tr[\hat{\rho}(\tau)\ket{q,p}\bra{q,p}] = |\psi(q,p,\tau)|^2,
\end{equation}
similarly to the quantum case \eqref{eq:to_consider_1}.
Also like the quantum case, in KvNM measuring the probability density right before and after will not alter the distribution:

$$\frac{\Tr[\hat{\rho}(0_-)\ket{q,p}\bra{q,p}]}{\Tr[\hat{\rho}(0_-)]}  =  \frac{\Tr[\hat{\rho}(0_+)\ket{q,p}\bra{q,p}]}{\Tr[\hat{\rho}(0_+)]}$$
However, a non-selective measurement can be shown to leave the KvN wavefunction unaltered, whereas a quantum wavefunction will be changed due to the non-selective measurement .

$$\rho(0_+) = \int dq_0 dp_0 |\psi(q_0,p_0,0)|^2 \ket{q_0,p_0}\bra{q_0,p_0}.$$
The above probability density, taken right after the measurement, can be shown to naturally evolve into \eqref{eq:future_kvn_function}, as if no measurement has taken place (\cite{kvn_thesis}). 

There is one final similarity worth mentioning between QM and KvNM. Under QM, when a measurement is taken for position $q$, it affects the outcome of the distribution of $p$. Perhaps, this is not surprising, considering the Heisenberg Uncertainty Principle. Under KvNM, a measurement taken for $(q,p)$ will alter the statistical distribution of the variable $\lambda$ (\cite{kvn_thesis}). This should also not be surprising because, as we discussed in section 5.2, there exists a classical uncertainty relationship between $p$ and $\lambda$. The effects of QM are still present in KvNM therefore, if we replace $q$ with $(q,p)$ and $p$ with $\lambda$.

\subsection{The Double Slit Experiment} 
The Double Slit Experiment is at the center of Quantum Theory. QM has photons, electrons, neutrons, and other quanta going through two tiny slits and forming an interference pattern on the other side. Even if we send one particle at a time through the double slit, the particle build up on the other side will reflect an interference pattern. Even if the particle does not seem to interfere with \textit{any} other particle, it \textit{still} is part of an interference pattern. This shows that the interference patterns are due to something more fundamental in nature. \footnote{Interestingly, if you cover up on of the two slits, you will recover particle travelling in straight trajectories without any interference pattern, just as in CM. This makes the relationship between CM and QM even more bizarre.}

The wavefunction mathematically describes this strange behavior of elementary particles, even though the exact interpretation of the wavefunction is not known. Feynman believed that the strange behavior must lie in the complex nature of the wavefunction (\cite{kvn_thesis}). KvNM also contains complex wavefunctions in Hilbert Space, so we might logically ask the obvious question: if KvNM has complex wavefunctions, will  it successfully preserve our knowledge of CM, without interference patterns arising? Here we will test the outcome of complex wavefunctions in classical phase space. This analysis is based on the work of (\cite{kvn_thesis}). 

\begin{figure}[h]
\centering
\includegraphics{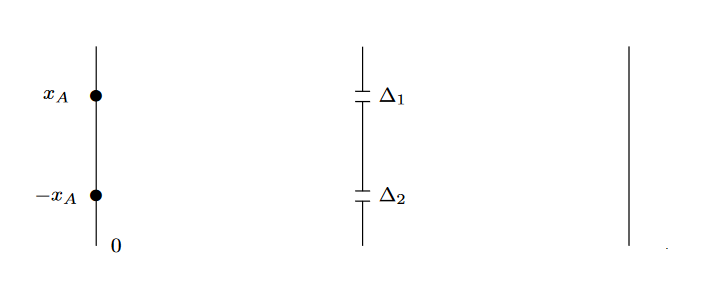}
\label{fig:sun1}
\caption{This is the set up of the double slit experiment. Middle wall is at $y= y_M$ and rightmost wall is at $y = y_R$. Adapted from \cite{DM}}
\end{figure}

The double slit experiment is set up in Figure 1 above. The y axis is aligned parallel to the direction of travel. The x axis will be the coordinate parallel to the back wall where the interference is supposed to occur for the quantum case. In the middle, at $y=y_M$ you have the wall with the two slits. At $y = y_F$ you have the wall where interference occurs in the quantum case.

Under KvNM, we should expect no interference build up at the backwall at $y = y_R$. Under QM (and in reality), we want there to be an interference build up at the backwall. We will start with analyzing the quantum case, and how it works. We want a gaussian wave emerge from $y = 0$ and travel right until it hits the double slit wall. Some of the wavefunction will travel through the two slits and eventually interfere at $y = y_R$ if the system behaves quantum.

A gaussian function has the following form:
\begin{equation}\label{eq:gaussian_eq}
f(u) = \frac{1}{\sqrt{2\pi\sigma^2}}e^{-(u-\mu)^2/{2\sigma^2}}
\end{equation}
where $\mu$ is the average value $\mu = \langle u \rangle$ and $\sigma$ is the standard deviation, given by $\sigma = \sqrt{\langle u^2 \rangle - \langle u \rangle ^2}$. 

Recalling the form of a quantum wavefunction \eqref{eq:configuration_space_wavefunction}, we can make it a gaussian wavefunction by setting $R$ equal to \eqref{eq:gaussian_eq} with a generic renormalization constant $N$.\footnote{In order to keep things nice and streamlined, the constant in front of \eqref{eq:gaussian_eq} will simply be absorbed into the Renormalization Constant.} Let $\mu = 0$ in order to center the wavefunction of the double slit experiment. The equation will have the form:

$$ \psi(x,t=0) = N\exp[-\frac{x^2}{2\sigma_x^2}]\exp[\frac{i}{\hbar}S]$$
Note that for this wavefunction, $S = 0$, however. This is obvious when you consider the definition of the Action \eqref{eq:action} for a free particle. Since we decided that the time at the leftmost wall is $t=0$, the Action at the leftmost wall would be:

\begin{equation*}
S = \int_0^0 dt T  = 0
\end{equation*}
Therefore, a better way to write the previous equation at $t = 0$ is:

\begin{equation}
\psi(x,t=0) = N\exp[-\frac{x^2}{2\sigma_x^2}]
\end{equation}
Following the Path Integral Formulism \eqref{eq:feynman_path_integral_formulism_1}, in order to know the know the wavefunction $\psi(x,t)$ at some later time $t$ we will need to utilize the Kernel \eqref{eq:discrete_quantum_propogator}.
The Kernel $\bra{x} e^{-\frac{i}{\hbar}t\hat{H}}\ket{x_0}$ will evolve the system to the time of interest. 

\begin{equation*}
\bra{x}\ket{\psi(t)} = \int dx_0 \bra{x} e^{-\frac{i}{\hbar}t\hat{H}}\ket{x_0}\psi(x_0,0)
\end{equation*}
We will utilize a very special trick in order to evaluate the above expression. Since we are dealing with a free particle, there are no potentials, or $V(x,y)=0$ (hence no forces or accelerations). Because of this, we do not have to use the full, intimidating form of the Feynman Kernel \eqref{eq:discrete_quantum_propogator}. In this semiclassical treatment there is no acceleration of the quantum particle in the y direction, and so the Kernel can be written in a much simpler form. 

Instead of taking on the form \eqref{eq:discrete_quantum_propogator}, it will instead have the form \eqref{eq:Kernel_portion}. This is because we can treat the final position of the wavefunction as $q_{j+1}$ and the initial position as $q_j$ \textit{if} there is no acceleration. The term $q_{j+1} - q_j$ simply becomes one large step instead of a miniscule step as before. This is analogous to how in kinematics, when there is no acceleration, the formula for \textit{instantaneous} velocity is $v =\Delta x/\Delta t$, even though in every other case the formula gives you \textit{average} velocity (as it is a slope formula). You can also think of this simplification as taking \eqref{eq:discrete_quantum_propogator} and setting $N = 1$ (one large step due to no acceleration). Please note that this is not an approximation, but this will give us an \textit{exact} expression.

Our Kernel $K$ (not to be confused with the Koopman Generator) is therefore:

\begin{equation}\label{eq:quantum_path_integral_1}
 K = \sqrt{\frac{m}{2\pi i \hbar (t - t_0)}}\exp[\frac{im(x(t)  - x_{0})^2}{2\hbar (t - t_0)}]
\end{equation}
And then we can write the following Path Integral:
\begin{equation*}
\bra{x}\ket{\psi(t)} = \int_{-\infty}^\infty dx_0 \sqrt{\frac{m}{2\pi i \hbar (t - t_0)}}\exp[\frac{im(x(t) - x_{0})^2}{2\hbar (t - t_0)}]N\exp[-\frac{x_0^2}{2\sigma_x^2}]
\end{equation*}
We want to evaluate the path integral up until the point it reached the middle wall with the double slits, at time $t_M$. Remembering that we have set the initial time $t_0$ equal to 0 and treating $t$ and $x(t)$ as constants, we can evaluate the integral. First, we will pull out all constants and absorb them into $N$, the renormalization constant (recall the renormalization condition described in section 2.3). Then we do the following algebraic manipulations, in preparation to evaluating the integral itself:

$$ \exp[\frac{im(x(t) - x_{0})^2}{2\hbar t}]\exp[-\frac{x_0^2}{2\sigma_x^2}] =  \exp[\frac{im(x(t)^2 - 2x(t) x_{0} + {x_{0}}^2)}{2\hbar t} -\frac{x_0^2}{2\sigma_x^2}] = $$
$$ = \exp[\Big(\frac{im}{2\hbar t}-\frac{1}{2\sigma_x^2}\Big)x_0^2 -\frac{im x(t)}{\hbar t}x_0 + \frac{im x(t)^2}{2\hbar t}]  $$
We can then use Theorem VIII of Appendix A to evaluate the integral. Comparing the two, we can recognize the following coefficients of the Gaussian integral:

$$ a = - \Big(\frac{im}{2\hbar t}-\frac{1}{2\sigma_x^2}\Big)$$
$$ b =  -\frac{im x(t)}{\hbar t}$$
$$ c = \frac{im x(t)^2}{2\hbar t}$$
And so by the Theorem, the integral turns out to be:
$$N\int_{-\infty}^\infty dx_0  \exp[\Big(\frac{im}{2\hbar t}-\frac{1}{2\sigma_x^2}\Big)x_0^2 -\frac{im x(t)}{\hbar t}x_0 + \frac{im x(t)^2}{2\hbar t}] = $$
$$= N\sqrt{\frac{\pi}{- \Big(\frac{im}{2\hbar t}-\frac{1}{2\sigma_x^2}\Big)}} \exp[\frac{(\frac{im x(t)}{\hbar t})^2}{-4 \Big(\frac{im}{2\hbar t}-\frac{1}{2\sigma_x^2}\Big)}] \exp[\frac{im x(t)^2}{2\hbar t}] =$$
$$ =  N\sqrt{\frac{2\pi \hbar t \sigma_x^2}{-im\sigma_x^2 + \hbar t}}\exp[\frac{m^2 x(t)^2 \sigma_x^2}{2 \hbar t (-im\sigma_x^2 + \hbar t)}] \exp[\frac{im x(t)^2}{2\hbar t}]$$

$$\therefore \bra{x}\ket{\psi(t)} = N\sqrt{\frac{2\pi \hbar t \sigma_x^2}{-im\sigma_x^2 + \hbar t}}\exp[-\frac{1}{2}\frac{mx(t)^2}{m\sigma_x^2 + i\hbar t}]$$
The above wavefunction $\bra{x}\ket{\psi(t)} = \psi(x,t)$ is true for anytime $0 < t \leq t_M$, where $t_M$ is the time to reach the middle barrier with the two slits. This is because the above wavefunction is that of a free particle, and once it encounters a barrier (or potential), the particle is obviously no longer free. 

Next, since the expression $\psi(x,t_M)$ gives you the quantum wavefunction at the double slits, we have to take into consideration the portion of the wavefunction that goes through the double slits. To describe the wavefunction going through the double slit we will utilize step functions. The step functions will closely model the fact that the wavefunction that hits the middle wall and stops will not be important in creating any effects on the other side of the wall. Only the wavefunction that goes through the two slits will be important in creating the diffraction pattern on the other side (\cite{QM}; \cite{kvn_thesis}). The step function, also called the $\theta$-Heaveside function, is given by:

\begin{equation}\label{eq:step_function}
 H(x) =  \left\{
        \begin{array}{ll}
            1 & \quad x ~\textgreater ~0 \\
            0 & \quad x \leq 0
        \end{array}
    \right.
\end{equation}
Each slit has a width of $2\delta$. The first slit has width $\Delta_1 = (x_A - \delta, x_A + \delta)$ and the second slit has spans $\Delta_2 = (-x_A - \delta, -x_A + \delta)$ (following the notation of \cite{kvn_thesis}). 

Then, the wavefunction right after the two slits can be written as 
\begin{equation}\label{eq:double_slit_quantum_case}
\psi(x,t_M + dt) = N\psi(x,t_M)[C_1(x) + C_2(x)] = N \exp[-\frac{1}{2}\frac{mx^2}{m\sigma_x^2 + i\hbar t}][C_1(x) + C_2(x)]
\end{equation}
where the following functions can be defined:

\begin{equation}\label{eq:C_function_1}
C_1(x) = H(x - x_A +\delta) - H(x - x_A -\delta) 
\end{equation}
\begin{equation}\label{eq:C_function_2}
C_2(x) = H(x + x_A +\delta) - H(x + x_A -\delta)
\end{equation}
This is done in order to filter out the effects of the wavefunction that actually goes through the slits. We do not care for the portion of the wavefunction that does not go through the double slit. They will produce no effects. The above equations make certain that only the wavefunction that goes through slits $\Delta_1$ and $\Delta_2$ will be studied. The $N$ in \eqref{eq:double_slit_quantum_case} is a renormalization constant, so that $\int dx \psi^*\psi = 1$ (the renormalization condition described in section 2.3). 

Using our Kernel for the free particle again, we will evaluate the wavefunction using the path integral for the region \textit{after} the two slits. We again utilize \eqref{eq:quantum_path_integral_1}. Notice that the initial wavefunction $\psi(x_0,0)$ in that equation starts at time zero, so we will need to adjust the time at the double slit wall so that it is set to zero. We can do this by substituting the variable $t$ in the Kernel \eqref{eq:quantum_path_integral_1} with $t - t_M$. This can be done because time has the unique property that any point can be arbitrarily set to $t = 0$, and by doing this swap we effectively set the time at the middle wall equal to zero (starts at $t_M$ and evolves from there). So the integral we need to evaluate will have the following form:

\begin{equation*}
\bra{x}\ket{\psi(t)} = N \int dx_M \exp[\frac{im(x(t)  - x_{M})^2}{2\hbar (t - t_M)}] \psi(x_M,t_M) = 
\end{equation*}
\begin{equation}
 = N \int dx_M \exp[\frac{im(x(t)  - x_{M})^2}{2\hbar (t - t_M)}] \exp[-\frac{1}{2}\frac{mx_M^2}{m\sigma_x^2 + i\hbar t_M}][C_1(x_M) + C_2(x_M)]
\end{equation}
where all constants, as is the custom, have been absorbed into $N$. 

This expression above cannot be evaluated explicitly. However, because of the properties of the step functions (\cite{kvn_thesis}), it can be rewritten as

$$ \bra{x}\ket{\psi(t)} = N \Big(\int_{x_A - \delta}^{x_A + \delta}  dx_M \exp[\frac{im(x(t)  - x_{M})^2}{2\hbar (t - t_M)}-\frac{1}{2}\frac{mx_M^2}{m\sigma_x^2 + i\hbar t_M}] +$$
\begin{equation}\label{eq:important_phase}
+\int_{-x_A - \delta}^{-x_A + \delta} dx_M \exp[\frac{im(x(t)  - x_{M})^2}{2\hbar (t - t_M)}-\frac{1}{2}\frac{mx_M^2}{m\sigma_x^2 + i\hbar t_M}] \Big)
\end{equation}
You can consider each of the two integrals to represent an independent wavefunction, so that you can rewrite the whole expression as

$$\bra{x}\ket{\psi(t)} = N[\psi_1(x) + \psi_2(x)]  $$
You can think of $\psi_1$ as being the wavefunction coming out of $\Delta_1$ and $\psi_2$ as the wavefunction associated with $\Delta_2$. 

The equation \eqref{eq:important_phase} shows us that the phases of the two wavefunctions $\psi_1$ and $\psi_2$ are out of phase. You can see this because the integration is over two different intervals, so the two phases will turn out different. This is significant, because wavefunctions out of phase is characteristic of interference. When two wavefunctions with different phases overlap, they form the characteristic interference pattern (\cite{QM}; \cite{kvn_thesis}). 

If we want the probability of finding a particle at the rightmost wall (the interference wall), we need to know the wavefunction $\bra{x}\ket{\psi(t)}$ at the rightmost wall. If it takes time $t_R$ to reach this wall, that wavefunction can be simply represented as $\bra{x}\ket{\psi(t_R)} = \psi(x, t_R)$ (equivalent to taking \eqref{eq:important_phase} and plugging in $t = t_R$). Then, according to the postulates of section 2.5, we just have to apply the Born Rule to find the probability density as a function of $x$:

$$ \rho(x, t = t_R) = \psi(x,t_R)^* \psi(x,t_R) = \bra{\psi(t_R)}\ket{x}\bra{x}\ket{\psi(t_R)} =$$
$$ = \psi_1(x,t_R)^* \psi_1(x,t_R) + \psi_2(x,t_R)^* \psi_2(x,t_R) + \psi_1(x,t_R)^* \psi_2(x,t_R) + \psi_2(x,t_R)^* \psi_1(x,t_R)$$
The last two terms are not identically zero (\cite{kvn_thesis}). If we plot $\rho$ vs $x$, we will get the typical interference pattern we expected, as seen in Figure 2b. 

\begin{figure}[!htb]
\centering
\includegraphics[width=\textwidth]{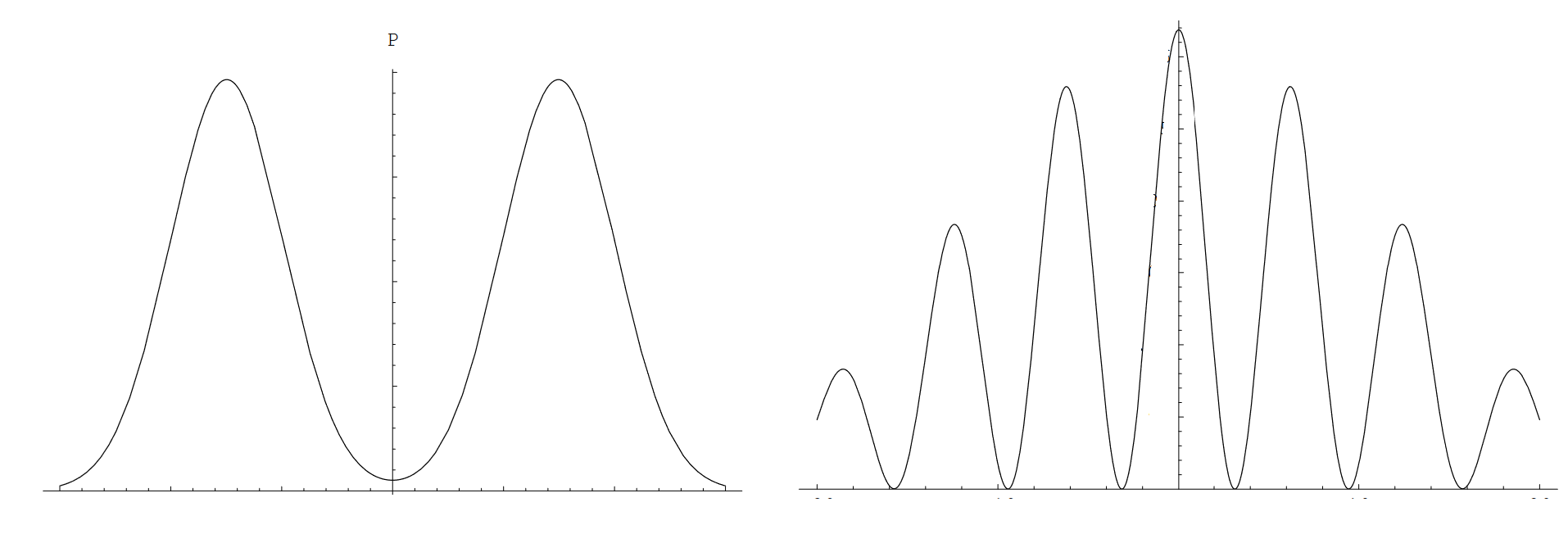}
\caption{a. Left: Graph of KvN probability distribution. b. Right: Graph of QM probability distribution. Adapted from \cite{DM}.}
\end{figure}

\begin{figure}[!htb]
\centering
\includegraphics[width=\textwidth]{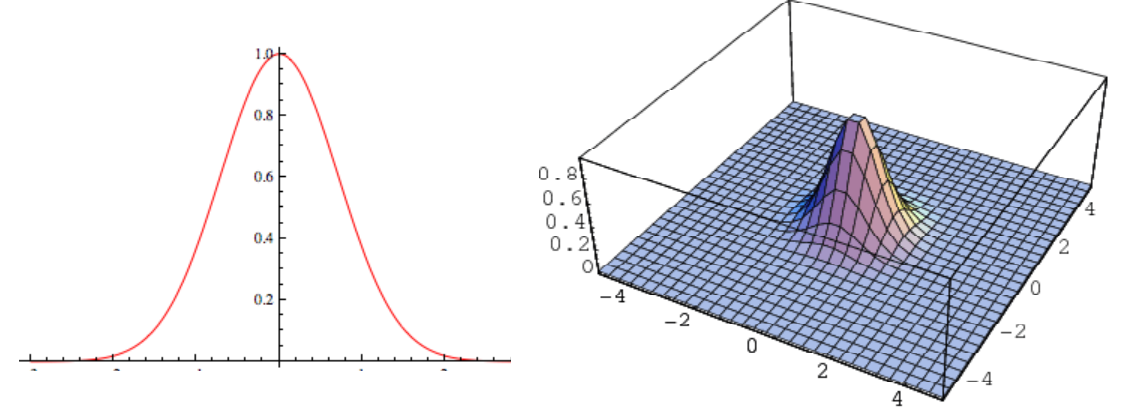}
\caption{a. Left: Example graph of a single variable gaussian curve. b. Right: Example graph of a double variable gaussian curve. Adapted from \cite{matlab}.}
\end{figure}

This, of course, comes as no surprise. We are still dealing with quantum theory. Following the same steps, though, will we get the same results with KvNM? We will follow a similar series of steps to evaluate classical KvN theory.

For the KvNM case, we will use a double gaussian function because we are making considerations with \textit{both} the position and momentum (KvNM is a phase space theory). A double gaussian (a gaussian in x and y coordinates; see Figure 3b) takes on the mathematical form:
\begin{equation}\label{eq:double_gaussian_eq}
f(u, v) = \frac{1}{\sqrt{\pi{\sigma_u}{\sigma_v}}}e^{-(u-\mu_u)^2/{2{\sigma_u}^2}}e^{-(v-\mu_v)^2/{2{\sigma_v}^2}}
\end{equation}
We are interested in the interference pattern that lies along the $x$ direction. Taking the KvN wavefunction $\psi(x,p_x,t) = R(x, p_x, t)e^{iG(x,p_x,t)}$ (like in section 5.1) and plugging in \eqref{eq:double_gaussian_eq} for $R$ at $t = 0$ we will have

\begin{equation}\label{eq:double_gaussian_koopman_wavefunction_1}
\psi(x,p_x, t = 0) = \frac{1}{\sqrt{\pi \sigma_x \sigma_p}}\exp[-\frac{x^2}{2{\sigma_x}^2}-\frac{p_x^2}{2{\sigma_p}^2}+iG(x,p_x,t=0)]
\end{equation}
Please note here that $\sigma_p$ refers to the momentum in the $x$ direction. Following the Path Integral Formulism, in order to know the know the wavefunction $\psi$ at some later time $t$ we will need to utilize the Kernel. For this situation, the travelling classical particle is free of any potential $V(x,y)$, so we will need the free particle wavefunction. As $V(x,y) = 0$, there will be no derivatives of the potential. We can take the results \eqref{eq:delta_enforcing_CM_1} and \eqref{eq:delta_enforcing_CM_2} to get a straightforward and concise Kernel, which will make the calculation we are about to do simple. 

We can take  \eqref{eq:delta_enforcing_CM_1} and rewrite it in the following way (dropping the constant $2\pi$, as it can be normalized away, and replacing $q$ with the desired variable $x$):

\begin{equation*}
\delta\Big(\frac{\Delta t}{m}(p_j -m\frac{x_{j+1} - x_j}{\Delta t})\Big) =  \delta\Big(\frac{p_j \Delta t}{m} - x_{j+1} + x_j \Big)
\end{equation*}
In the continuum limit, we can just simply write this as:
\begin{equation*}
 =  \delta\Big(\frac{p_0  t}{m}  + x_0 - x\Big)
\end{equation*}
where $x_0$ is the initial position of the particle, $p_{0x}$ is the initial momentum, and $m$ is the mass as before. This comes straight from elementary kinematics, as since there is no potential, the derivative of the potential must be 0, so there is no acceleration:

$$ F_x = - \frac{\partial V}{\partial x} = ma_x \Rightarrow a_x = 0 $$

$$ x(t) = x_0 + v_{0x} t + \frac{1}{2}a_x t^2 \Rightarrow x(t) = x_0 + v_{0x} t = x_0 + \frac{p_{0x} t}{m}$$
Using the momentum relation $p_x = mv_x$ on the last step. In other words, the Dirac delta functional is enforcing classical position evolution, $x(t)$. 

The same can be done with \eqref{eq:delta_enforcing_CM_2} to get:
\begin{equation*}
 \frac{1}{\Delta t} \delta\Big(\frac{p_{j+1} - p_j}{\Delta t} +V'(q_j)\Big) = \delta\Big(p_{j+1} - p_j\Big)
\end{equation*}
In the continuum limit remembering the direction is in $x$:

$$ = \delta\Big(p_x - p_{0x}\Big) $$
This makes sense, considering that there are no forces present (because potential is zero), and hence the momentum should be constant. This Dirac delta functional enforces classical momentum conservation. 

The Kernel $K$ can therefore be succinctly represented as 

\begin{equation}\label{eq:koopman_Kernel}
K = \delta\Big(p_x - p_{0x}\Big)\delta\Big(x - \frac{p_{0x}  t}{m}  - x_0 \Big)
\end{equation}
And plugging this Kernel into the original equation \eqref{eq:koopman_path_integral_1} shows us that indeed it provides us with the correct result that we desire, forcing classical evolution of the KvN wavefunction:
\begin{equation*}
\bra{x, p_x}\ket{\psi(t)} =  \int dx_0 ~ dp_{0x} \delta\Big(p_x - p_{0x}\Big)\delta\Big(x - \frac{p_{0x}  t}{m}  - x_0  \Big)\psi(x_0,p_{0x}, 0) = 
\end{equation*}
$$= \psi(x - \frac{p_{x}  t}{m}, p_x, 0) = \psi(x,p_x,t),$$
utilizing the property \eqref{eq:int_dirac_delta_functional} . This statement checks out, as we know that $\bra{x, p}\ket{\psi(t)} = \psi(x, p, t)$. 

Returning to \eqref{eq:double_gaussian_koopman_wavefunction_1}, we will evaluate the same integral utilizing the Kernel. By equation \eqref{eq:koopman_path_integral_1}, the evolution of the wavefunction from its original state when $t = 0$ is given by

\begin{equation*}
\bra{x, p_x}\ket{\psi(t)} =   \frac{1}{\sqrt{\pi \sigma_{x} \sigma_{p}}} \int dx_0 ~ dp_{0x} \delta\Big(p_x - p_{0x}\Big)\delta\Big(x - \frac{p_{0x}  t}{m}  - x_0\Big) \times
\end{equation*}
$$ \times \exp[-\frac{x_0^2}{2{\sigma_{x}}^2}-\frac{p_{0x}^2}{2{\sigma_{p}}^2}+iG(x_0,p_{0x},0)] = $$
\begin{equation}\label{eq:double_slit_wall_kvn_wavefunction}
=   \frac{1}{\sqrt{\pi \sigma_{x} \sigma_{p}}} \exp[-\frac{1}{2\sigma_{x}^2}\Big({x - \frac{p_{x}  t}{m}}\Big)^2 -\frac{p_{x}^2}{2{\sigma_{p}}^2}+iG\Big(x - \frac{p_{x}  t}{m} ,p_x\Big)] 
\end{equation}
Note that the above is strictly in terms of the position and momentum in the $x$ direction. The above equation is true at least until it hits the middle wall with the slits, because up until that point the wavefunction is that of a free particle. Using geometric considerations, we can figure out the wavefunction at the double slit wall. Since there is no acceleration in the direction of travel (as $-\partial V/\partial y = 0$), we know from kinematics:

$$ v_y = \frac{p_{0y}}{m} = \frac{\Delta y}{\Delta t} = \frac{y_f - y_0}{t_f - t_0}$$
Under the assumption that $y(0)=0$ we can write:

\begin{equation}\label{eq:time_to_double_slit_wall}
\frac{y_M}{t} = \frac{p_{0y}}{m} \Rightarrow t_M = \frac{y_M m}{p_{0y}},
\end{equation}
where $t_M$ is the time to the double slit wall. We can substitute this value into \eqref{eq:double_slit_wall_kvn_wavefunction} to know what is the classical wavefunction at that point in time.  

Then, the wavefunction right after the two slits can be written as 

\begin{equation}\label{eq:right_after_double_slit}
\psi(x,p_x, t_M + dt) = N\psi(x,p_x,t_M) [C_1(x) + C_2(x)]
\end{equation}
Utilizing, as before, the functions \eqref{eq:C_function_1} and \eqref{eq:C_function_2} we had defined in the quantum picture above.
This is done in order to filter out the effects of the wavefunction that actually goes through the slits. We do not care for the portion of the wavefunction that does not go through the double slit. They will produce no effects. The above equations make certain that only the wavefunction that goes through slits $\Delta_1$ and $\Delta_2$ will be studied. The $N$ in \eqref{eq:right_after_double_slit} is a renormalization constant, similar to before, so that $\int dx ~dp_x \psi^*\psi = 1$ (the renormalization condition for the KvN wavefunction.) 

Using our Kernel for the free particle again, we will evaluate the wavefunction using the path integral for the region \textit{after} the two slits. We again utilize \eqref{eq:koopman_path_integral_1}. Notice that the initial wavefunction $\psi(q_0,p_{0},0)$ in that equation starts at time zero, so we will need to adjust the time at the double slit wall so that it is set to zero. We can do this by substituting the variable $t$ in the Kernel \eqref{eq:koopman_Kernel} with $t - t_M$. This can be done because time has the unique property that any point can be arbitrarily set to $t = 0$, and by doing this swap we effectively set the time at the middle wall equal to zero (starts at $t_M$ and evolves from there). So the integral we need to evaluate has the following form:

\begin{equation*}
\bra{x, p_x}\ket{\psi(t - t_M)} =  \int dx_M ~ dp_{Mx} \delta\Big(p_x - p_{Mx}\Big)\delta\Big(x - \frac{p_{Mx}  (t - t_M)}{m}  - x_M  \Big) \times
\end{equation*}
$$ \times N\psi(x_M,p_{Mx}, t_M)[C_1(x_M) + C_2(x_M)] $$
Plugging everything in:

\begin{equation*}
\bra{x, p_x}\ket{\psi(t - t_M)} =  N\int dx_0 ~ dp_{0x} \delta\Big(p_x - p_{0x}\Big)\delta\Big(x - \frac{p_{0x}  (t - t_M)}{m}  - x_0  \Big) \times
\end{equation*}
$$\times\exp[-\frac{1}{2\sigma_{x}^2}\Big({x_0 - \frac{p_{0x}  t_M}{m}}\Big)^2 -\frac{p_{0x}^2}{2{\sigma_{p}}^2}+iG\Big(x_0 - \frac{p_{0x}  t_M}{m} ,p_{0x}\Big)] [C_1(x_0) + C_2(x_0)]$$
And then using \eqref{eq:int_dirac_delta_functional} to evaluate:

\begin{equation*}
\bra{x, p_x}\ket{\psi(t - t_M)} =  N exp\Big[-\frac{1}{2\sigma_{x}^2}\Big({x - \frac{p_{x}  (t - t_M)}{m} - \frac{p_{x}  t_M}{m}}\Big)^2 -\frac{p_{x}^2}{2{\sigma_{p}}^2}+
\end{equation*}
$$ +iG\Big(x - \frac{p_{x}  (t - t_M)}{m} - \frac{p_{x}  t_M}{m} ,p_{x}\Big)\Big]\Big[C_1\Big(x - \frac{p_{x}  (t - t_M)}{m}\Big) + C_2\Big(x - \frac{p_{x}  (t - t_M)}{m}\Big)\Big]  = $$
\begin{equation*}
 = N\exp[-\frac{1}{2\sigma_{x}^2}\Big({x - \frac{p_{x}  t }{m} }\Big)^2 -\frac{p_{x}^2}{2{\sigma_{p}}^2}+iG\Big(x - \frac{p_{x}  t }{m} ,p_{x}\Big)] \times
\end{equation*}
\begin{equation}\label{eq:final_koopman_wavefunction_1}
\times \Big[C_1\Big(x - \frac{p_{x}  (t - t_M)}{m}\Big) + C_2\Big(x - \frac{p_{x}  (t - t_M)}{m}\Big)\Big] 
\end{equation}
Similar to \eqref{eq:time_to_double_slit_wall}, under geometric considerations one can show that the time to the back wall is $t_R = my_R/p_{0y}$, and so we can substitute this and \eqref{eq:time_to_double_slit_wall} into the above expression to get the expression for the wavefunction at the back wall (wall where we should see the interference pattern emerge in QM).

We want to know the probability distribution for finding a particle at coordinate $x$ at this back wall (\cite{kvn_thesis}). We have no need of the momentum information, as the interference pattern is a shape that appears in space. So what we can do is get the probability distribution in terms of $x$ is the following:

$$P(x) = \int_{-\infty}^\infty dp_x   |\psi(x,p_x,t_R - t_M)|^2 = \int_{-\infty}^\infty dp_x  \bra{\psi(t_R - t_M)}\ket{x, p_x}\bra{x,p_x}\ket{\psi(t_R - t_M)}$$
Plugging in \eqref{eq:final_koopman_wavefunction_1} for $\bra{x,p_x}\ket{\psi(t_R - t_M)}$ we get:
\begin{equation*}
 P(x) = N \int_{-\infty}^\infty dp_x \exp[-\frac{1}{2\sigma_{x}^2}\Big({x - \frac{p_{x}  t_R }{m} }\Big)^2 -\frac{p_{x}^2}{2{\sigma_{p}}^2}]^2 \exp[-iG\Big(x - \frac{p_{x}  t_R }{m} ,p_{x}\Big)]   \times
\end{equation*}
$$ \times \exp[iG\Big(x - \frac{p_{x}  t_R }{m} ,p_{x}\Big)] \Big[C_1\Big(x - \frac{p_{x}  (t_R - t_M)}{m}\Big) + C_2\Big(x - \frac{p_{x}  (t_R - t_M)}{m}\Big)\Big]^2 = $$
$$  = N \int_{-\infty}^\infty dp_x \exp[-\frac{1}{2\sigma_{x}^2}\Big({x - \frac{p_{x}  t_R }{m} }\Big)^2 -\frac{p_{x}^2}{2{\sigma_{p}}^2}]^2 \times $$
$$ \times \Big[C_1\Big(x - \frac{p_{x}  (t_R - t_M)}{m}\Big) + C_2\Big(x - \frac{p_{x}  (t_R - t_M)}{m}\Big)\Big]^2$$
We will use a very special property of the $\theta$-Heavyside step function at this point. We know that the following important property is true (\cite{kvn_thesis}):

$$(C_1 + C_2)^2 = C_1 + C_2 $$
Therefore:

$$P(x) = N \int_{-\infty}^\infty dp_x \exp[-\frac{1}{2\sigma_{x}^2}\Big({x - \frac{p_{x}  t_R }{m} }\Big)^2 -\frac{p_{x}^2}{2{\sigma_{p}}^2}]^2 \times $$
$$ \times \Big[C_1\Big(x - \frac{p_{x}  (t_R - t_M)}{m}\Big) + C_2\Big(x - \frac{p_{x}  (t_R - t_M)}{m}\Big)\Big]$$
This last equation is very important, because it shows that no interference pattern will emerge from the classical KvN wavefunction. If you look at the above equation closely, you can see that the probability distribution of $x$ is a superposition of two probability distributions:
$$P(x) = N \int_{-\infty}^\infty dp_x \exp[-\frac{1}{2\sigma_{x}^2}\Big({x - \frac{p_{x}  t_R }{m} }\Big)^2 -\frac{p_{x}^2}{2{\sigma_{p}}^2}]^2 C_1\Big(x - \frac{p_{x}  (t_R - t_M)}{m}\Big)  + $$
$$+ N \int_{-\infty}^\infty dp_x \exp[-\frac{1}{2\sigma_{x}^2}\Big({x - \frac{p_{x}  t_R }{m} }\Big)^2 -\frac{p_{x}^2}{2{\sigma_{p}}^2}]^2 C_2\Big(x - \frac{p_{x}  (t_R - t_M)}{m}\Big) $$
The first integral in the above expression is the probability at $x$ if only the first slit $\Delta_1$ was opened and the second slit $\Delta_2$ was closed, and the second integral in the above expression is the probability at $x$ if only the second slit $\Delta_2$was opened and the first slit $\Delta_1$ was closed. This can be seen from how $C_1$ filters out any portion of the wavefunction that does not go through $\Delta_1$ and $C_2$ filters out any portion of the wavefunction that does not go through $\Delta_2$. This means that the total probability distribution is the sum of the probability distribution of each slit, which is exactly what you would expect for particle instead of wave behavior (no interference effects).

Notice also that the phase of the wavefunction has completely disappeared from the above equation. You will see this is important, because for interference effects we will want the phase to play an active role. Waves must be out of phase in order to produce interference effects, not have the phase cancel each other out like in the above equations.

If you graph the above function, you will see that indeed no interference fringes were produced (see Figure 2a). Koopman wavefunctions therefore successfully predict the behavior of classical particles. Despite their complex nature in a Hilbert space, they do no produce any interference effects. All the interference effects completely disappear. It shows how wavefunctions can describe classical physics.

\section{Miscellaneous KvN Topics of Interest}
\subsection{Operational Dynamic Modeling}
The history of physics is full of failed models, models that did not stand the test of time. Physics is often carried out in a trial and error fashion, a process that can be repetitive, messy, and slow. This raises the question if discovery and model building in physics can be optimized, for the sake of speeding up the progress of discovery. 

One such optimization model is called Operation Dynamical Modeling (ODM) and utilizes KvNM as an essential ingredient (\cite{ODM}). The two main pillars of modern physics is QM and CM. \footnote{Note, General Relativity and Special Relativity are here considered to be part of CM.} Since we have seen that both QM and CM must obey the Ehrenfest Theorems, any particular model must conform to either QM or CM to the best of our knowledge, and if it falls outside either pillar, the model is quite likely false. If repeated measurements influence the outcome of the experiment, one knows that there must be a noncommutative algebra involved (i.e., there must be a nonzero commutator involved, like in QM.) If repeated measurements do not generally alter the outcome of the experiment in the fashion of QM, you can assume a commutator of 0 between all variables involved. 

Since the Ehrenfest Theorems deal with dynamical weighted averages (expectation values), this is a statistical test of different models. If a model does not obey the Ehrenfest Theorems, it is quite likely to be false, and progress can be saved by either fixing the model so that it is inline with the fundamental Ehrenfest Theorems or scrapping the model. 

Not only does ODM allow quantum and classical physics, it also explains the nexus of the two. Semi-classical systems are incorporated into ODM, as they must also obey the Ehrenfest Theorems. By construction, one could create a commutator that encompasses \textit{both} the classical and quantum worlds, such as [$\hat{q}$, $\hat{p}$] = $i\hbar\kappa$ where $\kappa$ is a constant that lies between 0 (classical world) and 1 (purely quantum world). By taking the limit of $\kappa$ goes to 0, one should be able to recover either CM. 

This melding, in fact, results in the derivation of the semi-classical Wigner quasiprobability distribution. This function is known to accurately describe the interface between the quantum world and the classical world. As the Schr\"{o}dinger equation is generally written in terms of configuration space, the Wigner quasiprobability distribution is important because it recasts QM into a phase space theory. The Wigner function has a few odd properties, including the fact that it can take on negative probabilities, but thanks to ODM some light has been shed to the reasons why this might be the case (as we will discuss later.)

We will begin with Stone's Theorem again like we did in section 3, i.e. $i \frac{\partial }{\partial t}\ket{\psi} = \hat{H}\ket{\psi}$, and try to create an operator $\hat{H}_{QC}$ which will encompass both the classical and quantum worlds using our newly constructed commutator [$\hat{q}$, $\hat{p}$] = $i\hbar\kappa$, $0 \le \kappa \le 1$.

We must have a unified quantum-classical algebra which gives us the Koopman-von Neumann Algebra \eqref{eq:koopman_algebra} when $\kappa = 0$ and a quantum algebra for $\kappa = 1$. We desire:

$$\lim\limits_{\kappa \to 0}\hat{q}_{QC} = \hat{q}_c$$
$$\lim\limits_{\kappa \to 0}\hat{p}_{QC} = \hat{p}_c$$
$$\lim\limits_{\kappa \to 0}\hat{H}_{QC} = \hbar\hat{K}$$
In order to reflect the KvN Algebra, \cite{wigner} proposed the following unified Classical-Quantum Algebra: 

\begin{equation}
[\hat{q}_{QC}, \hat{\theta}_{QC}] = [\hat{p}_{QC}, \hat{\lambda}_{QC}] = i
\end{equation}
All other commutators will equal 0.
For the unified dynamics, we first begin with the Ehrenfest Theorems (equations \eqref{eq:ehrenfest_theorem_1} and \eqref{eq:ehrenfest_theorem_2} for reference) as before:
\begin{equation*}
\frac{d}{dt}\langle q \rangle  = \frac{d}{dt}\bra{\psi(t)}\hat{q}_{QC}\ket{\psi(t)} = \frac{1}{m}\langle p \rangle = \frac{1}{m} \bra{\psi(t)}\hat{p}_{QC}\ket{\psi(t)}
\end{equation*}
\begin{equation*}
\frac{d}{dt} \langle p \rangle = \frac{d}{dt}\bra{\psi(t)}\hat{p}_{QC}\ket{\psi(t)} = \langle - V'(q) \rangle = \bra{\psi(t)}-V'(\hat{q}_{QC})\ket{\psi(t)}
\end{equation*}
Following the same series of steps as in section 3, we will obtain the differential equations:
$$\kappa m\frac{\partial H}{\partial p_{QC}} + \frac{m}{\hbar}\frac{\partial H}{\partial \theta_{QC}} = p_{QC}$$
$$\kappa \frac{\partial H}{\partial q_{QC}} - \frac{1}{\hbar}\frac{\partial H}{\partial \lambda_{QC}} = V'(q_{QC})$$ 
The general solution $H$ to the two above differential equation is simply:

\begin{equation*}
H(\hat{q}_{QC}, \hat{p}_{QC} , \hat{\theta}_{QC} , \hat{\lambda}_{QC} ) = \frac{1}{\kappa}\Big[\frac{\hat{p}_{QC} ^2}{2m} + V(\hat{q}_{QC})\Big] + F(\hat{p}_{QC} - \hbar\kappa\hat{\theta}_{QC}, \hat{q}_{QC} +\hbar\kappa\hat{\lambda}_{QC})
\end{equation*}
where the function $F$ is a function undetectable to experiment because one can show that for any observable $A$, you have $[\hat{F},\hat{A}] =0$ (\cite{ODM}). 

The quantum operators can be constructed from classical ones, so \cite{ODM} propose that 

$$\hat{q}_{QC} = \hat{q}_c - \hbar\kappa\frac{\hat{\lambda}_c}{2}$$
$$\hat{p}_{QC} = \hat{p}_c + \hbar\kappa\frac{\hat{\theta}_c}{2} $$
$$\hat{\theta}_{QC} = \hat{\theta}_{c}$$
$$\hat{\lambda}_{QC} = \hat{\lambda}_{c}$$
It is clear that the above expressions have a nice, smooth classical limit as $\kappa \rightarrow 0$. All the variables operators become classical operators in that limit. 

To make the limit smooth as you go from the quantum Hamiltonian $\hat{H}$ to $\hbar\hat{K}$ as $\kappa$ goes from 1 to 0, \cite{ODM} dervied:

\begin{equation}
\hat{H}_{QC} = \frac{1}{\kappa}[\frac{\hat{p}_{QC} ^2}{2m} + V(\hat{q}_{QC})] - \frac{1}{2m\kappa}(\hat{p}_{QC} - \hbar\kappa\hat{\theta}_{QC})^2 - \frac{1}{\kappa}V( \hat{q}_{QC} +\hbar\kappa\hat{\lambda}_{QC})
\end{equation}
Equivalently, you can write by plugging in the quantum-classical operators:

$$ \hat{H}_{QC} = \frac{\hbar}{m}\hat{p}_c\hat{\theta}_c + \frac{1}{\kappa}V(\hat{q}_c - \frac{\hbar\kappa}{2}\hat{\lambda}_c) - \frac{1}{\kappa}V(\hat{q}_c + \frac{\hbar\kappa}{2}\hat{\lambda}_c)$$
The above expression can be shown to lead to the phase space formulation of QM, the Wigner quasiprobability distribution. This derivation will be left for the next section. The Wigner distribution is significant because it is known to accurately model the `nexus' between quantum and classical worlds (\cite{ODM}; \cite{wigner}). It also has a nice classical limit (\cite{ODM}; \cite{wigner}). Independently deriving this distribution indicates that the Ehrenfest Theorems apply at all times, including in the regime of quantum/classical boundary. ODM, therefore, is a very successful melding of the two worlds. 

ODM suggests that any physical theory will follow KvNM, QM, or the above melding of the two. KvNM is fundamentally important because any novel classical theory proposed must be consistent with it. We will see certain applications of ODM and its successes in the pages that follow. Without KvNM, there would be no unifying paradigm based in the Ehrenfest Theorems. ODM is a methodology that will pave for more efficient discovery in Physics.

\subsection{Wigner Quasiprobability Distribution}

The Wigner Quasiprobability Distribution connects our understanding of the quantum wavefunction to phase space. It is an important area of study developed by Eugen Wigner in 1932 and independently formulated by another eminent mathematical physicist, José Enrique Moyal. The Wigner Distribution has a variety of applications, from quantum optics to quantum computing (\cite{wigner}). It has one significantly odd feature though. This distribution allows for negative probabilities, which is axiomatically impossible by mainstream probability theories (\cite{wigner}). ODM sheds some light on this odd feature. The Wigner Distribution can be derived from the unified dynamics of ODM, and we will proceed to do so below. 

It is very simple to derive the Wigner quasiprobability distribution from ODM's unified dynamics. Again, we hark back to Stone's Theorem, in order to try to figure out the generator of motion:

$$i\hbar \frac{d}{dt}\ket{\psi(t)_\kappa} = \hat{H}_{QC} \ket{\psi(t)_\kappa}$$
We will sandwich it from the left with  $\ket{q_c, \lambda_c}$ to get the $\lambda_c, q_c$ representation. Referring back to  the rules for operators acting on vector kets, we can see that:

\begin{equation}
\bra{q, \lambda}\hat{q}_c \ket{\psi(t)} = [\bra{q, \lambda}\hat{q}_c]\ket{\psi(t)} = q \braket{q, \lambda}{\psi(t)}
\end{equation}
\begin{equation}
\bra{q, \lambda}\hat{\lambda}_c \ket{\psi(t)} = [\bra{q, \lambda}\hat{\lambda}_c]\ket{\psi(t)} = \lambda \braket{q, \lambda}{\psi(t)}
\end{equation}
where $q$ and $\lambda$ are the eigenvalues. From Theorem IV of Appendix A, we can say that $B = p$, $A = \lambda$, $\kappa = -1$ (since $[\hat{\lambda}, \hat{p}] = -i$) \footnote{This is because $[\hat{\lambda}, \hat{p}] = - [\hat{p}, \hat{\lambda}]$ and as indicated earlier, $[\hat{p}, \hat{\lambda}] = i$.} and keep in mind that the Classical Commutator is $[\hat{q}_c, \hat{p}_c] = 0$. Therefore:

\begin{equation}
\bra{q, \lambda}\hat{p}_c \ket{\psi(t)} = -i(-1)\frac{\partial}{\partial \lambda} \braket{q, \lambda}{\psi(t)} = i\frac{\partial}{\partial \lambda} \braket{q, \lambda}{\psi(t)}
\end{equation}
There is no partial derivative with respect to $q$ as the Classical Commutator between $\hat{q}$ and $\hat{p}$ is 0. Theorem IV can also be applied in the situation where $B = \theta$ in order to derive:
\begin{equation}
\bra{q, \lambda}\hat{\theta}_c \ket{\psi(t)} = -i\frac{\partial}{\partial q} \braket{q, \lambda}{\psi(t)}
\end{equation}

Now we create the actual `sandwich' from Stone's Theorem:

$$\bra{q,\lambda}i\hbar \frac{d}{dt}\ket{\psi(t)_\kappa} = \bra{q,\lambda}[\frac{\hbar}{m}\hat{p}_c\hat{\theta}_c + \frac{1}{\kappa}V(\hat{q}_c - \frac{\hbar\kappa}{2}\hat{\lambda}_c) - \frac{1}{\kappa}V(\hat{q}_c + \frac{\hbar\kappa}{2}\hat{\lambda}_c)] \ket{\psi(t)_\kappa}$$

$$i\hbar \frac{d}{dt}\bra{q,\lambda}\ket{\psi(t)_\kappa} = \bra{q,\lambda}\frac{\hbar}{m}\hat{p}_c\hat{\theta}_c\ket{\psi(t)_\kappa} + \bra{q,\lambda}\frac{1}{\kappa}V(\hat{q}_c - \frac{\hbar\kappa}{2}\hat{\lambda}_c)\ket{\psi(t)_\kappa} + $$
$$  -\bra{q,\lambda} \frac{1}{\kappa}V(\hat{q}_c + \frac{\hbar\kappa}{2}\hat{\lambda}_c) \ket{\psi(t)_\kappa}$$
Utilizing the previous results for Theorem I (see footnote 13), we can see:

$$i\hbar \frac{d}{dt}\bra{q,\lambda}\ket{\psi(t)_\kappa} = \frac{\hbar}{m}\bra{q,\lambda}\hat{p}_c\hat{\theta}_c\ket{\psi(t)_\kappa} + \frac{1}{\kappa}V(q - \frac{\hbar\kappa}{2}\lambda)\bra{q,\lambda}\ket{\psi(t)_\kappa} -\frac{1}{\kappa}V(q + \frac{\hbar\kappa}{2}\lambda) \bra{q,\lambda}\ket{\psi(t)_\kappa}$$
Finally utilizing the results from Theorem IV above:
$$i\hbar \frac{d}{dt}\bra{q,\lambda}\ket{\psi(t)_\kappa} =-i^2\frac{\hbar}{m}\frac{\partial^2}{\partial \lambda \partial q} \bra{q,\lambda}\ket{\psi(t)_\kappa}+ \frac{1}{\kappa}V(q - \frac{\hbar\kappa}{2}\lambda)\bra{q,\lambda}\ket{\psi(t)_\kappa} +$$
$$ -\frac{1}{\kappa}V(q + \frac{\hbar\kappa}{2}\lambda) \bra{q,\lambda}\ket{\psi(t)_\kappa}$$
Let us move all the terms to the left hand side of the equation, so that the left hand side equals 0 pm the right hand side. 
Introducing two new variables, $u = q - \hbar\kappa\lambda/2$ and $v = q + \hbar\kappa\lambda/2$, we will rewrite the whole equation so it has a prettier form. To eliminate the double partial derivative, we can take the following steps:

$$u - v = q - \frac{\hbar\kappa\lambda}{2} - q - \frac{\hbar\kappa\lambda}{2} = - \hbar\kappa\lambda$$
$$u + v = q - \frac{\hbar\kappa\lambda}{2} + q + \frac{\hbar\kappa\lambda}{2} = 2q $$
Therefore the partial derivative can be rewritten:

$$\frac{\partial^2}{\partial \lambda \partial q} = \frac{\partial^2}{\partial (\frac{u-v}{-\hbar\kappa})\partial (\frac{u + v}{2})}  = -2\hbar\kappa \frac{\partial^2}{\partial (u-v)\partial (u + v)}= -2\hbar\kappa \Big[\frac{\partial^2}{\partial u^2} - \frac{\partial^2}{\partial v^2}\Big]$$
Recasting the entire equation in terms of $u$ and $v$:
\begin{equation}
\Big[i\hbar\kappa \frac{\partial}{\partial t} - \frac{(\hbar\kappa)^2}{2m}\Big(\frac{\partial^2}{\partial u^2} - \frac{\partial^2}{\partial v^2}\Big) - V(u) + V(v)\Big]\braket{q,\lambda}{\psi(t)_\kappa} = 0
\end{equation}
Here, $\braket{q,\lambda}{\psi(t)_\kappa} $ is equal to the density matrix $\rho_\kappa(u,v,t)$ up to a constant of proportionality $C$, i.e., $\rho_\kappa(u,v,t) = C\braket{q,\lambda}{\psi(t)_\kappa} $. 

We can transition to $q,p$ representation to derive the Wigner distribution from the above using Theorem II with $A = (q, \lambda)$:

$$ \braket{q,p}{\psi(t)_\kappa} = \bra{q,p}I\ket{\psi(t)_\kappa} = \int d\lambda \bra{q,p}\ket{q, \lambda}\bra{q, \lambda}\ket{\psi(t)_\kappa} $$  
Note that we do not integrate over $q$ due to the commutator relationships $[\hat{q}, \hat{q}] =[\hat{q}_c, \hat{p}_c] = 0$, whereas $[\hat{p}_c, \hat{\lambda}_c] = i$. Since $[\hat{q}, \hat{q}] =[\hat{q}_c, \hat{p}_c] = 0$ it makes no sense to integrate over $q$, as might be initially implied by looking at the expressions akin to \eqref{eq:basic_properties_KvN}. Since the other commutator is nonzero, we will retain integration over $\lambda$, however. We know that $\rho_\kappa(u,v,t) = C\braket{q,\lambda}{\psi(t)_\kappa} $, so we will plug that in. 

The question now is to figure out what $\bra{q, p}\ket{q, \lambda}$ is. We will utilize Theorem IV and the commutator relationships in order to deduce this. Since $[\hat{q}, \hat{q}]=[\hat{q}_c, \hat{p}_c] =  0$, solving this expression is equivalent to solving $\bra{ p}\ket{\lambda}$, as the other terms commute (i.e., the commutator of their operators is equal to 0) and are therefore irrelevant to this problem. By Theorem IV, we set $A = p$, $B = \lambda$, and $\kappa = 1$. We know that this expression is therefore 

\begin{equation*}
\braket{q,p}{q, \lambda} = \braket{p}{\lambda} = \frac{1}{\sqrt{2\pi}}e^{ip\lambda} 
\end{equation*}
Putting it all together:
\begin{equation}
\braket{q,p}{\psi(t)_\kappa} = \int d\lambda \frac{C}{\sqrt{2\pi}}\rho_\kappa(q - \frac{\hbar\kappa\lambda}{2},q + \frac{\hbar\kappa\lambda}{2},t)e^{ip\lambda} 
\end{equation}
The above expression shows us that the wavefunction $\braket{q,p}{\psi(t)_\kappa}$ is equal to the Wigner quasi-probability distribution up to a constant of proportionality $C$. The Wigner distribution is very accurate when it comes to describing the boundary of Quantum and Classical (\cite{ODM}; \cite{wigner}), so this is no trivial derivation. 

What ODM reveals is that the Wigner ``distribution" is therefore a type of wavefunction for a quantum particle at a point in classical phase space. The strange property of having negative probabilities can by explained in the ODM framework as being due to the fact that the Wigner ``distribution" is not a distribution at all! It is a wavefunction. Wavefunctions, unlike probability distributions, can take on negative values (\cite{wigner}). The KvN wavefunction can have negative or imaginary weights to it, whereas traditional Louiville theory gives only positive, real weights to the same trajectories (\cite{wigner}). 

It is worth to highlight one more interesting result of the above analysis. Although not directly physically observable, the operators $\hat{\theta}$ and $\hat{\lambda}$ can be indirectly observed through the Wigner wavefunction. They are a critical ingredient as the so-called `Bopp Operators'. 

\subsection{Time Dependent Harmonic Oscillator}

The Harominc Oscillator is one of the most thoroughly studied models in all of physics. Its application arises in many situations. We are especially interested in the Time Dependent Harmonic Oscillator, where factors like the mass or frequency might change with time. Because of the common mathematical basis for QM and KvNM, it should come to no surprise that common sets of tools can be used to investigate both. The Harmonic Oscillator is no exception. The operatorial tools to investigate quantum oscillators can be used to explore classical (KvN) oscillators and vise versa. 

For a classical or quantum Time Dependent Harmonic Oscillator we need a quadratic potential of the following form:

$$ V(\hat{q},t) = k(t)\frac{\hat{q}^2}{2}$$
The above potential is modelled off of the classical spring potential. An ideal block-mass and spring system can undergo harmonic oscillation. The spring potential is given by $V = (1/2) kq^2$ with spring constant $k$ and displacement $q$. Here, we add an extra layer of complexity by making the spring constant time dependent. This is our model potential for all Time Dependent Harmonic Oscillation, whether quantum or classical. 

In the quantum case, we would plug in the above expression for the potential term into the Schr\"{o}dinger equation to solve for the Quantum Oscillator. In the classical case, we plug in the \textit{derivative} of the above expression into the Koopman Generator \eqref{eq:koopman_generator}:

$$\frac{\partial}{\partial \hat{q}}   V(\hat{q},t) = \frac{\partial}{\partial \hat{q}} k(t)\frac{\hat{q}^2}{2} = k(t)\hat{q} = V'(\hat{q})$$
$$ i\frac{\partial}{\partial t}\ket{\psi(t)} = \Big[\frac{\hat{p}\hat{\theta}}{m} - V'(\hat{q})\hat{\lambda}\Big]\ket{\psi(t)} = \Big[\frac{\hat{p}\hat{\theta}}{m} - k(t)\hat{\lambda}\hat{q}\Big]\ket{\psi(t)}.$$
Similar to what we did before \eqref{eq:L} in section 3, we can attack both sides with the bra vector $\bra{q,p}$ to get:

$$\bra{q,p}i\frac{\partial}{\partial t}\ket{\psi(t)} = \bra{q,p}\Big[\frac{\hat{p}\hat{\theta}}{m} - k(t)\hat{\lambda}\hat{q}\Big]\ket{\psi(t)},$$

$$i\frac{\partial}{\partial t}\bra{q,p}\ket{\psi(t)} = \bra{q,p}\frac{\hat{p}\hat{\theta}}{m}\ket{\psi(t)} - \bra{q,p}k(t)\hat{\lambda}\hat{q}\ket{\psi(t)},$$

\begin{equation}\label{eq:convenient_form_for_transformation}
i\frac{\partial}{\partial t}\psi(q,p,t) = \Big[\frac{\hat{p}\hat{\theta}}{m} - k(t)\hat{\lambda}\hat{q}\Big]\psi(q,p,t)
\end{equation}
where we colloquially say that $\hat{q} = q$, $\hat{p} = p$, $\hat{\lambda} = -i\frac{\partial}{\partial p}$, and $\hat{\theta} = -i\frac{\partial}{\partial q}$, just as we derived in section 3. Notice as for the Path Integral (section 4.2), we choose to set the constant $C$ to zero in the Koopman Generator for simplicity's sake. 

One way to solve for the equations of motion of a Harmonic Oscillator is to use \textit{Ermakov-Lewis invariants} (\cite{lewis_1}). 
If you have a Hamiltonian of the form

$$\hat{H} = \frac{1}{2}[\hat{p}^2 + \Omega^2(t)\hat{q}^2],$$
where $\Omega(t)$ is an arbitrary function with time dependence, you can use the invariant to write a transformation where the time dependent term  $\Omega(t)$ completely disappears. This makes solving the problem much simpler. Like it sounds, the Ermakov-Lewis invariant $\hat{I}$ is an operator that does not change (stays constant in time). The invariant $\hat{I}$ obeys the following rule for the quantum system (\cite{lewis_1}; \cite{lewis_2}):

\begin{equation}\label{eq:invariant_equation_governing_motion}
\frac{d\hat{I}}{dt} = \frac{\partial \hat{I}}{\partial t} + \frac{1}{i\hbar}[\hat{I},\hat{H}] = 0.
\end{equation}
$\hat{H}$ is the generator of motion, usually the quantum Hamiltonian, but in the classical case it will be the Koopman Generator. Based on the Koopman Algebra \eqref{eq:koopman_algebra}, \cite{harmonic_oscillator} deduced the following expression for $\hat{I}$:

$$ \hat{I} = \alpha_0 \hat{q}^2 + \alpha_1 \hat{\theta}^2 + \alpha_2 \hat{p}^2 + \alpha_3 \hat{\lambda}^2 + \alpha_4 \hat{q}\hat{p} + \alpha_5 \hat{\theta}\hat{\lambda} $$
where all the $\alpha_j$ are time dependent. This was a scientific guess basically, as there is no simple way to derive this fact. Plugging the above expression into \eqref{eq:invariant_equation_governing_motion}, we get:

\begin{equation}\label{eq:important_invariant_1}
\frac{\partial \alpha_0}{\partial t} \hat{q}^2 + \frac{\partial\alpha_1}{\partial t} \hat{\theta}^2 + \frac{\partial\alpha_2}{\partial t} \hat{p}^2 + \frac{\partial\alpha_3}{\partial t} \hat{\lambda}^2 + \frac{\partial\alpha_4}{\partial t} \hat{q}\hat{p} + \frac{\partial\alpha_5}{\partial t} \hat{\theta}\hat{\lambda} + \frac{1}{i\hbar}[\hat{I},\hat{K}] = 0
\end{equation}
Evaluating $[\hat{I},\hat{K}]$, we first can do \eqref{eq:comm_rule_5}:

\begin{equation}\label{eq:important_comm_relationship_a1}
[\hat{I},\hat{K}] = \Big[\hat{I},\frac{\hat{p}\hat{\theta}}{m} - k(t)\hat{\lambda}\hat{q}\Big] = \Big[\hat{I},\frac{\hat{p}\hat{\theta}}{m}\Big] - \Big[\hat{I}, k(t)\hat{\lambda}\hat{q}\Big]
\end{equation}
Taking the first term and applying \eqref{eq:comm_rule_3} and \eqref{eq:comm_rule_5}:
$$\Big[\hat{I},\frac{\hat{p}\hat{\theta}}{m}\Big] = -\Big[\frac{\hat{p}\hat{\theta}}{m}, \hat{I}\Big] =-\Big[ \frac{\hat{p}\hat{\theta}}{m}, \alpha_0 \hat{q}^2 + \alpha_1 \hat{\theta}^2 + \alpha_2 \hat{p}^2 + \alpha_3 \hat{\lambda}^2 + \alpha_4 \hat{q}\hat{p} + \alpha_5 \hat{\theta}\hat{\lambda}\Big] = $$
$$= -\Big[ \frac{\hat{p}\hat{\theta}}{m}, \alpha_0 \hat{q}^2\Big] - \Big[\frac{\hat{p}\hat{\theta}}{m},\alpha_1 \hat{\theta}^2 \Big] - \Big[\frac{\hat{p}\hat{\theta}}{m},\alpha_2 \hat{p}^2 \Big] - \Big[\frac{\hat{p}\hat{\theta}}{m},\alpha_3 \hat{\lambda}^2 \Big] - \Big[\frac{\hat{p}\hat{\theta}}{m},\alpha_4 \hat{q}\hat{p}\Big]- \Big[\frac{\hat{p}\hat{\theta}}{m},\alpha_5 \hat{\theta}\hat{\lambda}\Big]=$$
$$= - \frac{\hat{p}}{m}\Big[\hat{\theta}, \alpha_0 \hat{q}^2 \Big] - \Big[\frac{\hat{p}}{m},\alpha_0 \hat{q}^2 \Big]\hat{\theta} - \frac{\hat{p}}{m}\Big[\hat{\theta}, \alpha_1 \hat{\theta}^2 \Big] - \Big[\frac{\hat{p}}{m},\alpha_1 \hat{\theta}^2 \Big]\hat{\theta} - \frac{\hat{p}}{m}\Big[\hat{\theta}, \alpha_2 \hat{p}^2 \Big] - \Big[\frac{\hat{p}}{m},\alpha_2 \hat{p}^2 \Big]\hat{\theta}  + $$
\begin{equation}\label{eq:this_is_part_1}
  - \frac{\hat{p}}{m}\Big[\hat{\theta}, \alpha_3 \hat{\lambda}^2 \Big] - \Big[\frac{\hat{p}}{m},\alpha_3 \hat{\lambda}^2 \Big]\hat{\theta} - \frac{\hat{p}}{m}\Big[\hat{\theta}, \alpha_4 \hat{q}\hat{p} \Big] - \Big[\frac{\hat{p}}{m},\alpha_4 \hat{q}\hat{p}\Big]\hat{\theta} - \frac{\hat{p}}{m}\Big[\hat{\theta}, \alpha_5 \hat{\theta}\hat{\lambda} \Big] - \Big[\frac{\hat{p}}{m},\alpha_5  \hat{\theta}\hat{\lambda} \Big]\hat{\theta} 
\end{equation}
Utilizing our nifty identities \eqref{eq:comm_rule_2} and \eqref{eq:comm_rule_4} and Koopman Algebra again the following can be said about the above commutators:

$$ \Big[\hat{\theta}, \alpha_0 \hat{q}^2 \Big] = \alpha_0 \hat{q} \Big[\hat{\theta}, \hat{q}\Big] + \Big[\hat{\theta}, \alpha_0 \hat{q}\Big]\hat{q}  = -2i\alpha_0 \hat{q}$$

$$\Big[\hat{\theta}, \alpha_1 \hat{\theta}^2 \Big] = \alpha_1 \hat{\theta} \Big[\hat{\theta}, \hat{\theta}\Big] + \Big[\hat{\theta}, \hat{\theta}\Big]\alpha_1 \hat{\theta}  = 0 $$

$$ \Big[\hat{\theta}, \alpha_2 \hat{p}^2 \Big]  =  \alpha_2 \hat{p} \Big[\hat{\theta}, \hat{p}\Big] + \Big[\hat{\theta}, \hat{p}\Big]\alpha_2 \hat{p} = 0 $$

$$ \Big[\hat{\theta}, \alpha_3 \hat{\lambda}^2 \Big] = \alpha_3 \hat{\lambda} \Big[\hat{\theta}, \hat{\lambda}\Big] + \Big[\hat{\theta}, \hat{\lambda}\Big]\alpha_3 \hat{\lambda} = 0$$

$$\Big[\hat{\theta}, \alpha_4 \hat{q}\hat{p} \Big] = \alpha_4 \hat{q} \Big[\hat{\theta}, \hat{p}\Big] + \Big[\hat{\theta}, \hat{q}\Big]\alpha_4 \hat{p} = -i\alpha_4 \hat{p}$$

$$\Big[\hat{\theta}, \alpha_5 \hat{\theta}\hat{\lambda} \Big] =\alpha_5 \hat{\theta} \Big[\hat{\theta}, \hat{\lambda}\Big] + \Big[\hat{\theta}, \hat{\theta}\Big]\alpha_5 \hat{\lambda} = 0$$
$$\Big[\frac{\hat{p}}{m},\alpha_0 \hat{q}^2 \Big] = \alpha_0 \hat{q} \Big[\hat{p}/m, \hat{q}\Big] + \Big[\hat{p}/m, \hat{q}\Big]\alpha_0 \hat{q} = 0$$

$$ \Big[\frac{\hat{p}}{m},\alpha_1 \hat{\theta}^2 \Big] = \alpha_1 \hat{\theta} \Big[\frac{\hat{p}}{m},\hat{\theta}\Big]  + \Big[\frac{\hat{p}}{m},\hat{\theta}\Big]\alpha_1\hat{\theta} = 0$$

$$\Big[\frac{\hat{p}}{m},\alpha_2 \hat{p}^2 \Big] = \frac{\alpha_2}{m}\hat{p}\Big[\hat{p}, \hat{p}\Big] + \Big[\hat{p}, \hat{p}\Big] \frac{\alpha_2}{m}\hat{p} = 0$$

$$\Big[\frac{\hat{p}}{m},\alpha_3 \hat{\lambda}^2 \Big] =  \frac{\alpha_3}{m}\hat{\lambda}\Big[\hat{p},\hat{\lambda}\Big] + \Big[\hat{p},\hat{\lambda}\Big] \frac{\alpha_3}{m}\hat{\lambda} = \frac{2i\alpha_3}{m}\hat{\lambda}$$

$$ \Big[\frac{\hat{p}}{m},\alpha_4 \hat{q}\hat{p}\Big] = \frac{\alpha_4}{m}\hat{q}\Big[\hat{p},\hat{p}\Big] + \Big[\hat{p},\hat{q}\Big] \frac{\alpha_4}{m}\hat{p} = 0$$

$$ \Big[\frac{\hat{p}}{m},\alpha_5  \hat{\theta}\hat{\lambda} \Big] =  \frac{\alpha_5}{m}\hat{\theta}\Big[\hat{p}, \hat{\lambda}\Big] + \Big[\hat{p},\hat{\theta}\Big]\frac{\alpha_5}{m}\hat{\lambda} = \frac{i\alpha_5}{m}\hat{\theta}$$
Ergo, the above expression \eqref{eq:this_is_part_1} reduces to:

\begin{equation}\label{eq:commutator_a_1}
\Big[\hat{I},\frac{\hat{p}\hat{\theta}}{m}\Big] = 2i\alpha_0 \frac{\hat{p}\hat{q}}{m} + i\alpha_4 \frac{\hat{p}^2}{m}  -2i\alpha_3 \frac{\hat{\lambda}\hat{\theta}}{m} -i\alpha_5 \frac{\hat{\theta}^2}{m}
\end{equation}
Now, we do the same for $[\hat{I}, k(t)\hat{\lambda}\hat{q}]$:

$$\Big[\hat{I}, k(t)\hat{\lambda}\hat{q}\Big] = - \Big[ k(t)\hat{\lambda}\hat{q}, \hat{I}\Big] = -\Big[ k(t)\hat{\lambda}\hat{q}, \alpha_0 \hat{q}^2 + \alpha_1 \hat{\theta}^2 + \alpha_2 \hat{p}^2 + \alpha_3 \hat{\lambda}^2 + \alpha_4 \hat{q}\hat{p} + \alpha_5 \hat{\theta}\hat{\lambda}\Big] = $$
$$= -\Big[ k(t)\hat{\lambda}\hat{q}, \alpha_0 \hat{q}^2\Big] - \Big[k(t)\hat{\lambda}\hat{q},\alpha_1 \hat{\theta}^2 \Big] - \Big[k(t)\hat{\lambda}\hat{q},\alpha_2 \hat{p}^2 \Big] +$$
$$- \Big[k(t)\hat{\lambda}\hat{q},\alpha_3 \hat{\lambda}^2 \Big] - \Big[k(t)\hat{\lambda}\hat{q},\alpha_4 \hat{q}\hat{p}\Big] - \Big[k(t)\hat{\lambda}\hat{q},\alpha_5 \hat{\theta}\hat{\lambda}\Big]=$$

$$= - k(t)\hat{\lambda}\Big[\hat{q}, \alpha_0 \hat{q}^2 \Big] - \Big[k(t) \hat{\lambda}, \alpha_0 \hat{q}^2 \Big]\hat{q} - k(t)\hat{\lambda}\Big[\hat{q}, \alpha_1 \hat{\theta}^2  \Big]  + $$
$$ - \Big[k(t) \hat{\lambda}, \alpha_1 \hat{\theta}^2  \Big]\hat{q} - k(t)\hat{\lambda}\Big[\hat{q}, \alpha_2 \hat{p}^2 \Big]  - \Big[k(t) \hat{\lambda}, \alpha_2 \hat{p}^2 \Big]\hat{q} +$$
$$- k(t)\hat{\lambda}\Big[\hat{q}, \alpha_3 \hat{\lambda}^2  \Big] - \Big[k(t) \hat{\lambda}, \alpha_3 \hat{\lambda}^2  \Big]\hat{q} - k(t)\hat{\lambda}\Big[\hat{q}, \alpha_4 \hat{q}\hat{p} \Big]+$$
$$ - \Big[k(t) \hat{\lambda}, \alpha_4 \hat{q}\hat{p} \Big]\hat{q} - k(t)\hat{\lambda}\Big[\hat{q}, \alpha_5 \hat{\theta}\hat{\lambda} \Big] - \Big[k(t) \hat{\lambda},\alpha_5 \hat{\theta}\hat{\lambda} \Big]\hat{q}$$
We can perform a similar series of steps to achieve: 
$$\Big[\hat{q}, \alpha_0 \hat{q}^2 \Big] = 0 $$

$$\Big[k(t) \hat{\lambda}, \alpha_0 \hat{q}^2 \Big] = 0 $$

$$\Big[\hat{q}, \alpha_1 \hat{\theta}^2  \Big] = \alpha_1 \hat{\theta}\Big[\hat{q}, \hat{\theta} \Big] + \Big[\hat{q},\hat{\theta}\Big]\alpha_1\hat{\theta} = 2i\alpha_1\hat{\theta}$$

$$ \Big[k(t) \hat{\lambda}, \alpha_1 \hat{\theta}^2  \Big] = 0$$

$$ \Big[\hat{q}, \alpha_2 \hat{p}^2 \Big] = 0 $$

$$  \Big[k(t) \hat{\lambda}, \alpha_2 \hat{p}^2 \Big] = k(t)\alpha_2 \hat{p} \Big[\hat{\lambda}, \hat{p}\Big] +  \Big[\hat{\lambda}, \hat{p}\Big]k(t)\alpha_2 \hat{p}= -2i k(t)\alpha_2 \hat{p} $$

$$\Big[\hat{q}, \alpha_3 \hat{\lambda}^2  \Big] = 0$$

$$ \Big[k(t) \hat{\lambda}, \alpha_3 \hat{\lambda}^2  \Big] =0$$

$$ \Big[\hat{q}, \alpha_4 \hat{q}\hat{p} \Big] = 0$$

$$\Big[k(t) \hat{\lambda}, \alpha_4 \hat{q}\hat{p} \Big] = k(t) \alpha_4 \hat{q} \Big[\hat{\lambda}, \hat{p}\Big] + \Big[\hat{\lambda}, \hat{q}\Big] k(t) \alpha_4 \hat{p} = -i k(t) \alpha_4 \hat{q}$$

$$\Big[\hat{q}, \alpha_5 \hat{\theta}\hat{\lambda} \Big] = \alpha_5 \hat{\theta} \Big[\hat{q}, \hat{\lambda}\Big] + \Big[\hat{q}, \hat{\theta}\Big]\alpha_5\hat{\lambda} = i\alpha_5\hat{\lambda}$$

$$\Big[k(t) \hat{\lambda},\alpha_5 \hat{\theta}\hat{\lambda} \Big]= 0$$
Therefore, putting it all together:

\begin{equation}\label{eq:commutator_second_part}
\Big[\hat{I}, k(t)\hat{\lambda}\hat{q}\Big] =  -2i\alpha_1 k(t) \hat{\lambda}\hat{\theta} +2ik(t)\alpha_2 \hat{p}\hat{q} +ik(t)\alpha_4 \hat{q}^2 - ik(t)\alpha_5 \hat{\lambda}^2
\end{equation}

Utilizing equations \eqref{eq:important_comm_relationship_a1}, \eqref{eq:commutator_a_1}, \eqref{eq:commutator_second_part} with \eqref{eq:important_invariant_1}, we can compute the invariant to be:

\begin{equation*}
\frac{\partial \alpha_0}{\partial t} \hat{q}^2 + \frac{\partial\alpha_1}{\partial t} \hat{\theta}^2 + \frac{\partial\alpha_2}{\partial t} \hat{p}^2 + \frac{\partial\alpha_3}{\partial t} \hat{\lambda}^2 + \frac{\partial\alpha_4}{\partial t} \hat{q}\hat{p} + \frac{\partial\alpha_5}{\partial t} \hat{\theta}\hat{\lambda} + \frac{1}{i\hbar}\Big(2i\alpha_0 \frac{\hat{p}\hat{q}}{m} + i\alpha_4 \frac{\hat{p}^2}{m}  -2i\alpha_3 \frac{\hat{\lambda}\hat{\theta}}{m} -i\alpha_5 \frac{\hat{\theta}^2}{m} +
\end{equation*}
\begin{equation}
+2i\alpha_1 k(t) \hat{\lambda}\hat{\theta} -2ik(t)\alpha_2 \hat{p}\hat{q} -ik(t)\alpha_4 \hat{q}^2 + ik(t)\alpha_5 \hat{\lambda}^2\Big) = 0
\end{equation}
Combining like terms:

\begin{equation*}
\Big(\frac{\partial \alpha_0}{\partial t} - \frac{k(t)\alpha_4}{\hbar} \Big)\hat{q}^2 +\Big(\frac{\partial \alpha_1}{\partial t}-\frac{\alpha_5}{\hbar m}\Big)\hat{\theta}^2 + \Big(\frac{\partial \alpha_2}{\partial t} + \frac{\alpha_4}{\hbar m}\Big)\hat{p}^2 +\Big(\frac{\partial \alpha_3}{\partial t}+\frac{k(t) \alpha_5}{\hbar}\Big)\hat{\lambda}^2 + 
\end{equation*}
\begin{equation}
+ \Big(\frac{\partial \alpha_4}{\partial t} + \frac{2\alpha_0}{\hbar m} - \frac{2 k(t) \alpha_2}{\hbar}\Big)\hat{q}\hat{p} + \Big(\frac{\partial \alpha_5}{\partial t} - \frac{2\alpha_3}{\hbar m} + \frac{2k(t)\alpha_1}{\hbar}\Big)\hat{\theta}\hat{\lambda}  = 0
\end{equation}
 It must equal zero, but each of the operators we know is nonzero, so the only way for this equation to be true is if all the coefficients above equal zero. Setting all the coefficients equal to zero and solving the system of equations will lead you to the dynamics of the (here classical) Harmonic Oscillator.

For instance, if we solve for $\alpha_1$ we will get the equation:

$$ \frac{\partial^2 \alpha_1}{\partial t^2} + 2k(t)\alpha_1 = \frac{1}{2\alpha_1}\Big(\frac{\partial \alpha_1}{\partial t}\Big)^2 + \frac{C}{\alpha_1}$$
And if we define an arbitrary function called $\rho (t)$ so that $\alpha_1 = \frac{\rho^2 (t)}{2}$, we will have 

$$ \frac{d^2 \rho}{dt^2} + k(t)\rho = C/\rho^3$$
which is an equation we call the Ermakov equation (\cite{lewis_1}; \cite{lewis_2}). Using the Ermakov equation, all other coefficients can be rewritten so that the invariant has the following form:

$$ \hat{I} = \frac{1}{2}\Big[\frac{\hat{q}}{\rho} +(\frac{d\rho}{dt}\hat{q} - \rho\hat{p})^2 + \frac{\hat{\lambda}^2}{\rho^2}+ (\frac{d\rho}{dt}\hat{\lambda}+\rho\hat{\theta})^2 \Big].$$ 
Notice that the above classical Ermakov-Lewis invariant cannot be directly observed. This is because operators like $\hat{\theta}$ and $\hat{\lambda}$ are not directly tied to physically observable quantities (\cite{harmonic_oscillator}). Nevertheless, it serves as an important tool in solving Harmonic Oscillation problems.  

We can define following the unitary transformation 

$$\hat{T} = \exp[i\frac{ln(\rho)}{2}(\hat{q}\hat{p} +\hat{p}\hat{q}) ]\exp[-\frac{i}{2\rho}\frac{d\rho}{dt}\hat{q}^2],$$
which for the quantum case transforms the time dependent problem described above to a time independent problem (\cite{lewis_1}; \cite{lewis_2}) by the following:

$$ \frac{1}{2}[\hat{p}^2 + \hat{q}^2] = \hat{T}I\hat{T}^\dagger.$$
In the classical case, we can define the following two unitary transformations (\cite{harmonic_oscillator}):

$$\hat{T}_1 = \exp[\frac{i}{\rho}\frac{d\rho}{dt}\hat{q}\hat{\lambda}], $$
$$\hat{T}_2 = \exp[i\frac{ln(\rho)}{2}(\hat{q}\hat{\theta}+\hat{\theta}\hat{q})]\exp[-i\frac{ln(\rho)}{2}(\hat{p}\hat{\lambda}+\hat{\lambda}\hat{p})]. $$
We can define the new wavefunction $\Psi(q,p,t) = \hat{T}_2 \hat{T}_1 \psi(q,p,t)$ and we can rewrite our equation of interest \eqref{eq:convenient_form_for_transformation}:

\begin{equation}
i\frac{\partial}{\partial t}\Psi(q,p,t) = \frac{1}{\rho^2}\Big[\frac{\hat{p}\hat{\theta}}{m} - \hat{\lambda}\hat{q}\Big]\Psi(q,p,t).
\end{equation}
where the values of the operators are exactly the same as in \eqref{eq:convenient_form_for_transformation}.
The original wavefunction equation got transformed into a wavefunction equation without the time dependent potential term in front of the position operator. This can be much more easily solved to deduce the statistical motion of the classical particles. 

Note that this method was originally used in the operatorial form for the Schr\"{o}dinger equation (\cite{lewis_1}). We are able to successfully implement this procedure for KvNM as well. If you graph the phase space diagrams for a selected $k(t)$, you can see that it is indeed the graph of an oscillator (\cite{harmonic_oscillator}). KvNM is successful at bridging methods that involve quantum-like operators with classical physics.

\subsection{Aharonov–Bohm Effect}

The Aharonov-Bohm Effect is a surprising phenomena that emerges from the quantum wavefunction interacting with the magnetic potential. In classical electromagnetic theory, one can take the electric $\vec{E}$ and magnetic $\vec{B}$ fields and decompose them into various derivatives of electric and magnetic potentials:

$$ \vec{E} = -\vec{\nabla}\phi, $$
$$\vec{B} = \vec{\nabla} \cross \vec{A},$$ 
where $\phi$ is the electric scalar potential and $\vec{A}$ is the magnetic vector potential. Both of these potentials have the unique property that they make $\vec{E}$ and $\vec{B}$ \textit{gauge invariant}, in other words, that a `shift' $\chi$ will not produce different magnetic or electric fields. In other words, the new potentials $\vec{A}'$ and $\phi'$ defined by

$$\vec{A}' = \vec{A} + \vec{\nabla}\chi,$$
and
$$ \phi' = \phi - \frac{\partial \chi}{\partial t},$$ 
will produce the same magnetic and electric fields as $\vec{A}$ and $\phi$.
These potentials have historically been seen as useful mathematical constructs with little physical bearing on reality in themselves. The Aharonov-Bohm effect soon put this assumption to the test. 

The Aharonov-Bohm effect is a a quantum interference, much like that in Young's Double Slit Experiment, produced by the magnetic potential changing the phases of the quantum wavefunctions. When different wavefunctions with difference phases meet, they interfere with themselves. The surprising thing is that the interference pattern does not depend on the magnetic field $\vec{B}$ but directly on the magnetic potential $\vec{A}$. In other words, if the magnetic field $\vec{B} = \vec{\nabla} \cross \vec{A}$ is zero but the potential is nonzero, an effect will \textit{still} be observed. This suggests that the magnetic potential $\vec{A}$ is more fundamental than the magnetic field $\vec{B}$, which have always been seen as being on equal footing in classical theory. 

Aharonov and Bohm described the theory behind the phenomena in 1959, but the idea was not experimentally verified for the magnetic potential until after their deaths. The experiment used a solenoid, which is the model we will use here for the magnetic potential. Recall that a solenoid is a cylindrical device made of winding wire which produces a magnetic field inside the solenoid and zero magnetic field outside the solenoid.

\begin{figure}[!htb]
\centering
\includegraphics[width=\textwidth]{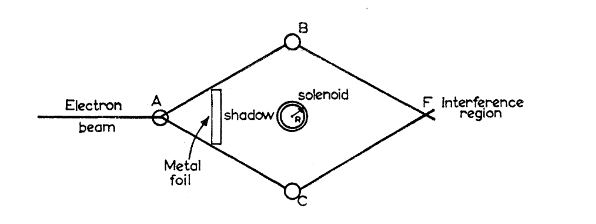}
\caption{Aharanov-Bohm Experiment set up. Reproduced from \cite{AB}.}
\end{figure}

The generic setup of the experiment for either type of potential is depicted in Figure 4. A coherent beam of electrons would be split into two branches, travel through an electric or magnetic potential where $\vec{E} = \vec{0}$ or $\vec{B} = \vec{0}$, and the two branches would then be recombined. Each branch's wavefunction took a slightly different path, so when the two branches are recombined an interference pattern should emerge because of a phase difference acumulated along each path. This is a semiclassical way to visualize the experiment (recall that in QM proper, the particle actually takes \textit{all} possible paths between source and sink in the allotted time.) 

To understand mathematically how the phase shift happens, first recall the polar form of the wavefunction \eqref{eq:configuration_space_wavefunction} and what the Action represents \eqref{eq:action}. We can write that:

\begin{equation}\label{eq:wavefunction_in_AB_exp}
 \psi = R\exp[\frac{i}{\hbar}\int_0^t T - V dt]. 
\end{equation}
The above wavefunction can be written for a region with no potential energy (free particle case) as:

$$ \psi_0 = R\exp[\frac{i}{\hbar}\int_0^t T dt]. $$
And so \eqref{eq:wavefunction_in_AB_exp} can be rewritten in the generic case as:

$$ \psi = \psi_0 \exp[-\frac{i}{\hbar}\int_0^t V dt]. $$
For a wavefunction going through an electric potential, from CM we know we can write that the potential energy is $V = e\phi$, where $e$ is the charge of the particle's wavefunction and $\phi$ is the scalar potential. And so the generic wavefunction can be written as the following for the case with electric potential:

\begin{equation}\label{eq:wavefunction_in_AB_exp_2}
\psi = \psi_0 \exp[-\frac{ie}{\hbar}\int_0^t \phi dt].
\end{equation}
The above formulas can be used to describe what should happen in the experiment in the presence of the electric potential. 

Refer to Figure 4. Starting off, the beam of coherent electrons has wavefunction $\psi_0$ in the absence of an external potential. It is split into two, and both branches travel through a region of potential $\phi$, so each wavefunction gains the form of \eqref{eq:wavefunction_in_AB_exp_2}. The branching could be represented as $\psi_1^0$ and $\psi_2^0$ upon reaching the area with electric potential at time $t_1$. After some evolution they leave the potential at time $t_2$, and they are then recombined. The first branch travelled along potential $\phi_1$ and the second branch travelled along $\phi_2$, in this semiclassical picture. You can write the following superposition for the final wavefunction:

$$ \psi_f = \psi_1 + \psi_2 = \psi_1^0 \exp[-\frac{i}{\hbar}C_1] + \psi_2^0 \exp[-\frac{i}{\hbar}C_2],$$
where we define the phases

$$ C_1 = e\int_{t_1}^{t_2} \phi_1 dt,$$
and

$$ C_2 = e\int_{t_1}^{t_2} \phi_2 dt.$$
Since $C_1$ and $C_2$ are different, we have gained a phase shift of $(C_1 - C_2)/\hbar$ \footnote{This comes straight from the Euler formula. Recall that:

$$e^{i(\theta + C)} = cos(\theta +C) + i sin(\theta + C)$$.
Here, the $C$ is the phase. If the phase of two waves are different, we should expect interference.} There is absolutely no classical force $\vec{F} = e\vec{E}$ exerted on the electrons as there is no electric field, but a measurable effect should still be produced on the electron's behavior (\cite{AB}).

 A similar analysis can be done with the magnetic vector potential. Following \cite{AB}, we can generalize \eqref{eq:wavefunction_in_AB_exp_2} to include the magnetic potential. In relativity, you can write the electromagnetic potential as the four vector $A^\mu = (\phi/c ~ \vec{A}) = (\phi/c ~ A_x ~ A_y ~A_z)$ where as before, $\phi$ is the electric potential and $\vec{A}$ is the magnetic potential. The position four vector can be represented is $x^\mu = (ct ~\vec{x}) = (ct ~ x ~ y ~z)$. Rewriting the position four vector as the differential $dx^\mu$, we can generalize the potential integral to be:

\begin{equation}
\frac{e}{\hbar}\oint\,A^\mu dx_\mu = \frac{e}{\hbar} \oint\,\Big( \phi dt - \frac{\vec{A}}{c}\cdot d\vec{x}\Big).
\end{equation}
The path of integration has to be closed in spacetime (\cite{AB}). Let us assume that the variable $t$ is a constant in order to understand how interference could arise strictly from the magnetic potential. The above integral suggests that we will expect the following phase shift to hold true:

$$ (C_1 - C_2)/\hbar = -\frac{e}{c\hbar}\oint\, \vec{A}\cdot d\vec{x} $$
We can set up a similar situation as to before. 

Starting off, the beam of electrons has wavefunction $\psi_0$ in the absence of an external magnetic potential. It is split into two, and both branches travel through a region of potential $\vec{A}$. Each wavefunction gains the form of \eqref{eq:wavefunction_in_AB_exp_2} except we replace $\exp[-(ie/\hbar)\int \phi dt]$ with $\exp[-(ie/\hbar c) \int \vec{A}\cdot d\vec{x} ]$ as described above. In other words:

\begin{equation}\label{eq:wavefunction_in_AB_exp_3}
\psi = \psi_0 \exp[-\frac{ie}{\hbar c}\int \vec{A}\cdot d\vec{x}].
\end{equation}
The branching could be represented as $\psi_1^0$ and $\psi_2^0$ upon reaching the area with magnetic potential at time $t_1$. After some evolution they leave the potential at time $t_2$, and they are then recombined. You can write the following superposition:

$$ \psi_f = \psi_1 + \psi_2 = \psi_1^0 \exp[-\frac{i}{\hbar}C_1] + \psi_2^0 \exp[-\frac{i}{\hbar}C_2],$$
where we define

$$ C_1 = \frac{e}{c}\int_{\vec{x}(t_1)}^{\vec{x}(t_2)} \vec{A}\cdot d\vec{x}',$$
and

$$ C_2 = -\frac{e}{c}\int_{\vec{x}(t_1)}^{\vec{x}(t_2)} \vec{A}\cdot d\vec{x}',$$
where we assume equal path length for each branch. If the path length is indeed kept the same, then you will have phase difference $(C_1 - C_2)/\hbar = (e/c \hbar) \int \vec{A}\cdot d\vec{x} $ in the end (\cite{AB}). Note that this is true even if the magnetic force $\vec{F} = q\vec{v}\times\vec{B}$ is zero because $\vec{B}=\vec{0}$. This is a purely quantum effect. 

The electric Aharonov-Bohm effect described above has been theorized, but no experimental data exists for it as of yet. We will focus on the magnetic potential therefore. No further space will be dedicated to the electric potential, but all considerations from here on out will be framed in the context of the magnetic potential. 

One may ask the obvious question, in the descriptions above, if all possible paths are taken in the Path Integral, would not some paths go through the magnetic field, explaining the outcome of the experiment? Since the Feynman Path Integral defines a set starting point and endpoint in a set time period, it is actually possible to show that you can set up a barrier (described in \cite{AB}) so that no wavefunction directly interacts with the inside of the solenoid in the magnetic case. The changes we see, therefore, must be due to the nonzero magnetic potential instead of the magnetic field itself.

KvNM, having the mathematical form of quantum theory, might be one of the best tools to show that the Aharonov-Bohm effect does not exist in CM (\cite{kvn_thesis}). This is a purely quantum phenomenon. Interference is not something expected in the domain of CM, but is expected under QM. This can be traced to the behavior of the generator of motion, the quantum Hamiltonian, having the following form under QM in the presence of the magnetic potential for a free particle:

$$ \hat{H} = \frac{1}{2m}\Big(\hat{p} - \frac{e}{c}\vec{A} \Big)^2$$
In the presence of the magnetic potential, the free particle Hamiltonian has the $\hat{p}$ operator replaced with $\hat{p} - (e/c)\vec{A}$. This is described as a \textit{minimal coupling rule} (\cite{minimal_coupling}). If KvNM describes CM precisely, we should expect that the generator of motion, the Koopman Generator \eqref{eq:koopman_generator}, will not be affected by the magnetic potential even if we treat it mathematically in the exact same ways as we treat the quantum Hamiltonian. In the following pages, we will put this to the test. 

First, imagine two cylinders, one within the other (following \cite{minimal_coupling}; \cite{kvn_thesis}). Let them be infinite in height. The smaller cylinder on the inside will represent the solenoid which houses a magnetic field. The outside of the solenoid (area inside the larger cylinder but outside of the smaller cylinder) will have no magnetic field, but there will be a magnetic potential. Since we are dealing with cylinders, it would be useful to rewrite the generators of motion in terms of cylindrical coordinates. The conversion of cartesian to cylindrical coordinates is given by:

 $$ x = r \cos(\theta),$$
$$y = r \sin(\theta),$$
and
\begin{equation}\label{eq:cart_to_coord}
z =z,
\end{equation}
where $r$ is the distance from the $z$ axis and $\theta$ is the angle in the $xy$ plane.

With this notation, the quantum Hamiltonian for a free particle can be written:

$$ \hat{H} = -\frac{\hbar^2}{2m}\vec{\nabla}^2 = -\frac{\hbar^2}{2m}\Big(\frac{\partial^2}{\partial x^2}+\frac{\partial^2}{\partial y^2}+\frac{\partial^2}{\partial z^2} \Big) = -\frac{\hbar^2}{2m}\Big(\frac{\partial^2}{\partial r^2}+\frac{1}{r}\frac{\partial}{\partial r}+\frac{1}{r^2}\frac{\partial^2}{\partial \theta^2}+\frac{\partial^2}{\partial z^2} \Big).$$
This is now in cylindrical coordinates. The eigenvalue problem we want to solve is:

$$\hat{H}\psi(r,\theta,z) = E\psi(r,\theta,z).$$
where $E$ are the energy eigenvalues. The above expression is often termed the time independent Schr\"{o}dinger equation, as it uses the quantum Hamiltonian to extract energy values with no explicit time derivative. 

Since the operators $\partial/\partial \theta$ and $\partial/\partial z$ commute with the above form of $\hat{H}$, we can choose solutions of the form (\cite{minimal_coupling}):

\begin{equation}\label{eq:wavefunction_AB}
\psi(r,\theta,z) = \frac{1}{2\pi}\exp[\frac{i p_z^0 z}{\hbar}]\exp[in\theta]R(r),
\end{equation}
where $p_z^0$ is a constant and $n$ is an integer. Plugging in this wavefunction into the time independent Schr\"{o}dinger equation will give you solutions in the form of Bessel Functions:

$$ \frac{d^2 R(r)}{dr^2} + \frac{1}{r}\frac{dR(r)}{dr} + \Big(\frac{2mE}{\hbar^2}-\frac{{p_z^0}^2}{\hbar^2}-\frac{n^2}{r^2} \Big)R(r) = 0.$$
Bessel Functions are a class of functions that govern cylindrical harmonics. 

If you turn on the magnetic potential, we will now have to solve
$$\hat{H}_A\psi(r,\theta,z)_A = E_A \psi(r,\theta,z)_A.$$
In order to achieve this, we have the following minimal coupling rule in cylindrical coordinates:
\begin{equation}\label{eq:mc}
 \hat{p}_\theta = -i\hbar \frac{\partial}{\partial \theta} \Rightarrow \hat{p}_\theta  - \frac{e}{c}\frac{\Phi}{2\pi} = -i\hbar \frac{\partial}{\partial \theta} - \frac{e}{c}\frac{\Phi}{2\pi}
\end{equation}
The quantum Hamiltonian therefore becomes:
$$ \hat{H}_A = -\frac{\hbar^2}{2m}\Big(\frac{\partial^2}{\partial r^2}+\frac{1}{r}\frac{\partial}{\partial r}+\frac{1}{r^2}\Big(\frac{\partial}{\partial \theta} - \frac{ie}{ch}{\Phi}\Big)^2+\frac{\partial^2}{\partial z^2} \Big),$$
where $\Phi$ is the flux (constant) through an area perpendicular to the axis of the smaller cylinder. This area will be larger than the smaller cylinder's cross-section.
Notice that $\hat{p}_\theta $ is not present in the wavefunction, so the form of \eqref{eq:wavefunction_AB} will remain unaltered. The energy values in the presence of the magnetic potential, as you will see, will be different. 

As before, substituting \eqref{eq:wavefunction_AB} into the magnetic potential version of the time independent Schr\"{o}dinger equation will give you the following Bessel function

$$ \frac{d^2 R(r)}{dr^2} + \frac{1}{r}\frac{dR(r)}{dr} + \Big(\frac{2mE}{\hbar^2}-\frac{{p_z^0}^2}{\hbar^2}-\frac{(n-\alpha)^2}{r^2} \Big)R(r) = 0,$$
where $\alpha = \frac{e\Phi_B}{ch}$. If you compare the two Bessel functions, you can immediately see that the presence of the magnetic potential shifts the function's behavior. For the quantum case, the behavior of the system is impacted by the presence of the vector potential. 

Now we will do the same procedure for $\hat{L}$ as the generator of motion \eqref{eq:L_operator}. For the free particle in 3 dimensions, we know that
$$ \hat{L} = -i \frac{p_x}{m}\frac{\partial}{\partial x} -i \frac{p_y}{m}\frac{\partial}{\partial y} -i \frac{p_z}{m}\frac{\partial}{\partial z},$$
based off of \eqref{eq:L_operator}. Just as in the quantum case, we want this to be in cylindrical coordinates. We will therefore need to convert the cartesian momenta $p_x$, $p_y$, and $p_z$ into $p_r$, $p_\theta$, and $p_z$. The latter are the canonical momenta, which can be calculated directly from the Lagrangian $L$.\footnote{Do not confuse the Lagrangian $L$ with the operator $\hat{L}$. These are two different mathematical entities.} As before, any cartesian coordinates can be converted to cylindrical coordinates via \eqref{eq:cart_to_coord}, which we will make use of. 

To start off, we write down the Lagrangian of motion. Recall from section 4.1, the Lagrangian is just the kinetic energy minus the potential energy, $L = T - V$. Since the potential is zero, the Lagrangian is just equal to the total kinetic energy, so that

\begin{equation}\label{eq:lagrangian_1}
L = T = \frac{1}{2}m(\dot{x}^2 + \dot{y}^2 + \dot{z}^2),
\end{equation}
where we use the standard dot notation, so that any variable $\gamma$ has a time derivative $\dot{\gamma} = d\gamma/dt$. We can take time derivatives of \eqref{eq:cart_to_coord} to get:

 $$ x = r \cos(\theta) \Rightarrow \dot{x} = \dot{r}\cos(\theta) - r \sin(\theta)\dot{\theta},$$
$$y = r \sin(\theta) \Rightarrow \dot{y} = \dot{r}\sin(\theta) + r \cos(\theta)\dot{\theta},$$
with the $z$ coordinate unaffected. Plugging these terms into \eqref{eq:lagrangian_1}, we get our Lagrangian in cylindrical coordinates:
$$ L = \frac{1}{2}m\dot{r}^2 + \frac{1}{2}m r^2 \dot{\theta}^2 + \frac{1}{2}m\dot{z}^2 $$
Recalling from Lagrangian Dynamics that the canonical momentum associated with coordinate $\sigma$ is given by
$$ p_\sigma = \frac{\partial L}{\partial \dot{q_\sigma}} $$
Therefore applying the above formula for coordinates $r$, $\theta$, and $z$, we have:
$$ p_r = \frac{\partial L}{\partial \dot{r}} =  m\dot{r}$$
$$ p_\theta = \frac{\partial L}{\partial \dot{\theta}} = mr^2 \dot{\theta}$$
$$ p_z = \frac{\partial L}{\partial \dot{z}} = m\dot{z}$$
If we apply the same rule to the cartesian Lagrangian \eqref{eq:lagrangian_1} we get:
$$ p_x = \frac{\partial L}{\partial \dot{x}} =  m\dot{x},$$
$$ p_y = \frac{\partial L}{\partial \dot{y}} = m\dot{y},$$
with the $z$ coordinate momentum unaltered. Using all the above information, we can rewrite the cartesian momenta in terms of the cylindrical momenta, using the momenta expressions above:
$$ p_x =  m\dot{x} = m[\dot{r}\cos(\theta) - r \sin(\theta)\dot{\theta}] = p_r \cos(\theta) - \frac{1}{r} p_\theta \sin(\theta),$$
$$ p_y = m[\dot{r}\sin(\theta) + r \cos(\theta)\dot{\theta}] = p_r \sin(\theta) + \frac{1}{r} p_\theta \cos(\theta). $$
We therefore can rewrite the generator of motion $\hat{L}$ using the above information in terms of cylindrical coordinates:

\begin{equation}\label{eq:hat_L_operator_cyl}
 \hat{L} = -i \frac{p_x}{m}\frac{\partial}{\partial x} -i \frac{p_y}{m}\frac{\partial}{\partial y} -i \frac{p_z}{m}\frac{\partial}{\partial z} = -i \frac{p_r}{m}\frac{\partial}{\partial r} - i \frac{p_\theta}{mr^2}\frac{\partial}{\partial \theta} -i \frac{p_z}{m}\frac{\partial}{\partial z} - i \frac{p_\theta^2}{mr^3}\frac{\partial}{\partial p_r}. 
\end{equation}
For the case with no magnetic potential, we will be solving the analogue to the time independent Schr\"{o}dinger equation, 
$$ \hat{L}\psi = \tilde{E} \psi,$$
where $\psi$ is the KvN classical wavefunction and $\tilde{E}$ is the eigenvalue of the above expression. Note that $\tilde{E}$ is certainly \textit{not} the same as the Energy like in the Schr\"{o}dinger equation. We observe that $\hat{L}$ commutes with $-i\frac{\partial}{\partial \theta}$, $-i\frac{\partial}{\partial z} $, $p_\theta$, and $p_z$. We can therefore deduce that the wavefunction will have the form

\begin{equation}\label{eq:KvN_wavefunction_magnetic_potential}
\psi = \frac{1}{2\pi}R(r,p_r)\delta(p_\theta - p_\theta^0)\delta(p_z - p_z^0) \exp[in\theta]\exp[i\lambda_z^0 z] .
\end{equation}
If we plug this wavefunction into \eqref{eq:hat_L_operator_cyl}, we will get (\cite{minimal_coupling}):

\begin{equation}\label{eq:mag_potential_equation_1}
 \Big( - \frac{i}{m}p_r \frac{\partial}{\partial r} + \frac{p_\theta^0 n}{mr^2}-\frac{i {p_\theta^0}^2}{mr^3}\frac{\partial}{\partial p_r} + \frac{\lambda_z^0 p_z^0}{m} - \tilde{E} \Big)R(r, p_r) = 0.
\end{equation}
We have a similar expression for the magnetic potential case:
$$ \hat{L}_A\psi_A = \tilde{E}_A \psi_A$$
Just as in the quantum case, we will utilize the minimal coupling rule \eqref{eq:mc}. The only difference is that the representation of the operator $\hat{p}_\theta$ is no longer going to be a derivative with respect to $\theta$, and instead we will have $\hat{p}_\theta = p_\theta$ for the KvNM case (\cite{minimal_coupling}). Making the substitution 

\begin{equation}
 \hat{p}_\theta  \Rightarrow \hat{p}_\theta  - \frac{e}{c}\frac{\Phi}{2\pi} ,
\end{equation}
we will get the following two expressions:

\begin{equation}\label{eq:L_operator_modified_with_A}
 \hat{L} = -i \frac{p_r}{m}\frac{\partial}{\partial r} - i \frac{1}{mr^2}\Big({p}_\theta  - \frac{e}{c}\frac{\Phi}{2\pi} \Big)\frac{\partial}{\partial \theta} -i \frac{p_z}{m}\frac{\partial}{\partial z} - i \frac{1}{mr^3}\Big( {p}_\theta  - \frac{e}{c}\frac{\Phi}{2\pi}\Big)^2 \frac{\partial}{\partial p_r},
\end{equation}

\begin{equation}\label{eq:KvN_wavefunction_magnetic_potential_2}
\psi = \frac{1}{2\pi}R(r,p_r)\delta(p_\theta - p_\theta^0 - \frac{e\Phi}{2\pi c})\delta(p_z - p_z^0) \exp[in\theta]\exp[i\lambda_z^0 z] .
\end{equation}
Plugging \eqref{eq:KvN_wavefunction_magnetic_potential_2} into \eqref{eq:L_operator_modified_with_A} we will get:

\begin{equation}\label{eq:KvN_outcome_1}
\Big( - \frac{i}{m}p_r \frac{\partial}{\partial r} + \frac{p_\theta^0 n}{mr^2}-\frac{i {p_\theta^0}^2}{mr^3}\frac{\partial}{\partial p_r} + \frac{\lambda_z^0 p_z^0}{m} - \tilde{E}_A \Big)R(r, p_r) = 0 
\end{equation}
If you compare \eqref{eq:KvN_outcome_1} (case with magnetic potential) to \eqref{eq:mag_potential_equation_1} (case without magnetic potential) you will notice they happen to be the \textit{same} equation. This shows us that the same solutions will exist for classical mechanics regardless of the presence or absence of the gauge potential. As demonstrated, the gauge potential will however affect quantum systems (\cite{minimal_coupling}; \cite{kvn_thesis}). 

On a philosophical level, the above analysis raises the question what do the gauge potentials physically mean? Historically, it was simply seen as a mathematical tool or convenience in the study of Electromagnetism. The electromagnetic field and its potentials were fully equivalent descriptions of reality, but the potentials were only seen as a mathematical construct with useful properties. This might be a historic example of physicists not taking the mathematics of their theory \textit{seriously enough}. For example, Albert Einstein famously invented the theories of special relativity by acknowledging that the speed of light $c$ is the same in all noninertial reference frames. This is a fact tacitly present in Maxwell's equations, but physicists at the time did not treat the mathematics seriously enough and thought the equations must only be true relative to a hypothetical `fluid' filling all of space and time, the aether, an idea Einstein showed to be wrong. \cite{AB} say in their paper ``a further interpretation of the potentials is needed in the quantum mechanics", so perhaps this is an area for further scientific investigation and discovery in the future.

\section{Other Research}
Wavefunctions in a complex Hilbert phase space can exactly model CM in the form of statistical physics.
Both QM and CM can be articulated in a common mathematical language utilizing Hermitian operators to recover expectation values of observables. In the pages above we covered a host of different applications of the KvN wavefunction inside classical physics. We hardly exhausted the possible applications of KvNM, however. There are many areas of research we have not even touched upon.

Some recent examples of the widespread applicability of the KvNM formulism:
\begin{enumerate}
\item Why the usage of finite-dimensional Hilbert Spaces in QM and KvNM is often ill-advised 

A recent trends in QM research has been to try utilize QM with finite dimensions instead of the regular infinite Hilbert space. This is done to aid in classical computer modeling of quantum systems and an interest in the possibility of rooting QM in a finite instead of an infinite space. For instance, approximations of the quantum Hamiltonian tend to be finite matrices (\cite{finite_hilbert_space}). \cite{finite_hilbert_space} demonstrate that this will unavoidably violate the Ehrenfest Theorems. This is a serious shortcoming of finite dimensional treatments of QM, but there are a small number of ways that the issues might be circumvented (\cite{finite_hilbert_space}). 

\item KvNM has been used to study aspects of String Theory and the Gravitational Principle of Equivalence

The KvN Path Integral is utilized in \cite{string_theory} to study the properties of Classical Yang-Mills Theories (a subset of String Theories.) The Path Integral provides geometrical tools like exterior derivatives, forms, etc. The Classical Path Integral might shed light on the geometrical aspects of these string theories (\cite{string_theory}). \cite{free_fall} investigates the validity of Einstein's principle of mass equivalence for both classical KvNM and quantum systems. Fisher information is discussed and how it relates to insights from KvNM and QM. It is demonstrated that the weak principle of equivalence holds in QM just as it does in KvNM (\cite{free_fall}). 

\item ODM can be used to describe relativistic particles

The ODM paradigm is upheld in relativistic settings. The following two Ehrenfest-like theorems were postulated (\cite{relativistic_kvn}) based on relativistic principles

$$\frac{d}{dt} \langle \hat{x}^k \rangle = \langle c\gamma^0 \gamma^k \rangle .$$
$$ \frac{d}{dt}\langle \hat{p}_k \rangle = \langle ce\partial_k \hat{A}_\nu \gamma^0 \gamma^\nu \rangle.$$
If the quantum commutator is chosen, you can follow a series of steps similar to those in section 3 to derive the famous Dirac equation, which governs relativistic quantum particles (\cite{relativistic_kvn}). If the Koopman Algebra is chosen and you constrain the system from making antiparticles, the same series of steps as deriving the KvNM will give you the classical Spohn equation (\cite{relativistic_kvn}). This is quite a surprising result - ODM can study spin 1/2 relativistic particles.

\item ODM has been used to study Time's Arrow and stochastic behavior from determinism

A lot of processes are reversible in physics, but our intuitions of the world are that it is irreversible and things tend towards decay. How does irreversibility arise out of the reversible? \cite{time_arrow} utilizes the KvN Path Integral to formulate a functional model for how irreversibility arises out of the reversible laws we have. 

\item Reformulation of Electromagnetism

\cite{EM} provides a reformulation of electromagnetic theory that draws inspiration from the KvN theory. 

\item KvN wavefunctions were utilized to understand the interaction of classical with quantum systems  

The problem of modeling quantum and classical system interaction is an old one. Utilizing the KvN classical wavefunction, however, had led to new models of how the interaction occurs. Not only does it describe the classical system's effect on the quantum system, but uniquely it models how the quantum system will impact the classical system in return (\cite{coupled_systems}). It also overcomes difficulty in some other models. For instance, the quantum density matrix will always be positive-definite in this model (\cite{coupled_systems}). 

\end{enumerate}
Among other areas of inquiry. Doubtlessly, many more areas of research that will benefit from KvNM and a paradigm like ODM. 

It is important to have many ``tools”  in science when investigating novel phenomena. Feynman famously said that the reason he had solved problems others could not, is that his toolbox had a different set of tools than everyone else. KvNM is yet an another novel tool for you, the reader, to have in your toolkit. KvNM has and will continue to shed light on new and exciting phenomena in science.  

\newpage
\printbibliography[heading=bibintoc]
\newpage

\begin{appendices}
\section{Useful Mathematics for Hilbert Space Physics} 
Note: A lot of the following useful mathematics was found in the Appendix of \cite{ODM}. 
\newline

\textit{Theorem I}: If you have a function of operator $\hat{A}$, i.e. $f(\hat{A})$, then the following eigenvalue expressions are true:
$$f(\hat{A})\ket{A} = f(A)\ket{A}$$
$$\bra{A}f(\hat{A}) = \bra{A}f(A)$$
This should purposefully resemble the prior eigenvalue problem you are familiar with, $\hat{A}\ket{A} = A\ket{A}$. This theorem is true for multivariable operators, for instance, $$A(\hat{q},\hat{p})\ket{q,p} = A(q,p)\ket{q,p}$$.
\newline\newline
\textit{Theorem II (Resolution of Identity)}: The following theorem says that 
$$I = \int dA \ket{A}\bra{A}$$
where $I$ is the identity operator, functionally equivalent to multiplying an object by 1. Since it is just equal to 1, you can introduce the Resolution of Identity in a variety of calculations. As always $A$ is an arbitrary placeholder variable, and you can make it any variable of interest.
\newline\newline
\textit{Theorem III (Cauchy-Schwartz Inequality)}: 
It simply states that the following is always true for complex vectors in Hilbert Space:
$$\braket{C}{C}\braket{D} \ge |\braket{C}{D}|^2 $$
where $C$ and $D$ are generic placeholder variables. 
\newline\newline
\textit{Theorem IV}:
If the commutator relationship between two generic operators $\hat{A}$ and $\hat{B}$ is $[\hat{A}, \hat{B}] = i\kappa$, then the following statements hold true:
$$\bra{A}\hat{B}\ket{\psi} = - i\kappa \frac{\partial}{\partial A} \braket{A}{\psi}$$
$$\bra{B}\hat{A}\ket{\psi} =  i\kappa \frac{\partial}{\partial B} \braket{B}{\psi}$$
Recall that $\braket{B}{\psi} = \psi(B)$ and $\braket{A}{\psi} = \psi(A)$.
\newline\newline
\textit{Theorem V}: If the commutator relationship between two generic operators $\hat{A}$ and $\hat{B}$ is $[\hat{A}, \hat{B}] = i\kappa$, then the following statements hold true:
\begin{equation}\label{eq:intrig_eq_1}
\braket{A}{B} = \frac{1}{\sqrt{2\pi\kappa}}e^{iAB/\kappa} 
\end{equation}
This is not difficult to demonstrate. First, take Theorem IV and realize that $\ket{\psi}$ is a general vector that can be substituted for a more specific vector. This is a common practice in QM. Let us say that $\ket{\psi} = \ket{B}$. Then by Theorem IV we have:
$$ \bra{A}\hat{B}\ket{B} = B\braket{A}{B}= - i\kappa \frac{\partial}{\partial A} \braket{A}{B}$$
You can solve this common differential equation for the function $\braket{A}{B}$ to get:

$$\braket{A}{B}=Ce^{iAB/\kappa}$$
where $C$ is a constant of integration. $C$ can be found in the following way: Start with equation \eqref{eq:dirac_delta_1}

$$\braket{A}{A'} = \delta(A-A')$$
and use Theorem II with $A = B$ \footnote{In other words, since $A$ is just a generic placeholder for Theorem II, we will use $I = \int dB \ket{B}\bra{B}$} to get:
$$\braket{A}{A'} = \delta(A - A') = \bra{A}I\ket{A'} = \int dB \bra{A} \ket{B}\bra{B}\ket{A'}$$ 
Since we know what  $\braket{A}{B}$ is, we plug it in to get:
$$\braket{A}{A'} = \delta(A - A') = \int dB [Ce^{iAB/\kappa}][Ce^{-iA'B/\kappa}] = C^2 \int dB e^{i(A-A')B/\kappa}$$ 
From equation \eqref{eq:C} of section 2, we can easily see that
$$\int dB e^{i(A-A')B/\kappa} = 2\pi\kappa \delta(A - A')$$
Ergo,
$$\delta(A - A') =  C^2 2\pi\kappa \delta(A - A')$$
$$C^2 2\pi\kappa = 1$$
$$C^2 = \frac{1}{ 2\pi\kappa}$$
$$C = \frac{1}{ \sqrt{2\pi\kappa}}$$
We now know $C$ and have successfully demonstrated equation \eqref{eq:intrig_eq_1}.
\newline\newline
\textit{Theorem VI}: 
As strange as it sounds, the mathematics of Hilbert Space defines derivatives with respect to \textit{operators}. If you have $[\hat{A}_k, \hat{B}] = Constant$, then a function of operators $f(\hat{A}_1, \hat{A}_2, ..., \hat{A}_n)$ will obey the following rule:
$$[f(\hat{A}_1, \hat{A}_2, ..., \hat{A}_n), \hat{B}] = \sum_{k = 1} ^n  [\hat{A}_k, \hat{B}] \frac{\partial f(\hat{A}_1, \hat{A}_2, ..., \hat{A}_n)}{\partial \hat{A}_k}$$
Notice the partial derivative with respect to an operator. The next theorem shows how to change this odd operator derivative to regular derivatives that we know and love. 
\newline\newline
\textit{Theorem VII}
The following is the relationship between derivatives with respect to operators and the regular derivatives we are most generally familiar with:
$$ \frac{\partial f(\hat{A}_1, \hat{A}_2, ..., \hat{A}_n)}{\partial \hat{A}_k} = \frac{1}{(2\pi)^n} \int \prod_{l = 1} ^n d\epsilon_l ~d\mu_l ~ exp[i\sum_{q =1} ^ n \mu_q (\hat{A}_q - \epsilon_q)]\frac{\partial f(\epsilon_1, \epsilon_2,..., \epsilon_n)}{\partial \epsilon_k}$$
where on the left you can see the derivative with respect to an operator and on the right you can see the derivative with respect to a scalar variable. 
\newline\newline
\textit{Theorem VIII:}
An integral of a Gaussian function has the following form:

\begin{equation*}
 \int_{-\infty} ^{\infty} e^{-ax^2 + bx +c} dx = \sqrt{\frac{\pi}{a}}e^{\frac{b^2}{4a}+c}
\end{equation*}
where $a$, $b$, and $c$ can be real or imaginary constants.

\section{The Principle of Causality and Stone's Theorem}

A unitary operator $\hat{U}(t_2, t_1)$ always has the following properties:

\begin{enumerate}
\item Group product: $\hat{U}(t_3,t_2)\hat{U}(t_2, t_1) = \hat{U}(t_3, t_1)$
\item Reversibility: $\hat{U}(t_2, t_1)^\dagger = \hat{U}(t_2, t_1)^{-1} = \hat{U}(t_1, t_2)$
\item Identity: $\hat{U}(t_1, t_1) = 1$
\end{enumerate}
The job of the unitary operator $\hat{U}(t_2, t_1)$ is to bring you from $t_1$ to $t_2$. The group product encapsulates the concept of cause and effect, as you move from cause to effect through time. Causality is central to the philosophy of science, and we should not be surprised that it pops up inside the mathematics of physics. 

The three properties of the unitary operator above imply Stone's Theorem:
\newline\newline
If the three properties hold true, then the following relationship is also true:
$$i\frac{\partial}{\partial t} \hat{U}(t_2, t_1) = \hat{G}\hat{U}(t_2, t_1)$$
where $\hat{G}$ is a Hermitian operator. If we assume that the wavefunction ket evolves via the unitary operator

$$\ket{\psi(t_2)} = \hat{U}(t_2, t_1)\ket{\psi(t_1)},$$ 
then we recover the Schr\"{o}dinger-like equation we spoke of in earlier sections:

$$i\frac{\partial}{\partial t} \hat{U}(t_2, t_1)\ket{\psi(t_1)} = \hat{G}\hat{U}(t_2, t_1)\ket{\psi(t_1)} \Rightarrow i\frac{\partial}{\partial t} \ket{\psi(t_2)} = \hat{G}\ket{\psi(t_2)} $$
This is the reason why Stone's Theorem becomes an important ingredient in section 3.

\section{Exponated Operators and the Propogator}
Although it might seem odd at first, exponated operators are allowed within Quantum Mechanics. 
The Maclaurin series for the $f(x) = e^x$ is given by:

$$ e^x = \sum_{n=0}^\infty \frac{x^n}{n!}= 1 + \frac{1}{1!}x + \frac{1}{2!}x^2 + \frac{1}{3!}x^3 + \frac{1}{4!}x^4 + \cdot\cdot\cdot $$
Let us say that you have a generic operator $\hat{A}$, like a square matrix. As long as $\hat{A}^n$ is defined for any real number $n$, then you can say:

$$ e^{\hat{A}} =\sum_{n=0}^\infty \frac{\hat{A}^n}{n!}= 1 + \frac{1}{1!}\hat{A} + \frac{1}{2!}\hat{A}^2 + \frac{1}{3!}\hat{A}^3 + \frac{1}{4!}\hat{A}^4 + \cdot\cdot\cdot $$
The Maclaurin series defines exponated operators as an infinite sum. 

If we have an expression such as $e^{\hat{A}t}$ where $\hat{A}$  is time-independent, we can also derive a derivative rule for it:

$$ e^{\hat{A}t} =\sum_{n=0}^\infty \frac{(\hat{A}t)^n}{n!}= 1 + \frac{1}{1!}\hat{A}t + \frac{1}{2!}(\hat{A}t)^2 + \frac{1}{3!}(\hat{A}t)^3 + \frac{1}{4!}(\hat{A}t)^4 + \cdot\cdot\cdot $$
 \newline
$$ \frac{d}{dt}e^{\hat{A}t} = \frac{d}{dt} 1 + \frac{d}{dt}\frac{1}{1!}\hat{A}t + \frac{d}{dt}\frac{1}{2!}(\hat{A}t)^2 + \frac{d}{dt}\frac{1}{3!}(\hat{A}t)^3  + \cdot\cdot\cdot =$$
$$ = \frac{1}{1!}\hat{A} + \frac{1}{2!}\hat{A}^2 (2t)+ \frac{1}{3!}\hat{A}^3 (3t^2)  + \frac{1}{4!}\hat{A}^4 (4t^3)+ \cdot\cdot\cdot  = \hat{A}\Big[1 + \frac{1}{2!}\hat{A} (2t)+ \frac{1}{3!}\hat{A}^2 (3t^2)  + \frac{1}{4!}\hat{A}^3 (4t^3)+\cdot\cdot\cdot \Big] = $$
$$  = \hat{A}\Big[1 + \frac{1}{2 \cdot 1}\hat{A} (2t)+ \frac{1}{3 \cdot 2 \cdot 1}\hat{A}^2 (3t^2)  +\cdot\cdot\cdot \Big] =  \hat{A}\Big[ 1 + \frac{1}{1!}\hat{A}t + \frac{1}{2!}(\hat{A}t)^2 + \frac{1}{3!}(\hat{A}t)^3 + \cdot\cdot\cdot \Big] =  \hat{A}e^{\hat{A}t}$$
\newline
\begin{equation}\label{eq:derivative_relationship_exp_operator}
\therefore \frac{d}{dt}e^{\hat{A}t} = \hat{A}e^{\hat{A}t}
\end{equation}
The above relation is true even if a real or complex constant is placed in front of the exponated term $\hat{A}t$. You can redo the exercise to convince yourself of this. We will use this important fact in order to demonstrate the important relationship \eqref{eq:propogator_relationship} is fully equivalent to the Schr\"{o}dinger equation:
\begin{equation*}
\ket{\psi(t)} = e^{-\frac{i}{\hbar}t\hat{H}}\ket{\psi(0)}
\end{equation*}
\begin{equation*}
i\hbar\frac{\partial}{\partial t}\ket{\psi(t)} = i\hbar\frac{\partial}{\partial t} e^{-\frac{i}{\hbar}t\hat{H}}\ket{\psi(0)} = i\hbar\frac{-i}{\hbar}\hat{H} e^{-\frac{i}{\hbar}t\hat{H}}\ket{\psi(0)} = \hat{H}\ket{\psi(t)}
\end{equation*}
\begin{equation*}
\therefore i\hbar\frac{\partial}{\partial t}\ket{\psi(t)} = \hat{H}\ket{\psi(t)}
\end{equation*}
A similar derivation can be done for \eqref{eq:koopman}, to demonstrate that it is fully equivalent to Stone's Theorem (Appendix B). 

\end{appendices}

\end{document}